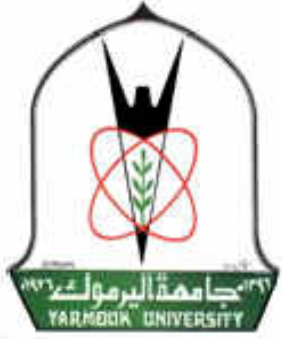

Yarmouk University
Hijjawi Faculty for Engineering Technology
Telecommunication Engineering Department

# "A DYNAMIC MULTI-CAST ROUTING ALGORITHM FOR OPPORTUNISTIC NETWORKS: IMPLEMENTING THE EXPECTED TRANSMISSION COUNT METRIC"

By
Rasha Ziyad Abu Samra

Advisor
Dr. Haythem Bany salameh

May, 2016

# "A Dynamic Multi-cast Routing Algorithm for Opportunistic Networks: Implementing the Expected Transmission Count Metric"

By

**Rasha Ziyad Abu Samra**

Thesis Submitted to the Department of Telecommunications Engineering in partial fulfillment of the requirements for the degree of Master of Science

At

Hijjawi Faculty for Engineering Technology

Yarmouk University

May, 2016

| Committee Member | | Signature and Date |
|---|---|---|
| Dr. Haythem Bany salameh | (Chairman) | 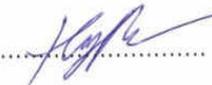 |
| Dr. Bassam Harb | (Member) | 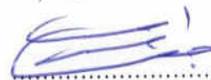 |
| Dr. Yaser Jararweh | (External Examiner) | 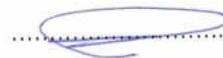 |

# Dedication

*Every challenging work needs self-effort as well as guidance of elders especially those who were every close to our heart. My humble effort I dedicated to my sweet and loving*

*Father & Mother,*

*And my beloved brothers and sisters,*

*Who are affection, loving, and encouragement me, to make me able to get such success and honor.*

*Along with all hard working and respected*

*Dr. Haythem Bany Salameh*

# Acknowledgment

I would like to express my special thanks and grateful to my thesis advisor Dr. Haythem Bany salameh who gave me the golden opportunity to do this thesis and, allowed this study to be my own work. Also, he helped and motivated me through sharing his knowledge with me, for his grace advice, encouragement, and patience.

I express my appreciation and very profound thanks to my family and to my dear friends for providing me with unfailing support and encouragement throughout my years of study and during the process of researching and writing this thesis.

I would also thanks to all of Hijjawi faculty members for their knowledge shared and taught.

This achievement would not have been possible without them. Thank you.

# Declaration

I'm **Rasha Abu-Samra**, recognize what plagiarism is and I hereby declare that this thesis, which is submitted to the department of **Telecommunications Engineering** at **Hijjawi Faculty for Engineering Technology**, for the partial fulfillment of the requirements for the degree of Master of Science, is my own work. I have not plagiarized from any sources. All references and acknowledgments of sources are given and cited in my thesis. I have used the conventional citation and referencing. Each significant contribution to and quotation in this report from work of other people has been attributed and referenced.

*Rasha Ziyad Mohammad Abu-Samra*

*May, 2016*

# Table of Contents

# Table of Figures

# List of Tables

# List of Abbreviations

| | |
|---|---|
| **AP** | **Access point** |
| **BER** | **Bit Error Rate** |
| **BW** | **Bandwidth** |
| **CCC** | **Common Control Channel** |
| **CR** | **Cognitive Radio** |
| **CRN** | **Cognitive Radio Network** |
| **CUs** | **Cognitive Users** |
| **DSA** | **Dynamic Spectrum Access** |
| **ETX** | **Expected Transmissions Count** |
| **FCC** | **Federal Communication Commission** |
| **GPS** | **Global Positioning System** |
| **IP** | **Internet Protocol** |
| **MAC** | **Media Access Control** |
| **MANETs** | **Mobile Ad hoc Networks** |
| **MASA** | **Maximum Average Spectrum Availability** |
| **MDR** | **Maximum Data Rate** |
| **MinTT** | **Minimum Transmission Time** |
| **MNs** | **Mobile Nodes** |
| **MST** | **Minimum Spanning Tree** |
| **NP** | **Non-deterministic Polynomial-time** |
| **PDA's** | **Personal Digital Assistances** |
| **PDR** | **Packet Delivery Rate** |
| **$P_I$** | **Idle Probability** |
| **POS** | **Probability of Success** |
| **PR** | **Primary Radio** |
| **PRNs** | **Primary Radio Networks** |
| **PU** | **Primary User** |
| **RF** | **Radio Frequency** |
| **RS** | **Random Selection** |
| **SDR** | **Software Defined Radio** |
| **SPT** | **Shortest Path Tree** |
| **SU** | **Secondary User** |

# Abstract

**A Dynamic Multi-cast Routing Algorithm for Opportunistic Networks: Implementing the Expected Transmission Count Metric.**
**By:  Rasha Ziyad Abu-Samra (2013976009).**
**Advisor: Dr.  Haythem Bany salameh.**

Cognitive radio (CR) technology enables an intelligent wireless communication system. CR provides an efficient solution for the inefficient spectrum utilization by allowing dynamic and opportunistic spectrum access. In designing CR networks, the main challenge is how to increase network throughput while protecting the performance of licensed primary radio networks (PRNs) and keeping the interference between primary users (PUs) and cognitive users (CUs) within a prescribed threshold. In this work, we develop a multi-cast routing algorithm that based on the expected transmission count metric (ETX), which implemented as a metric combined with minimum spanning tree (MST) and shortest path tree (SPT) schemes according to the various traffic loads in CRN to determine the path selection method and used the probability of Success (POS) metric for the channel assignment that used the required transmission time and the channel availability time in choosing unified channel. The main objective of our algorithm is to reduce the total number of the expected packet transmissions (with retransmissions) needed for successfully forwarding a data packet to a specific group of destinations and provide guarantees on the chances of a successful transmission over a given channel. This metric is capable to capture the CRNs environment, in which the channel availabilities are diversity and dynamically changing due to the dynamic and uncertainty activity of PUs. Specifically, a dynamic multi-cast routing protocol is proposed to maximize CR network (CRN) throughput by minimizing the required transmission time on multi-layer multi-hop CRN, by selecting the best path from all available paths that are given between the source and destinations.  Our proposed protocol achieves high-throughput and packet delivery rate (PDR) through a joint channel assignment and path selection to the specific destinations. Simulation results is used to demonstrate the effectiveness of our proposed algorithm in terms of throughput and packet delivery rate compared to other existing multi-cast routing protocols over different network conditions by using matlab as a simulations program.

# Chapter 1: Overview

## 1.1 Introduction

Wireless communications have been experienced an exponential growth in the last decade. It is the fastest growing sector of communication industry. We have noticed a great improvement in network infrastructures. Cellular network users and applications grew up very quickly, including wireless sensor network, automated factories, remote telemedicine and smart home and appliances. All these applications emerged from research area to concrete systems. The wireless devices are becoming smaller, cheaper, more powerful, and more convenient. So, it has spread quickly, they are rapidly supplanting wired system in many countries, including homes, campus, and business. Wireless devices include portable or handheld computers, personal digital assistants (PDAs), and cellular phones. For example, a mobile device is multifunctional devices. Users can make phones calls, browse the Internet, check e-mail, and determining the location during internal and international roaming by using the global positioning system (GPS) [1-2]. Also, the development includes the access point (AP) in traditional wireless networks. The AP is a device that connects wireless devices together to create a wireless network. Many APs can be connected together to create a larger network.

Ad hoc network is the type of wireless network, which consists of a group of wireless nodes, which are capable of communicating with each other without infrastructure [3]. There are two basic types of infrastructure-less wireless networks: (1) Static Ad-Hoc Network, it has fixed backbone wireless model consists of a large number of Mobile Nodes (MNs) and a small number of fixed nodes. They are communicating using wireless environment within its range, and (2) Mobile Ad-Hoc Network (MANET). MANETs are self-organizing and configuring a group of MNs without using any existing network infrastructure. It forms a network through radio links. The aspect of using MANETs that can be deployed in areas, which wired network is not available and for dynamic environments that usually need dynamic and very fast configurations. These include military battlefields, rescue sites, and emergency search. MNs dynamically form the routes among themselves. In this type of networks, nodes can directly communicate with the other nodes within its wireless range (single-hop) or indirectly with other nodes in a network (multi-hop) [4-9].

The technique that it used in these types of network to disseminate information to a given group of nodes is multicast. Multicasting is the process to distributing data (e.g., Audio/video streaming) to multiple recipients as destinations, by determining the path from the source to the destinations. Several multicast routing protocols were proposed for wireless networks to improve and increase the quality of communication links, to effectively exploit the available bandwidth, and to reduce the cost of communications in the network. Multicast Ad-hoc On demand Distance Vector protocol (MAODV) is the most popular multicast routing protocol for MANETs [11-15].

As a result of the rapid growth in wireless communications, which introduced multiple technologies that require higher data rate, the demand for more radio frequency (RF) spectrum has increased. Since the majority of spectrum resources have already allocated, the main challenge that needs to be tackled is the low utilization of those allocated spectrum bands [16]. The FCC and other organizations (e.g., XG Darba initative) have measured the spectrum utilization of licensed spectrum. They reported that the allocated spectrum used only in a limited geographical area for a limited period, (low utilization ranging from 15% to 85% [17], with average utilization of 17.4%). Therefore, the need of reliable communication systems with programmable radios (that can enable dynamic spectrum access, interference sensing, and environment learning) is essential to satisfy the increased spectrum demands and efficiently utilize the spectrum. The new technology is dynamic spectrum access (DSA) is known as CR. CR is defined as a radio that can change its operating parameters based on the interaction with the operating RF environment. This means that the CR has two main characteristics: Cognitive capability that describes the interaction of CR users with their surrounding RF environment that is randomly changing; (i.e., identifying spectrum opportunities). The CR reconfigurability describes the capability of CRs to manage and reprograms themselves, to exploit the idle channel efficiently. The CR technology has been proposed to solve the inefficient spectrum utilization. CR enables secondary radio users (SUs) to share the spectrum with primary radio users (PUs) without affecting the reception quality of PUs. Spectrum sensing, spectrum sharing, spectrum management, and spectrum mobility, are the main functions of a CR device [16-19]. On the other hand, the basic idea of CR is to be able to accurate sense the RF spectrum and detect the available spectrum through which the available idle channels are existed (e.g., white or gray holes). Also, it has the capability to choose the best available channels and appropriate technology used for transmission to share the spectrum based on DSA. Finally, it has the ability to quickly decide when to vacate the spectrum to another one. The cognitive radio cycle is shown in Figure 1.1.

In CRNs, there are several challenges in achieving the optimal performance. For example, finding the path with maximum multicast flow in the network is an NP-hard problem. We note that because of the unique characteristics of CRNs, traditional protocols and technologies for wireless networks are not efficient for CRNs. Thus, new protocols are needed that have the following attributes:-

- They need to be transparent to PR users, (no coordination between them).

- They should not affect the performance of primary radio network (PRN). By controlling the CR transmission powers, and vacating the channels that are reoccupied by PUs.

- They need to allow for cooperation between CR neighboring users to increase spectrum efficiency. This can be done by effectively share the available spectrum among the CR users.

- They need to improve fairness between CR users.

- They need to have common control channel (CCC) to effectively organize the CRNs, which is a dilemma for CRNs. Because of channel heterogeneity, and the unexpected, and dynamic allocation of users.

Therefore, extensive research has carried out to improve the effectiveness of the deployment of CR technologies in a large scale.

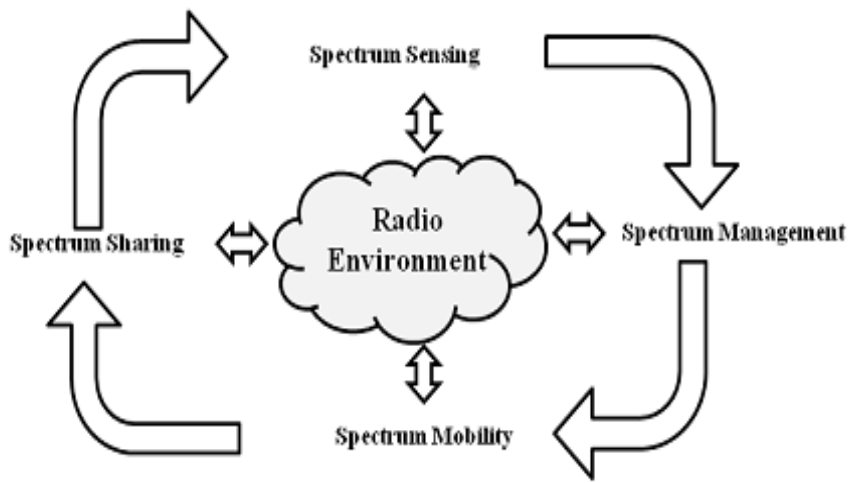

**Figure 1.1:** Cognitive radio cycle.

## 1.2 Literature Survey

Most of the fundamental networks use multicast in various applications. Including CRNs. CRNs have been designed as a solution to the fixed traditional wireless networks. A main application of CRNs is dynamic spectrum access, which improves spectrum utilization through opportunistic spectrum access by allowing unlicensed users to exploit the licensed spectrum. Recently, several schemes have been proposed to improve the performance of CRNs. Most of them dealt with only the spectrum sensing and dynamic spectrum access issues at the MAC and physical layers without considering the routing layer (the network layer) [16-21]. Also, very few studies have focused on the multi-hop routing without consideration the multicast routing case. The main challenge in this domain is how to find a high throughput and efficient distribution for the data traffic such that the delay is minimized by selecting the path based on different metric such as: the path with smallest transmission delay, minimum hop count, energy, and stability in [19-28]. The multi-cast problem is very challenging in dense traffic conditions. The work in [31] and [32], illustrated the attribute of communications in multi-hop CRNs and the challenges that face network implementation (e.g., how to enable opportunistic medium access control in multi-hop CRs). Also, they discussed MAC design protocols that have done in CRNs, and some challenges for future design.

We note here that multicast routing can significantly improve network performance by reducing the traffic in the network. The main idea of multicast is to transmit packets from one source/multiple sources to specific receptions. Multicasting is performed either by using a multicast tree structure, or a multicast meshes structure [11], [33]. Most of the existing multicast protocols are mesh structure-based. The main deployment challenge that faces the tree structure deployment is that the tree needs to be reconstructed based on channels dynamic. We note here that the tree-based structure needs high control overhead [34]. [34] Provides an overview of parameters that have been investigated in designing multicast routing protocols, which included mobility, instability of paths, limited energy, available bandwidth, and dynamic topology. In [35], the authors discussed the main challenges that should be taken into consideration when designing CRN protocols (e.g., spectrum heterogeneity and frequency switching latency). One of the main challenges is the video multicasting in CRNs. This problem was investigated in [36-37] based on different factors such as: video rate control, spectrum sensing, dynamic access, modulation, retransmission, scheduling, and primary user protections. The main objectives of these works are to achieve fairness and/quality among multicast users and reducing the interference with PU. In [38], the authors summarized several techniques, protocols, and algorithms designed for multicast in multi-hop CRNs, and its application, including network coding, optimization theory, and heuristic techniques. In addition, they provided an open research issues and the directions for future work. In [39], the authors defined routing protocol that focused on cooperative transmission techniques to expand the capacity of relay link by using spatial diversity rather than using multiple antennas at every node. To construct a minimum energy multicast tree-based, two factors should have been taken into account: the topology of the secondary users and the traffic load of PUs. The work in [40] used a fixed multicast tree. In [42-43], the authors proposed multicast routing for multi-level channel assignment for a mesh network with the objective of maximizing throughput, reducing transmission delay, and minimizing the number of relay nodes. Finding the path from the source to the destinations based on probability of success (POS) have been investigated in [44]. This work considered the channel availability time and required transmission time to improve throughput. The POS have been also investigated as a metric to fulfill multicast in single-hop CRNs by finding the best common channels for all destinations that improves the overall quality of the received video by all destinations. The authors of [44] studied the performance of their protocol by comparing its performance with three multicasting protocols, Maximum Average Spectrum Availability time (MASA), Minimum Transmission Time (MinTT), and Randomly Selection (RS). In [45], used different network conditions such as packet delay, control overhead, and the ratio of bad frame used to improve the performance of protocol. The authors in [44] studied the mechanism of channel assignment based on the maximum POS for multi-hop CRNs. This metric focuses on the dynamic changes of channel availability due to dynamic and uncertainty of PU activity, also, this mechanism is compared with different schemes (e.g., MASA, and MinTT). In [47-49], the authors used the expected transmission count metric (ETX) for routing in multi-hop CRNs to find the path with higher stability and higher throughput. ETX is also been used to minimize hop count in the routing path, which improves the network performance.

In [50], the authors used a the distance between each two nodes in the CR as a metric, to design the multicast routing in multi-hops CRNs. Based on the distances, it is able to construct the shortest path tree (SPT), and the minimum spanning tree (MST). This work has used the POS for the channel assignment. The POS outperforms the other schemes, and the SPT outperforms than the MST in terms of improved throughput under different network conditions. However, this work did not fully consider the unique characteristics of CRNs.

## 1.3 Motivation

CRNs are heterogeneous networks that enhance data rates and capacity by reusing the unused spectrum bands of PUs. The coexistence of PUs and SUs in the same environment is the main challenging in designing CRNs. The challenge is how to eliminate the interference between two networks while protecting the performance of PRNs and improving the performance of CRN.

As discussed in the literature survey section, several approaches have been designed to maximize the CRNs throughput and to increase the overall spectrum efficiency. However, some of them dealt with the spectrum sensing and the dynamic spectrum access in MAC and physical layers. Also, others dealt with network layer. The most significant approach that provided a useful solutions in wireless communications is multicast routing (It has several applications such as video conferencing, data disseminations, and military purposes). It reduces communication cost, improves channel efficiency, provides an effective usage of energy and bandwidth, minimizes the transmitter and receivers processing, and minimizes delivery delay. Mainly, the existing protocols were based on either the path selection process (e.g., [19-28]), or the channel assignment mechanisms (e.g., [42-44]), but there are very few studied that jointly consider the both factors.

Our proposed algorithm in this thesis considers jointly the path selection process based on the MST and SPT trees and the POS channel assignment approach. It differs from the work in [50] with the metric used to construct the trees. In [50], it used the distance between nodes and our scheme's uses the ETX. The ETX and POS, which is used in our protocol, significantly improve network throughput by reducing the total packets needed for successfully transmitting data packets to the destinations. Also, both of the required transmission time and availability time of each channel are jointly used on choosing the best channels. So, it reduces the overhead and increases the stability of the network.

To illustrate this idea, we consider the network shown in Figure 1.2. We assume a random topology with multi-hop CRN environment. We have a network of PUs and eight nodes of SUs. One source, seven nodes (M=7), and three destinations ($M_r$=3). Also, PUs and SUs share three common channels: CH1, CH2, and CH3.

In this system, if the channel is occupying by a PU, SU cannot access this channel. Also, if the channel is being used by a SU and a PU wants to access this channel, the SU should terminate it transmission and switch to another channel. The problem is how to send video messages from the source to the multi-destinations, by choosing the best channels from the source to all the destinations.

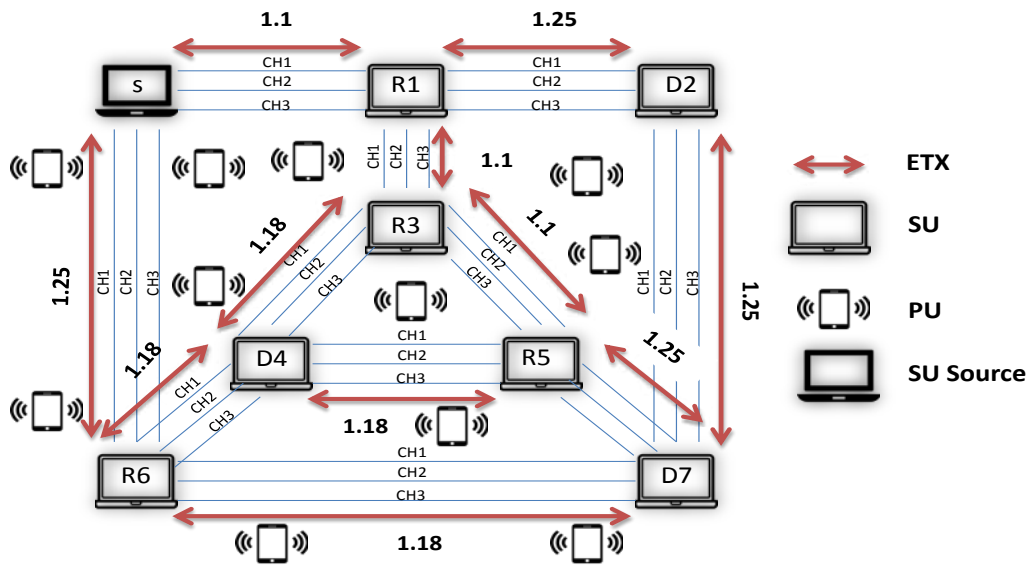

**Figure 1.2:** Example of CRN with ETX computed for each link.

Under these conditions and given the POS between each two nodes over all available channels. ETX metric, (ETX=1/Max (POS)) among each two nodes can be computed. Given the ETXs, we need to determine the path structure from the source to the destinations (D2, D4, and D7).

By using Dijkstra's algorithm, we find the shortest path tree (SPT). On the other hand, bu using Kruskal's algorithm, we can find the minimum spanning tree (MST). For more details, see Chapter 2. Figure 1.3 illustrates the execution of the Dijkstra's and the Kruskal's algorithms on the topology shown in Figure 1.2 based on the ETX metric.

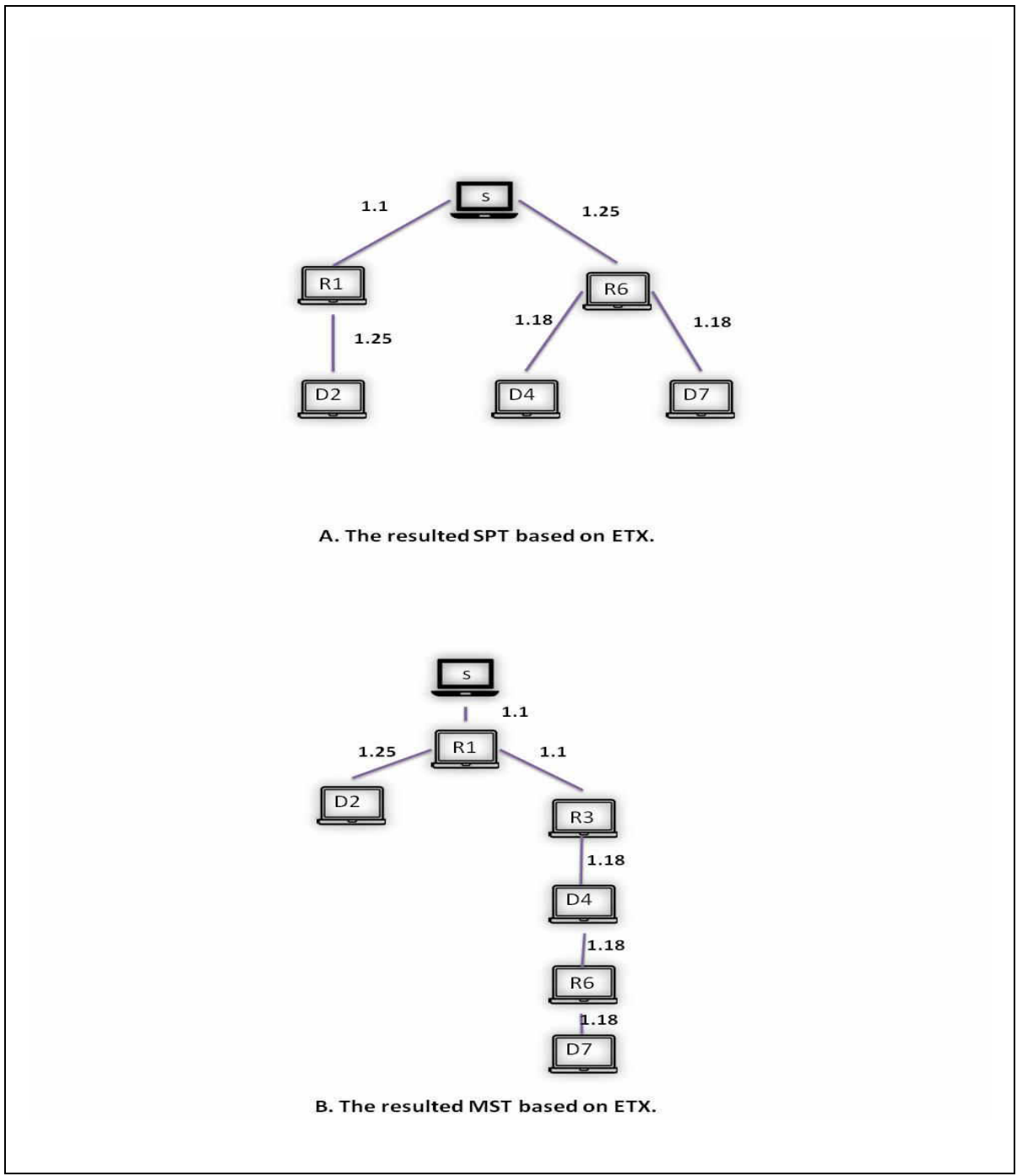

**Figure 1.3:** The resulted SPT & MST based on ETX.

Also, we use the utility Max (Min POS) to choose the channels in both trees. We explain this algorithm's by using the of POS values that shown in Table 1.1.

Table 1.1: The selected channels for SPT method.

|  |  | Channel<br>Receiver | CH1 | CH2 | CH3 | Max POS |
|---|---|---|---|---|---|---|
| Layer One |  | S-R1 | 0.9 | 0.3 | 0.8 |  |
|  |  | S-R6 | 0.75 | 0.8 | 0.4 |  |
|  |  | Min POS | 0.75 | 0.3 | 0.4 | 0.75 |
| Layer One | Part 1 | R1-D2 | 0.75 | 0.45 | 0.8 | 0.8 |
|  | Part 2 | R6-D4 | 0.75 | 0.85 | 0.55 |  |
|  |  | R6-D7 | 0.35 | 0.5 | 0.85 |  |
|  |  | Min POS | 0.35 | 0.5 | 0.55 | 0.55 |

As mentioned before, we have CR source S that wants to perform multicast group to three CR destinations (i.e., D2, D4, and D7), and they have three available channels. Given the POS values, the ETX values are computed in Table 1.1 for the SPT tree. The SPT tree is shown in Figure 1.2 contains only two layers:

- **The First Layer:** The transmission from CR source to nodes R1 and R6 is done using multicast. To choose the best channel, the source computes the minimum values of POS for each channel, which is clearly appeared in the final row of the first Layer in Table 1.1. The minimum values are 0.75, 0.3, and 0.4 that are related to CH1, CH2, and CH3, respectively. Then, the source chooses the maximum of these values, which is 0.75. As a result, CH1 will be chosen.

- **The Second Layer:** Consists of two parts:

**Part One:** The transmission from node R1 to node D2 is done using unicats. Therefore, node R1 chooses the channel that has the Maximum POS for it is transmission (in this case CH3).

**Part Two:** The transmission from node R6 to nodes D4 and D7 is done using multicast. It is similar to the case shown in the first layer. The minimum values of POS for CH1, CH2, and CH3 are 0.35, 0.5, and 0.55, respectively. The Maximum is 0.55. Accordingly, CH3 will be chosen.

Table 1.2 contains the value of POS for each channel and for each destination in the MST tree. As shown in Figure 1.2, MST contains five layers. The transmission from CR source to node R1, R3 to D4, D4 to R6, and R6 to D7 of Layer One, Layer Three, Layer Four, and Layer Five are unicast, respectively transmissions. As mentioned before, in this case, each node in the sub-layer of this tree chooses the best channel available for it, which has the maximum POS. By referring to Table 1.2, CH1, CH2, CH1, and CH3 will be selected. According to Layer Two, the transmission from node R1 to nodes R3 and D2 is done using multicast. Also, the same procedure is used to select the best channel for each transmission. Node R1 has the minimum values of POS that shown in Table 1.2, at the last line of Layer Two. The values are 0.75, 0.2, and 0.8. By selecting the maximum one, CH3 will be chosen.

**Table 1.2:** The selected channels for MST method.

| | Channel<br>Receiver | CH1 | CH2 | CH3 | Max POS |
|---|---|---|---|---|---|
| **Layer One** | S=R1 | 0.9 | 0.3 | 0.8 | 0.9 |
| **Layer Two** | R1-D2 | 0.75 | 0.45 | 0.8 | |
| | R1-R3 | 0.8 | 0.2 | 0.9 | |
| | Min POS | 0.75 | 0.2 | 0.8 | 0.8 |
| **Layer Three** | R3-R4 | 0.45 | 0.85 | 0.3 | 0.85 |
| **Layer Four** | R4-R6 | 0.85 | 0.75 | 0.45 | 0.85 |
| **Layer Five** | R6-R7 | 0.35 | 0.5 | 0.85 | 0.85 |

Therefore, our study provides an efficient solution for the multicast problem, which is considered the challenging problem in CRNs.

## 1.4 Contributions

In this thesis, multicast with multi-hop routing protocol over a single session is investigated. A new protocol that integrates the path selection process and the channel assignment mechanism is proposed in order to enhance and increase network throughput and packet delivery rate (PDR) under various network conditions. Several design variants are considered by our protocol:

1) A protocol that considers the path selection process based on the shortest path tree (SPT), which finds the shortest path with minimum weight (metric) from the source to each destination individually. It implemented for both metrics ETX and distance. Also, we compare the throughput performance among both metrics.

2) A protocol that considers the path selection process based on the minimum spanning tree (MST), which finds the shortest path with minimum weight (metric) for all destinations in the network. Also, it is implemented for both metrics ETX and distance and compared the performance between two metrics.

For both variants that mentioned before, the available channels between any two users (nodes) will be selected based on four multicast metrics, which are different only in the channel assignment mechanisms used to choose the best unified channels from all available channels:

1) The Maximum Probability of Successful (Max-POS) assignment that selects the best channels for transmissions. This assignment is based on the factors: the availability time of each channel and the time required for transmission. It means that if the required transmission time is greater than the average availability time, the channel cannot be used. The two factors should be used to improve performance..

2) Maximum Average Spectrum Availability (MASA), which selects the channel with the maximum average spectrum availability, irrespective of the other conditions.

3) Random Selection (RS) that randomly chooses the channel without any restriction.

4) Maximum Data Rate (MDR) that selects the channel with the maximum data rate.

The main idea of our algorithm is to dynamically construct the trees in CRN according to the varying traffic loads, which increased the throughput and the PDR in the network. To compare these protocols, the throughput performance is expected to increase when using ETX for (POS, MASA, RS, and MDR) protocols rather than using Distance metric in both trees. Also, the POS protocol is expected to outperform the other schemes in both trees.

## 1.5 Thesis layout

The thesis consists of four chapters. In Chapter two, the system models and problem formulation are described. In Chapter three, we present and analyze the simulation setup, parameters, and results. Finally, in Chapter four, we summarize the presented work and provide recommendations for future work.

## Chapter 2: Models and Problem Formulation

### 2.1 Transmission Methods in Multi-hop CRNs

In wireless networks, the crucial task is to manage the data distributions through the network, especially managing video streaming. There are three transmission methods: Multicast, Unicast, and Broadcast. The network application decides which one to use based. In this section, we describe the definition of each method and give brief comparisons between them [33-41].

#### 2.1.1 Multicast Transmission

In this method, the source transmits the data for a group of destinations, not all the nodes in the network. Multicast is not limited to one source only, multiple sources are possible. It is also called point to multi-point transmissions, or multi-point to multi-point transmissions. It integrates the performance for both unicast and broadcast in one type of transmission, which can reduce communication cost such as: bandwidth and energy losses, computational processing, and delivery delay. It uses the same internet protocol (IP) address to send the same data packet to a group of destinations. Also, it has a feedback connection between the sender and the destinations.

#### 2.1.2 Unicast Transmission

Unicast is used when the data are transmitted from one specific source to one specific destination. Also, it is called point-to-point transmission (one sender and one receiver). If the sender needs to send for multiple receptions, in this case it sends multiple unicast messages, each is addressed to a specific receiver. This type of transmission needs to know the IP address for the destinations. It increases the complexity of processing among the sender and the receiver, bandwidth and energy losses. Unicast is useful to achieve secure transmission, as compared to other methods. In addition, it forces a feedback connection between the source and destinations.

### 2.1.3 Broadcast Transmission

This method, the same data is to be sent to all destinations at the same time from single source without including a feedback connection between them, (e.g., radio streaming). If the packet has a broadcast address, it can send the same message for all destinations in the same region with a single IP address. So, all destinations will process the received packet. This type of transmission increases the bandwidth and energy losses but it reduces the processing between the sender and the destination. It is useful to be used for unreliable communications.

## 2.2 The Shortest Path Tree (SPT) and Minimum Spanning Tree (MST)

We now describe the SPT and MST that are used in our algorithms. The definition and the construction of such trees are described, with simple example to illustrate their operation. Both trees provide important role in designing routing algorithms. SPT separately finds the shortest path from the source to each destination in the network. On the other hand, MST finds the path that has the smallest weight and serves all the destinations. specifically, it finds the shortest path for all destination nodes in the network, it is not necessary to choose the shortest path between any two nodes. For both trees, the number of the formed links in the tree equals to (m = n – 1), where m is the number formed links in the tree, and n is the number of nodes in the undirected graph [51-53].

### 2.2.1 The Minimum Spanning Tree

**Definition**

This tree can be formed using the undirected graph that has all vertexes (nodes) in the undirected graph. It finds the path that has the minimum sum of the total weights of all edges (links) for all destinations, without forming loops. We may have several MST trees, for the same undirected graph with the same total weight, which is not unique. Also, it does not depend on the starting point (source). It is flexible, which is available for any node in the network that is known as a source. So, it does not need to reconstruct the tree if the source is changed. The MST can be found using Kruskal's algorithm with the total running time that equals to (m log m).

**Kruskal's algorithm**

Kruskal's algorithm constructs the MST algorithm. It is a greedy algorithm in graph theory, which finds the edges with the least weight or cost between any two nodes, where the weights of the links are non-negative. The following example illustrates the idea of this algorithm. We have a random topology of seven nodes (A, B, C, D, E, F, and G). The links of this topology have positive values that take the values from 1 to 9. These values are shown in Table 2.1. We need to construct the tree from the node A to node G as follows:

1. We sort the links between nodes in the undirected graph with ascending order see Table 2.1.

Table 2.1: The sorted links of undirected graph

| Order | 1 | 2 | 3 | 4 | 5 | 6 | 7 | 8 | 9 | 10 |
|---|---|---|---|---|---|---|---|---|---|---|
| Links | C-E | A-B | A-D | A-C | B-E | B-C | C-D | D-F | F-E | F-G |
| Weights | 1 | 2 | 3 | 3 | 3 | 4 | 5 | 7 | 8 | 9 |

2. For each link in the sorted list, keep the loop-free links as shown in Figure 2.1.

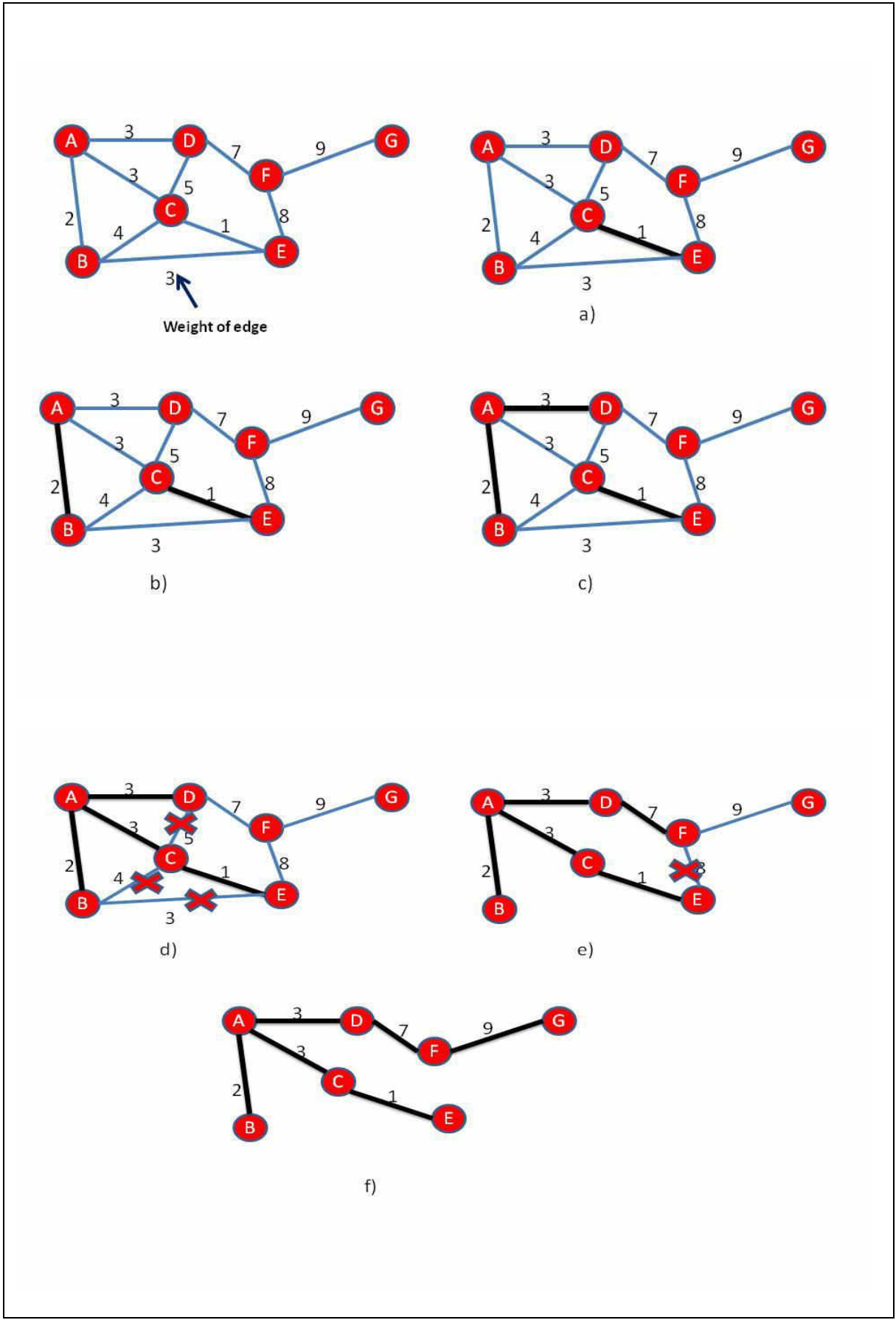

**Figure 2.1:** A simplified example describing the Kruskal's algorithm.

### 2.2.2 The Shortest Path Tree

**Definition**

It is another type of tree that is formed from undirected graph. Undirected graph contains non-negative weights, where a special node is called source or root. SPT finds the path with minimum weight from the source to every destination in the network. If the source is changed, the tree should be re-built, which increases the complexity of SPT. The SPT can be found using Dijkstra's algorithm with the total running time that equals to $(m + n \log n)$.

**Dijkstra's algorithm**

It can be used to find the shortest path tree between nodes in an undirected graph. To illustrate the idea of Dijkstra's algorithm, we consider the example in Figure 2.2. We have a random topology of seven nodes (A, B, C, D, E, F, and G), the links of this topology have non-negative weights. Also, we consider A as the source and need to construct the tree from the source node A to node G. Formally,

1. Initially, we set the cost to infinity for all nodes, except the source node. We set the cost of the source node to zero.
2. We need to update the neighbor cost by computing the cost of nodes that is directly connected to the source. The new cost equals to (source cost + the weight of edges between two nodes).
3. We repeated the second step until reached the final destination.

This process is shown in Figure 2.2.

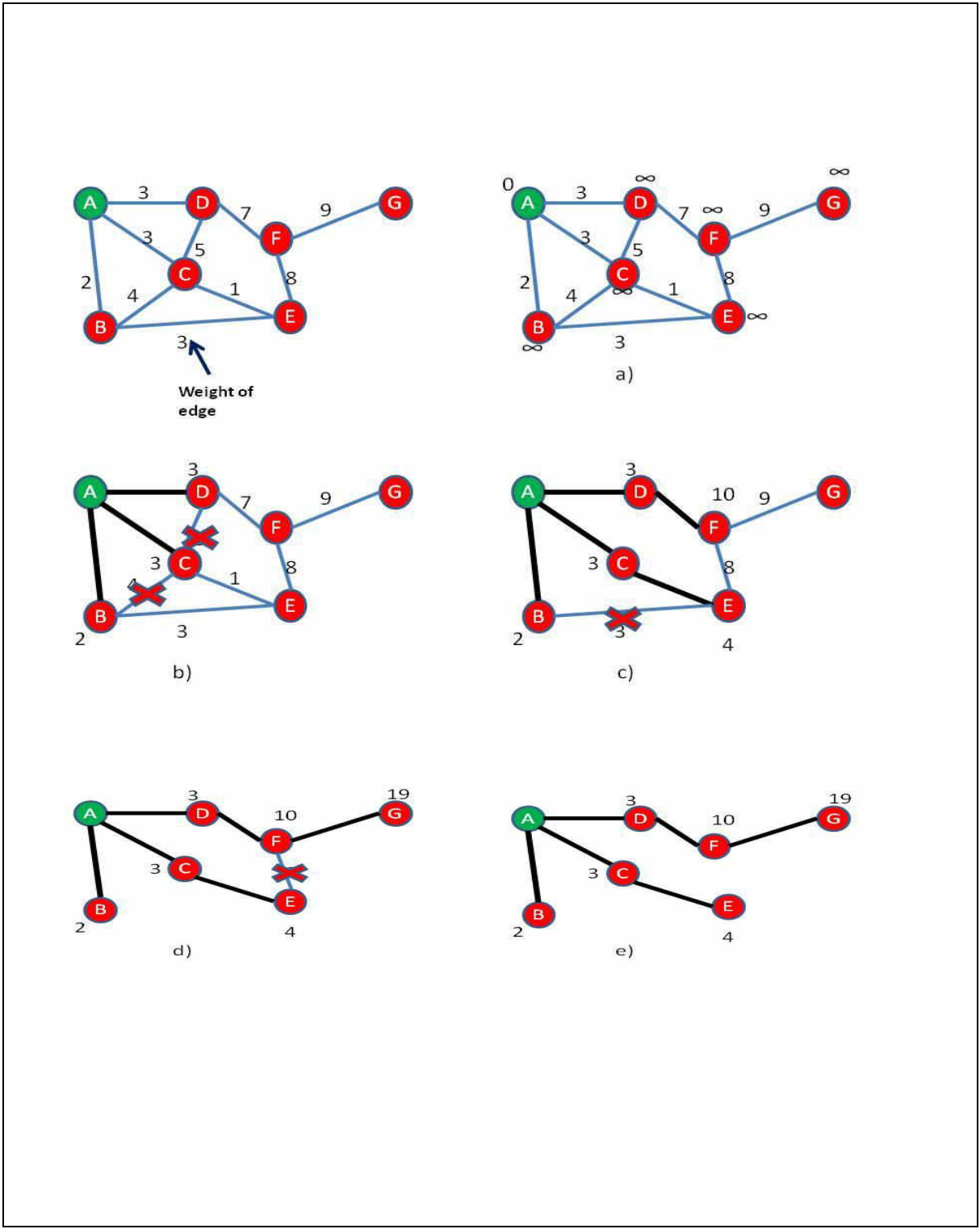

**Figure 2. 2**: A simplified example describing the Dijkstra's algorithm.

## 2.3 System Model

We consider a multi-hop CRN with |K| PU channels, each channel have the same bandwidth (BW), which is used in [31] as a common descriptions for the PU spectrum band. Also, we consider |M| CR users. Each CR user has a sensing capability to find the idle channels, individually. We assume a common control channel (CCC) is available to organize the transmissions in the CRN. In general, each CR user has two half-duplex transceivers that to be used simultaneously. They are used for data and control transmissions [23 and 31] for both MAC and routing protocols in CRN. In addition, we assume CR users have information about the average availability time and quality of each channels in the network from the history of PU activates. We assume a small fixed transmission power $P_t$ is used by CR over each channel j (where $1 \leq j \leq |M|$). We also use the Shannon capacity formula to examine the rate of each CR user over each channel. Also, we assume that the fading channel gain between any two CR users is Rayleigh [45].

### 1) Primary Radio Network

We assume the |K| channels are allocated to K primary networks. The Markov model is used to describe the state of a given channel state idle or busy. The channel is idle when it is not used by a PU and is busy, if the channel is occupied by a PU.

### 2) CR Network

We assume a CRN that coexists with the |K| PRNs in the same geographical area. We consider that there is one CR source tries to transmit data to $M_r$ destinations (multicast group) over the set of available channels K. So, this channel can opportunistically used by SUs. We assume infrastructure-less ad hoc multi-hop CRN.

Let $POS_j^{(i-L)}$ represent the probability of a successful transmission between nodes i and L over channel j∈ K, which depends on the required transmission time and the average spectrum a availability time of link j that is variant to randomness and unexpected activities of primary users. Let ETX represents the total expected number of packet transmissions (with retransmissions) needed to transmit the packets successfully to the specific destinations, (this metric should be minimized as much as possible). Also, a given channel can be used by a CR user if the actual transmission time needed for the CR transmission is less than that or equals to the average spectrum availability time.

Table 2.2 summarizes the main symbols used in this thesis.

**Table 2.2: Summary of symbols used in our algorithm.**

| Parameter | Description |
|---|---|
| $P_t$ | CR transmission power |
| $M_r$ | Total number of destinations |
| M | Total number of CR nodes |
| $POS_j^{(i-L)}$ | Probability of success (POS) between nodes i and L over channel j |
| $R_j^{(i-L)}$ | Transmission rate between nodes i and L over channel j |
| $Tr_j^{(i-L)}$ | Required Transmission time to transmit a packet between nodes i and L over channel j |
| $\mu_j$ | Average availability time of channel j |
| $Pr_j^{(i-L)}$ | Power received between nodes i and L over channel j |
| $\gamma_j^{(i-L)}$ | Channel power gain between nodes i and L over channel j |
| D | Data packet size |
| $N_0$ | Thermal power spectral density |
| BW | Channel bandwidth |
| n | Path loss exponent |
| N | Total number of channels for PRNs |
| K | Available channels for CRNs |
| $d$ | The distance between any two nodes in the network topology |
| $\lambda$ | Wavelength |

## 2.4 Problem Definition and Analysis

Given the network model, we consider the problem of video streaming from one CR source to a group of destinations in a CRN. The objective of our study is to design a multi-hop multicast routing protocol that effectively improves the overall network performance. By using both ETX and Max-POS as metrics, which aim to choose the path that reduces the total number of expected packet transmissions (with retransmissions) needed to transmit a packet successfully to a specific number of destinations, and to select the best common channel from the available channels based on the average availability time of the channels and required time needed for successful transmission. Our protocol finds the effective available path from the source to the destinations based on SPT and MST tree for a given topology that consists of multi-layer routes. At each layer, the transmitter selects a set of nodes that should receive the data and this will continue until the last destination in the tree is reached. Our algorithm uses a probabilistic approach to assignment the channels between any two CR users during the transmission.

In [47], the expected transmission count metric is given as: -

$$ETX_j^{i-L} = \frac{1}{MAX\left(POS_j^{i-L}\right)} \qquad (1)$$

where $POS_j^{(i-L)}$ is the POS between any two nodes i and L over each channel j ($j \in K$), which guarantees that the life-time of channel j is greater than the required transmission time over that channel. The authors in [43] and [44] presented a closed-form expression for the $POS_j^{(i-L)}$ as follows:-

$$POS_j^{(i-L)} = exp\left(\frac{-T_{r_j}^{(i-L)}}{\mu_j}\right) \qquad (2)$$

where $T_{r_j}^{(i-L)}$ is the required transmission time to send a packet from node i to L over channel j (in sec/packet), the term $\mu_j$ is the average availability of channel j (in sec). Note that $T_{r_j}^{(i-L)}$ can be calculated as shown in [43] and [44] as follows:-

$$T_{r_j}^{(i-L)} = \frac{D}{R_j^{(i-L)}} \qquad (3)$$

where D is the packet size (in bits/packet), and $R_j^{(i-L)}$ is the data rate between nodes i and L over channel j (in bit/sec), which is given by [43], [44]: -

$$R_j^{(i-L)} = BW \times log_2\left(1 + \frac{Pr_j^{(i-L)}}{Bw * N_0}\right) \qquad (4)$$

where $N_0$ represents the thermal power density in (Watt/Hz), $Bw$ is the channel bandwidth, and $Pr_j^{(i-L)}$ represents the received power from transmitter $i$ to receiver $L$ which is given by [44]:-

$$Pr_j^{(i-L)} = \frac{p_t}{d^n}\left(\frac{\lambda}{4\pi}\right)^2 (\gamma_j^{(i-L)}) \qquad (5)$$

note that $\gamma_j^{(i-L)}$ is the channel power gain between nodes $i$ and $k$ over channel $j$. For Rayleigh fading, $\gamma_j^{(i-L)}$ is exponentially distributed with mean 1 [45].

## 2.5 The Proposed Solution

The main idea of our scheme is to transmit the information from a given CR source to a group of destinations (multicast) through multi-hop transmissions using a minimum spanning tree (MST) and/or a shortest path trees (SPT), which determine loop free paths from the source to the destinations. Then, the best available set of channels from all available channels will be selected based on the Max-POS between nodes during transmissions. This improves the quality of reception for all destinations in the CRN.

The proposed protocol consists of two stages: (1) constructing the trees based on the ETXs for all link. (The SPT uses dijkstra's algorithm and the MST uses kruskal's algorithm), and (2) selecting a common channel at each layer in the tree that maximizes the delivery rate from the source to the destinations with probabilistic approaches. The proposed algorithm is performed as follows:-

1- Create a routing table for the CR source and its destinations by each CR user. This includes all receivers and their POS over all available channels in each layer by using equations (2) to (5).

2- Specify the channel with the max POS between each two pair of nodes.

3- Find the ETX for each destination by using equation (1).

4- Construct the required trees from the original topology, using dijkstra's / kruskal's algorithms.

5- At each layer in the constructed tree, each source will determine its destinations in this step, Unicast or Multicast transmissions can be used.

6- In unicast transmissions, the channel with Max POS will be selected for the transmission.

7- In multicast transmission, the proposed algorithm specifies the min POS over each channel and for all receivers in the same layer. Then, the channel with Max (Min POS) will be selected for the transmission. The flow-chart of the proposed scheme is given in Figure 2.1

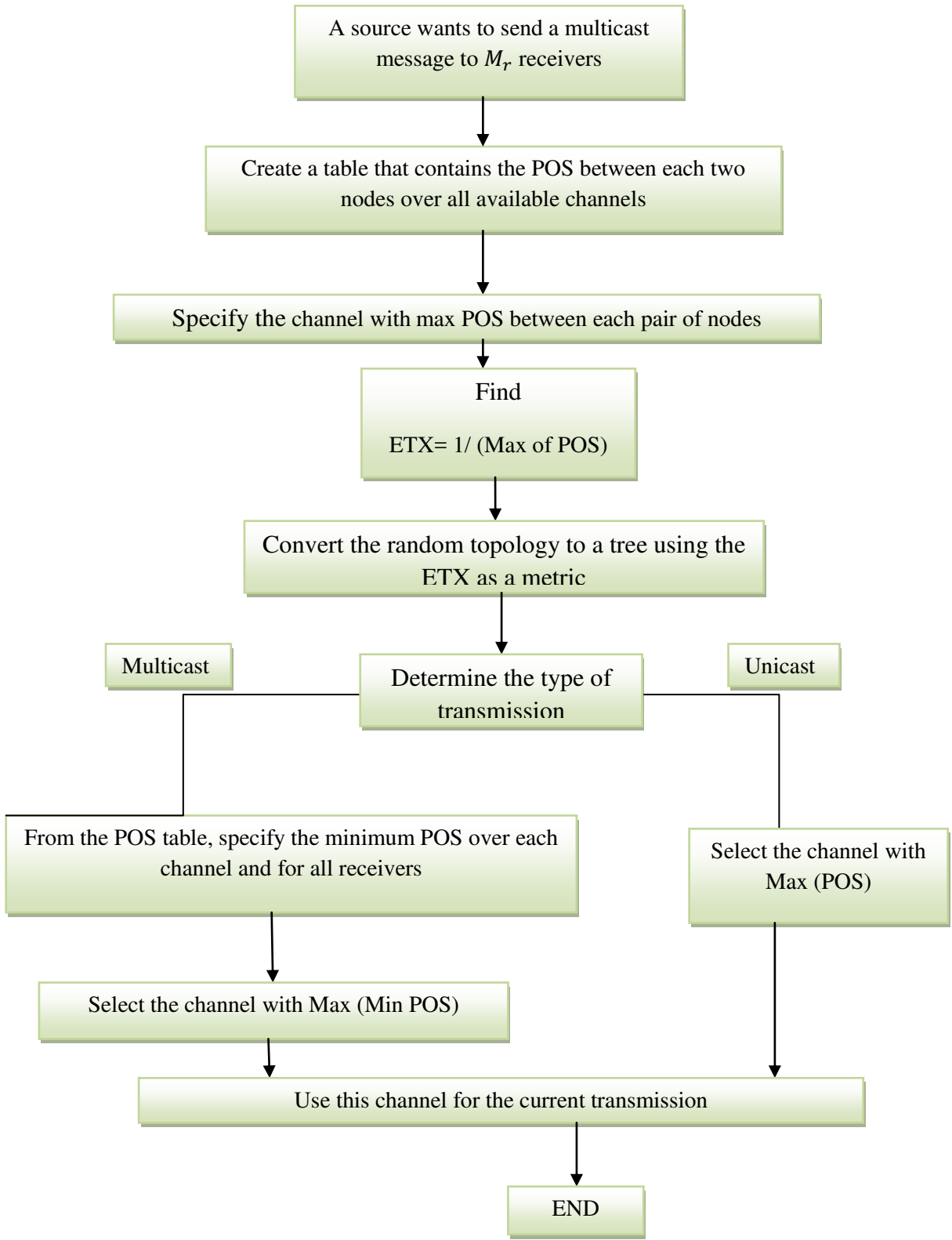

**Figure 2. 3:** A simplified flow-charts describing the proposed algorithm.

# Chapter 3: Simulation and Performance Evaluation

## 3.1 Simulation Setup

We consider a CRN that coexists with several PRNs, where no connection between the two networks. The CRN consists of M=40 CR users and $M_r$=16 CR destinations, which located randomly in a 200 X 200 field, and N=20 PU networks that are in the same geographical area. We set the transmission range to R=100 m, and the data packet size to D= 4 KB. Also, we set the bandwidth to BW=1 MHZ, and the thermal noise $N_o$ to (10^-18) W/HZ. We use the Markov model to describe the status of PU channel (e.g., ON /OFF). We consider the path loss exponent to n=4, which is used for indoor applications. The channel fading is Rayleigh. The power transmission of each CR is set to 0.1 W to minimize the interference with PUs. The average time that is allocated for primary users channels is ranging from (2-70) ms. Matlab is used for simulations.

## 3.2 Simulation Results and Discussion

In this section, we study the impact of PR activity on different traffic loads (idle probability 0.1, 0.5, and 0.9) on CRN performance (e.g., throughput, and PDR). The performance of the CRN increases for all protocols as the idle probability increases; because it minimizes the number of terminated connections through transmission and increases the chance to find appropriate channels.

As mentioned earlier, the purpose of this thesis is to identify the significant improvement could be achieved by the proposed protocol. So, we create new four different schemes; the probabilistic approaches (POS), MASA, MDR, and RS, which are constructed for both trees based on the ETX metric in the process of selecting routing path from the source to the destinations.

The proposed protocol considers the path routing process based on the MST and SPT and the POS uses for the channels assignment. POS is compared with MASA, MDR, and RS that are implemented over both trees under different network parameters (e.g., bandwidth, packet size, number of channel, idle probability, number of nodes, number of destinations, range, area, and power transmission). POS achieves the best performance in both trees for all channel condition because it uses a better channel assignment.

SPT tree outperforms MST tree performance over all network parameters. Therefore, SPT has become the focus of attention to be studied.

The result section includes the brief comparison between MST and SPT trees over POS, and the performance of SPT tree over four schemes: POS MASA, MDR, and RS. In order to compare the performance of the POS that uses ETX as a metric with the POS that uses the distance as a metric over SPT tree.

## 3.3 Performance Comparison between SPT and MST

### 3.3.1 Impact of Channel Bandwidth

We investigate the throughput and PDR performance as a function of channel bandwidth. We consider the following network conditions: - M=40, $M_r$=16, N=20, $P_t$=0.1 W and D=4 KB, for the different PU traffic loads.

### 3.3.1.1 Throughput Performance versus Channel Bandwidth

Figure 3.1 shows network throughput as a function of the channel bandwidth (BW). This figure shows that the throughput increases as the BW increases for all traffic loads. This is a result of increasing the channels availability. POS over SPT outperforms POS over MST with 6% when $P_I$=0.1, and achieves 11% over $P_I$=0.5 and 0.9.

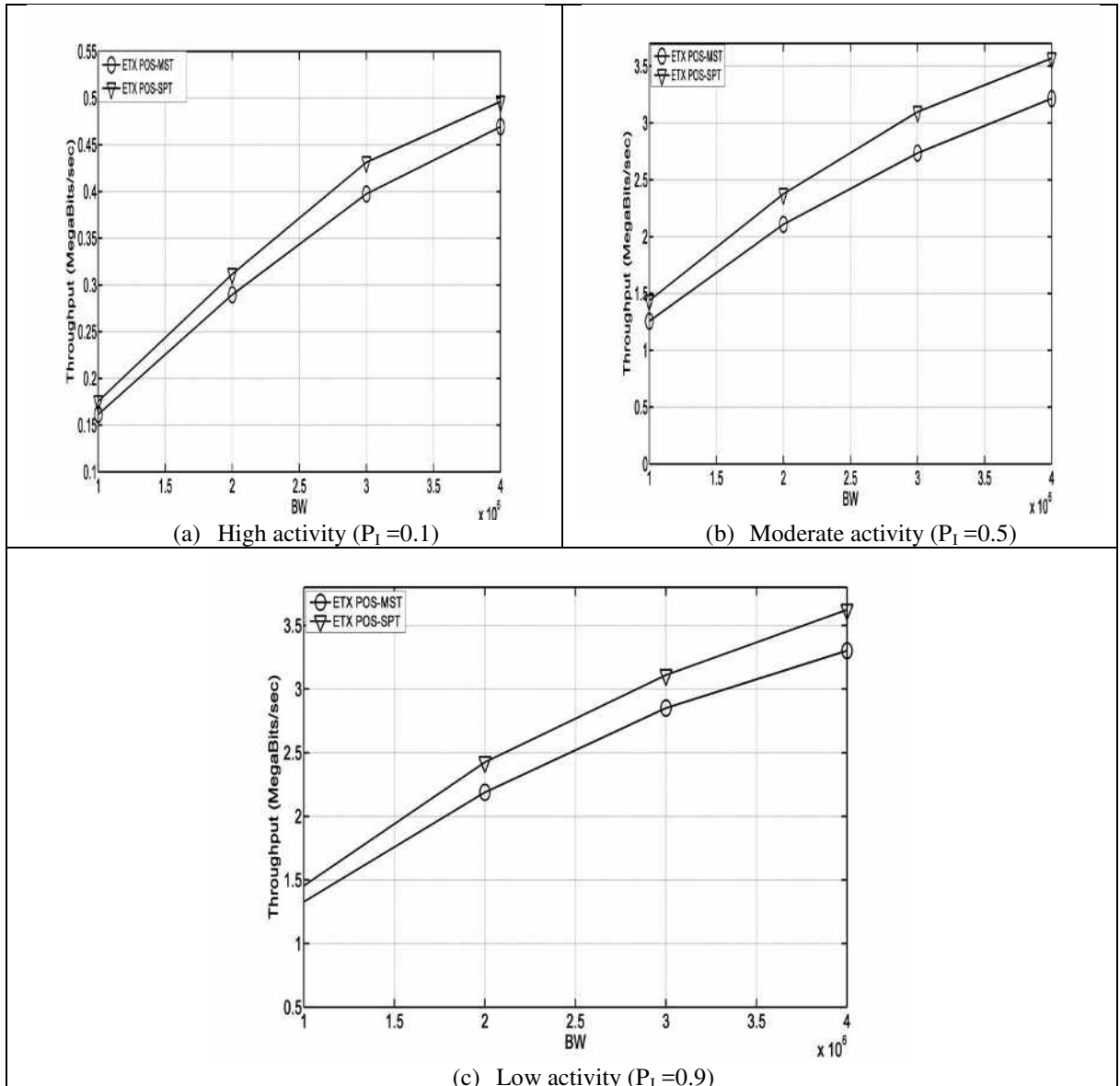

(a) High activity ($P_I$ =0.1)
(b) Moderate activity ($P_I$ =0.5)
(c) Low activity ($P_I$ =0.9)

**Figure 3.1:** Throughput vs. channel bandwidth under different PUs traffic (SPT vs. MST).

### 3.3.1.2 The PDR Performance versus Channel Bandwidth

Figure 3.2 shows the PDR as a function of the channel Bandwidth. As can be noticed, the PDR performance of both protocols increases as the BW increases. POS-SPT achieves 25%, 21%, and 12%, respectively.

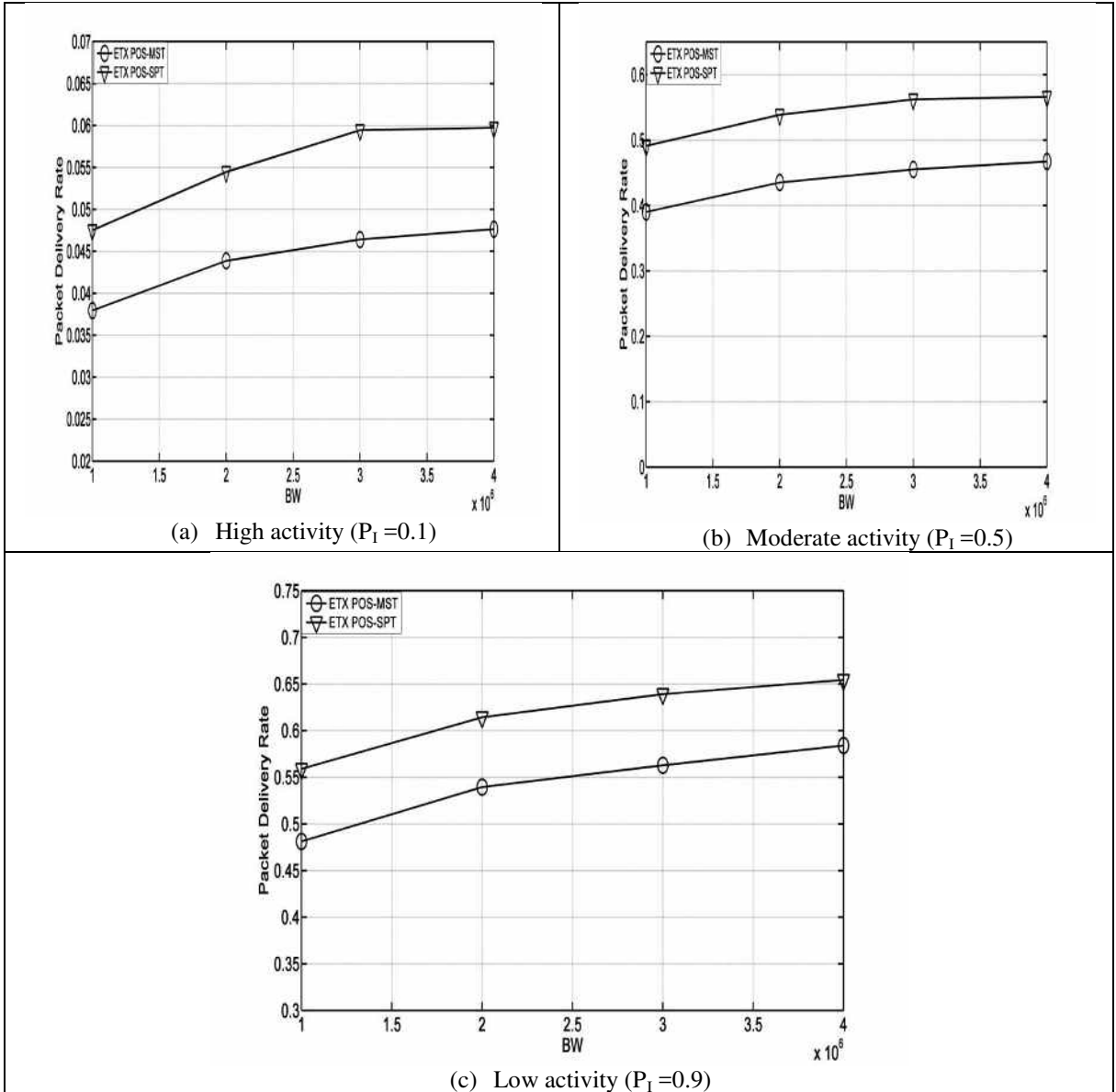

(a) High activity ($P_I$ =0.1)  (b) Moderate activity ($P_I$ =0.5)

(c) Low activity ($P_I$ =0.9)

**Figure 3.2:** The PDR vs. channel bandwidth under different PU traffic (SPT vs. MST).

### 3.3.2 Impact of Packet Size

We investigate the effects of increasing the packet size in terms of throughput and PDR. We consider the following network conditions: - M=40, $M_r$=16, N=20, $P_t$=0.1 W and BW=1 MHZ.

### 3.3.2.1 Throughput Performance versus the Packet Size

Figure 3.3 shows network throughput as a function of the packet size D. As the packet size increases, the throughput decreases. This is expected, which is in line with equations 3.

The improvement of POS-SPT as compared to POS-MST is 40%, 16%, and 8% over $P_I=0.1$, $P_I=0.5$, and $P_I=0.9$, respectively.

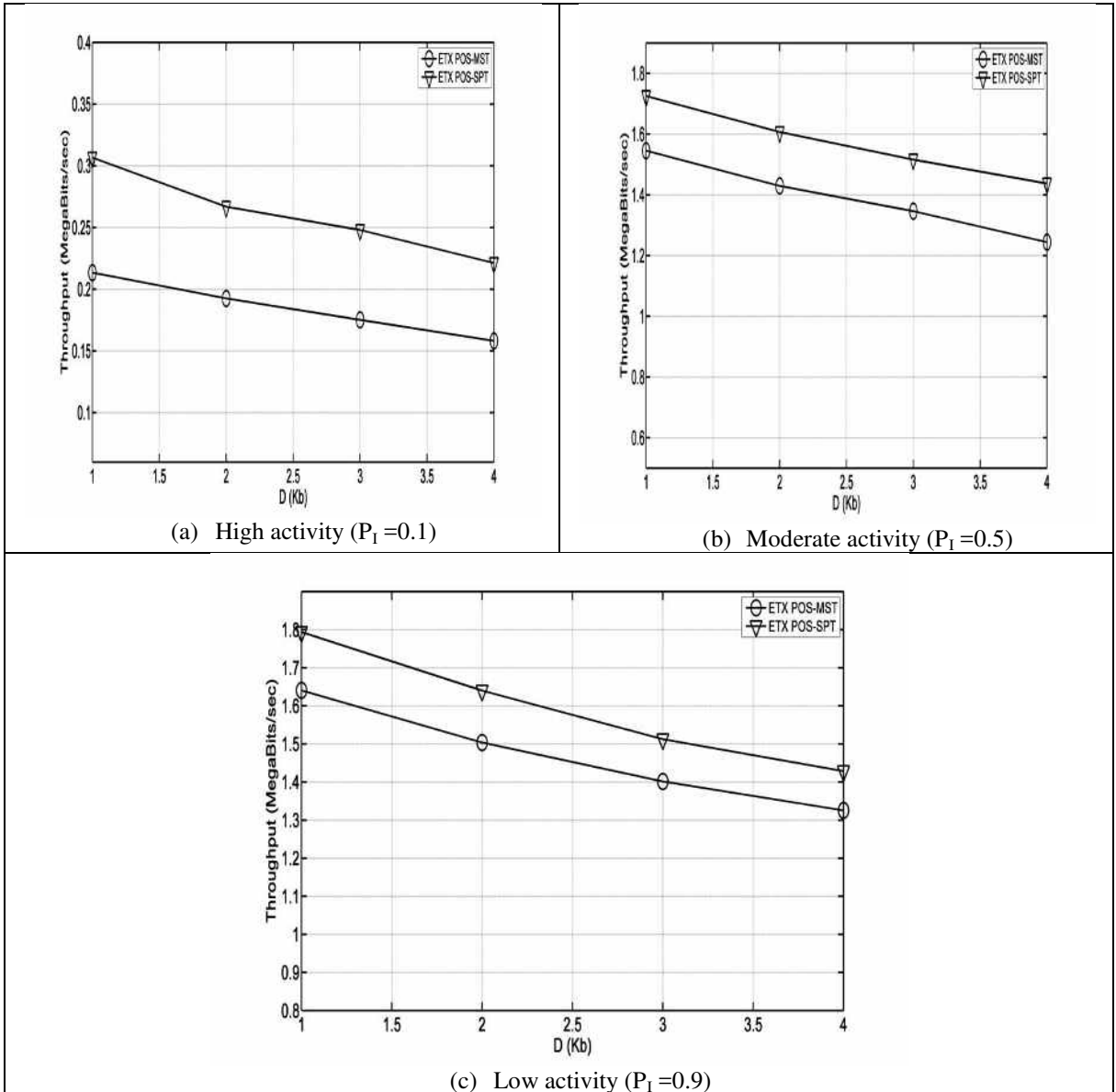

(a) High activity ($P_I =0.1$)  (b) Moderate activity ($P_I =0.5$)

(c) Low activity ($P_I =0.9$)

**Figure 3.3:** Throughput vs. channel packet size under different PU traffic (SPT vs. MST).

### 3.3.2.2 The PDR Performance versus the Packet Size

Figure 3.4 shows the PDR as a function of D. As the packet size increases, the throughput decreases. POS-SPT outperforms POS-MST with 117.5% at high activity, 25% at moderate activity, and 12% at low activity.

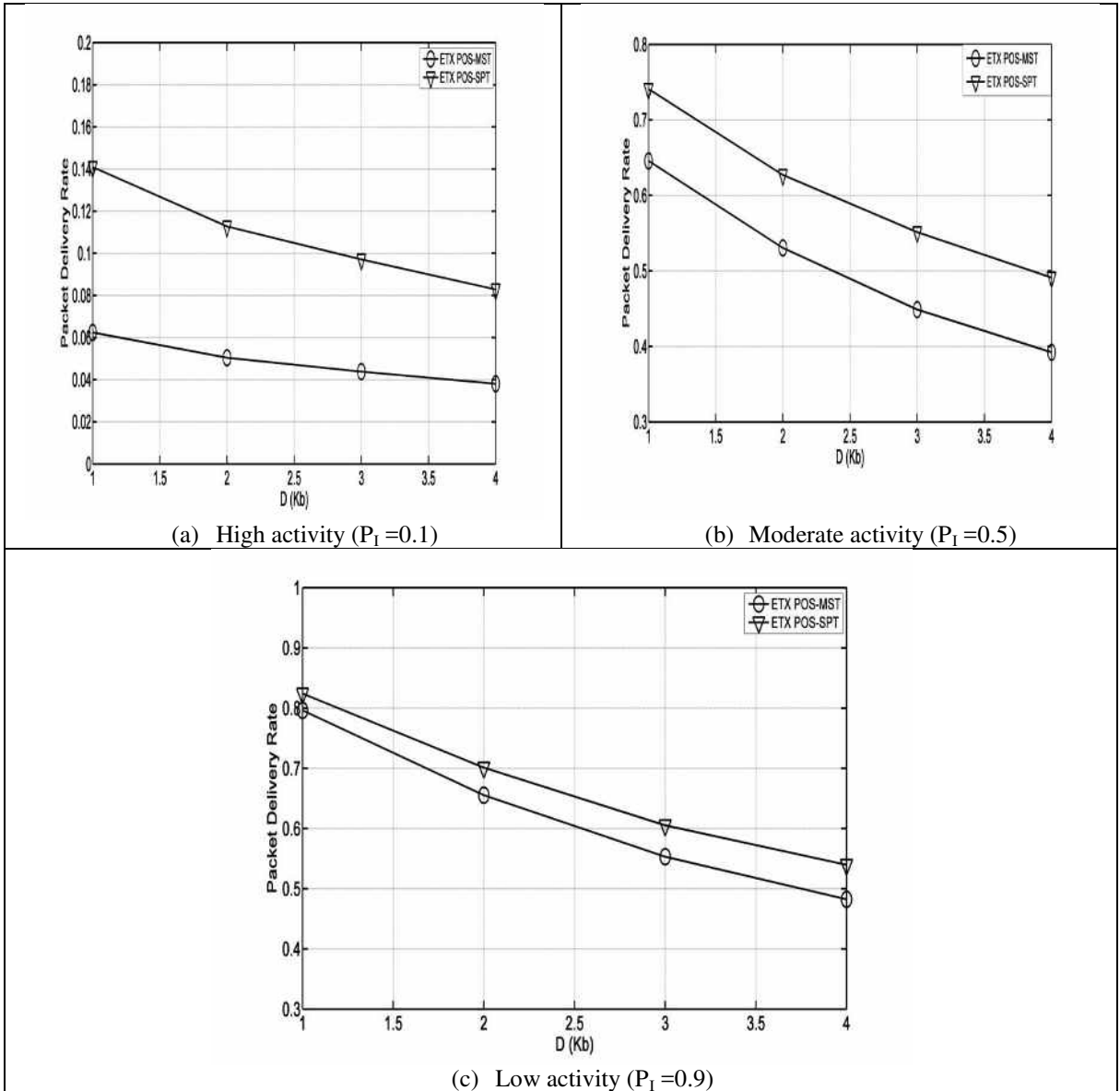

**Figure 3.4:** The PDR vs. Packet size under different PU traffic (SPT vs. MST).

### 3.3.3 Impact of the Number of PU Channels

In this section, we investigate the performance of the number of PU channels, in terms of throughput and PDR. We consider the following network conditions: M=40, $M_r$=16, $P_t$=0.1, BW=1 MHZ, W and D=4 KB, under different PUs activity.

### 3.3.3.1 Throughput Performance versus the Number of PU Channels

Figure 3.5 illustrates the effect of increasing the number of primary channels in network performance in terms of throughput, which increases as the channel availability increases. Thus, increasing the number of idle channels provides more chances to select the appropriate channels for transmissions. It achieves up to 28%, 9%, and 19.5% as the traffic load decreased.

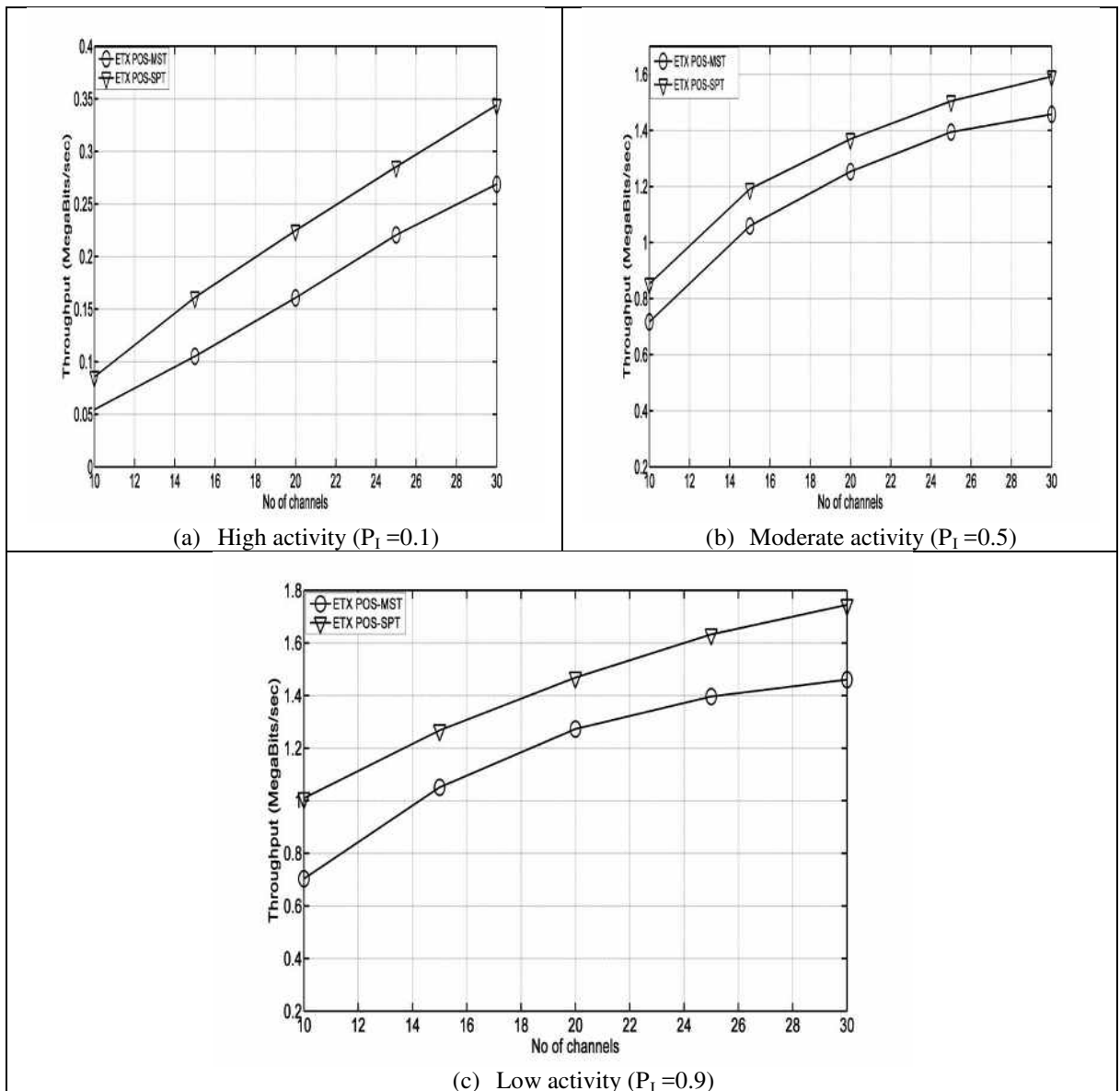

(a) High activity ($P_I$ =0.1)  (b) Moderate activity ($P_I$ =0.5)

(c) Low activity ($P_I$ =0.9)

**Figure 3.5:** Throughput vs. number of PU channels under different PU traffic (SPT vs. MST).

## 3.3.3.2 The PDR Performance versus the Number of PU Channels

Figure 3.6, illustrates the effect of increasing the number of primary channels in network performance in terms of PDR. It achieves up to 95.7%, 16%, and 36% as the traffic load decreased.

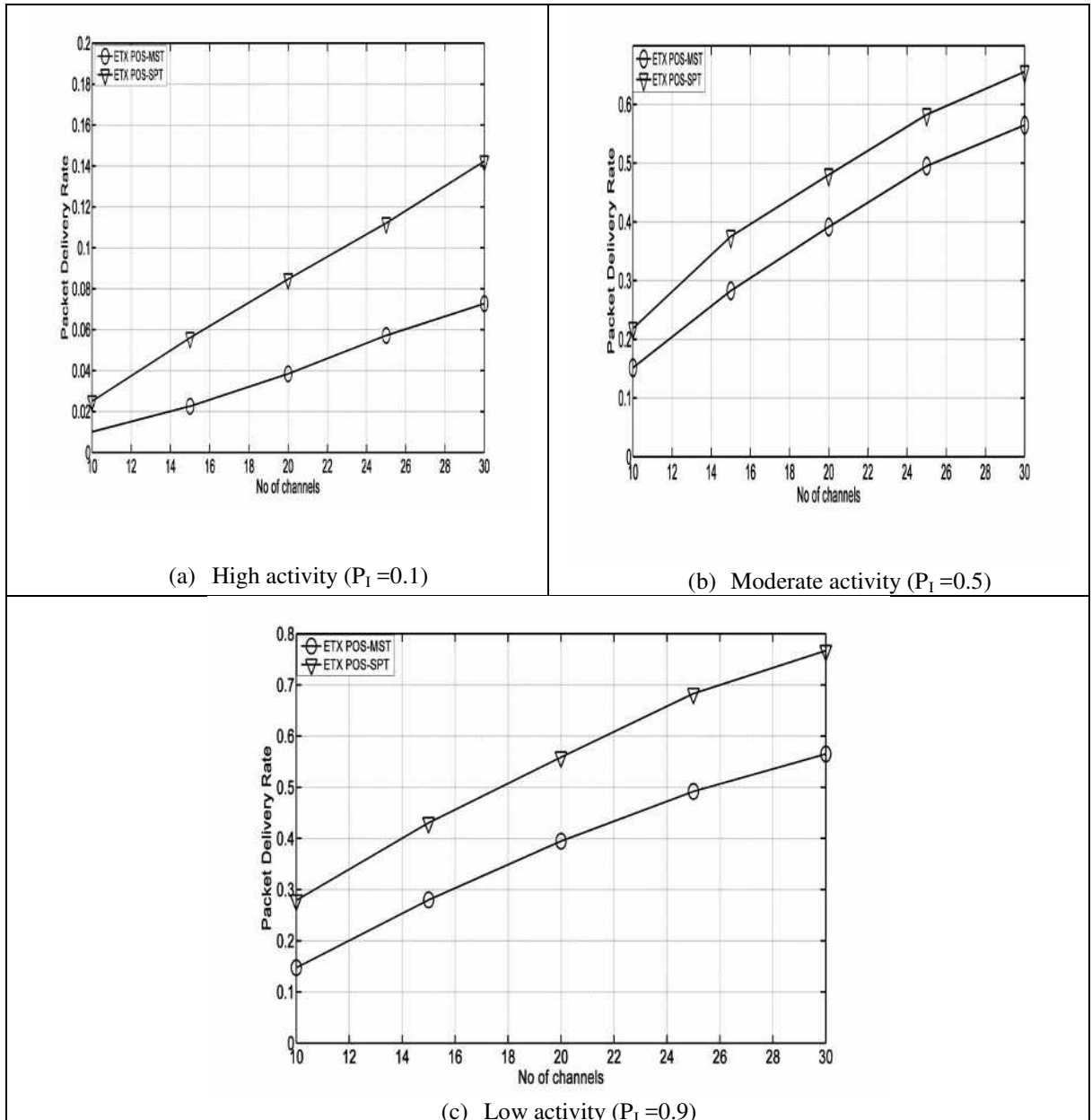

(a) High activity ($P_I$ =0.1)

(b) Moderate activity ($P_I$ =0.5)

(c) Low activity ($P_I$ =0.9)

**Figure 3.6:** The PDR vs. number of PU channels under different PU traffic (SPT vs. MST).

### 3.3.4 Impact of PUs Traffic Load

As the $P_I$ increases, the performance of the POS-SPT outperforms POS-MST in terms of network throughput and PDR. Figure 3.7 shows the improvement at lower rate ($P_I<=0.4$) is greater than higher rate of $P_I$; because it decreases the opportunities of cutting the transmission as long as the idle-probability increases,. It enhances the performance up to 10% over network throughput and 12% over the PDR.

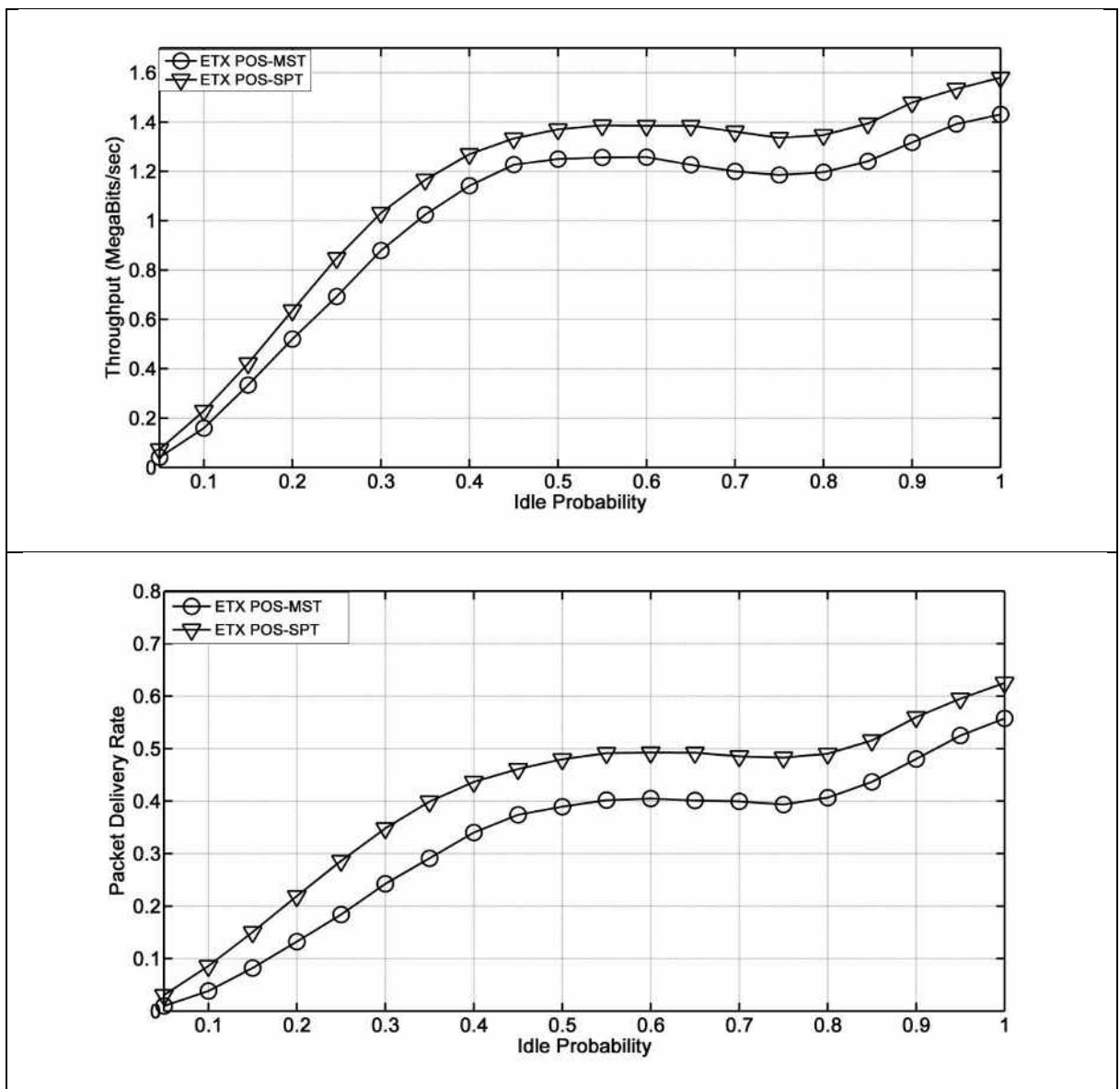

**Figure 3.7:** The PDR and Throughput vs. PU traffic (SPT vs. MST).

## 3.4 Performance Evaluation for Shortest Path Tree (SPT)

### 3.3.1 Impact of Channel Bandwidth

In this section, we investigate the throughput and PDR performance as a function of channel bandwidth. We consider the following network conditions: M=40, $M_r$=16, N=20, $P_t$=0.1 W and D=4 KB, at the different PU traffic loads.

### 3.4.1.1 Throughput Performance versus Channel Bandwidth

Figure 3.8 shows network throughput as a function of the channel Bandwidth (BW). This Figure shows that, the throughput increases as the BW increases for all traffic loads. However, the POS protocol outperforms the other protocols for all channel conditions because it uses the better channel assignment than the other protocols.

The throughput value of POS achieves 0.5 Mbps at BW 4 MHZ and a high activity of primary users. In this case, a small number of channels is available for SUs. Approximately, it is 10% from the total channels (20 channels). It means that SU has only one or two chances for transmission at each time. The achieved improvements are 106% and 118% for (POS and MASA) as compared to MDR, and RS, respectively. At moderate activity of primary users, the POS performance achieves 5%, 152%, and 173% improvement over MASA, MDR, and RS, respectively. Due to increasing the channels availability, this is shown in Figure 3.8(b).

For $P_I$=0.9, most of the channels are available to be used by SUs. POS achieves 3.6 Mbps at 4 MHZ, (i.e., it achieves up to 4.6%, 31%, and 51.7% improvement over MASA, MDR, and RS, respectively). Therefore, POS achieves the best performance at lower activity of PUs in compare to moderate and high activity. We note that the performance of MASA is closer to POS than the other schemes at $P_I$=0.1; due to the fewer number of availability channels. The improvement increases as the number of channels availability increases.

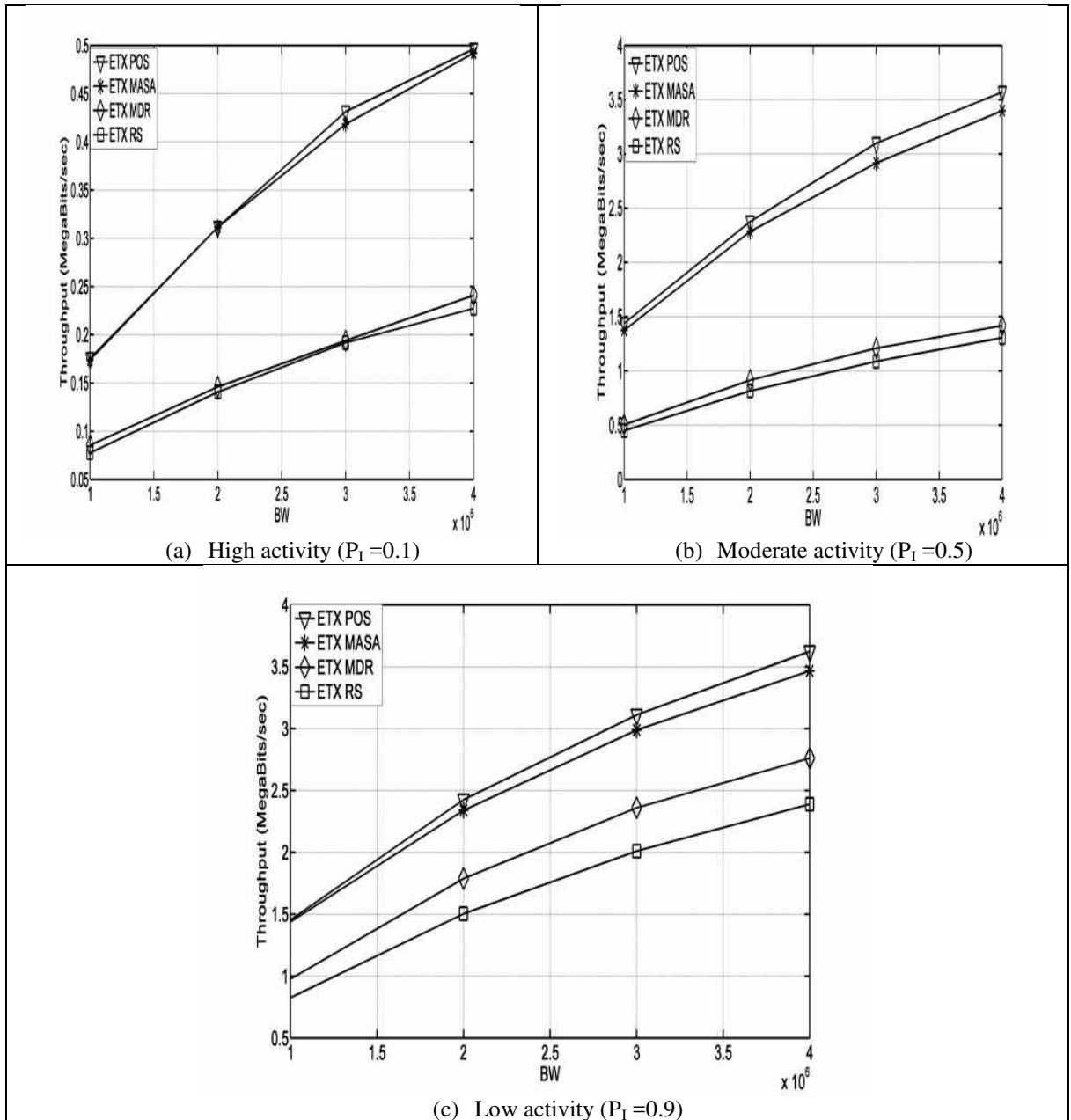

**Figure 3.8:** Throughput vs. channel bandwidth under different PUs traffic (SPT).

### 3.4.1.2 The Packet Delivery Rate Performance versus Channel Bandwidth

Figure 3.9 shows the PDR as a function of the channel Bandwidth. As can be noticed, all protocols result in increasing the PDR performance as BW increases. POS protocol outperforms the other protocols. At high activity of primary users, (POS and MASA) achieve 122%, and 148% as compared to MDR, and RS, respectively that shown in Figure 3.9(a). At moderate activity of primary users, POS achieves 2.45%, 202%, and 277% in compare to MASA, MDR, and RS, respectively. Regarding the low activity, the improvement is 5.7%, 59.1%, and 120%, by referring to both Figures 3.9(b) and 3.9(c). The best performance of the proposed protocol achieves in Figure 3.9(c) at idle probability of 0.9.

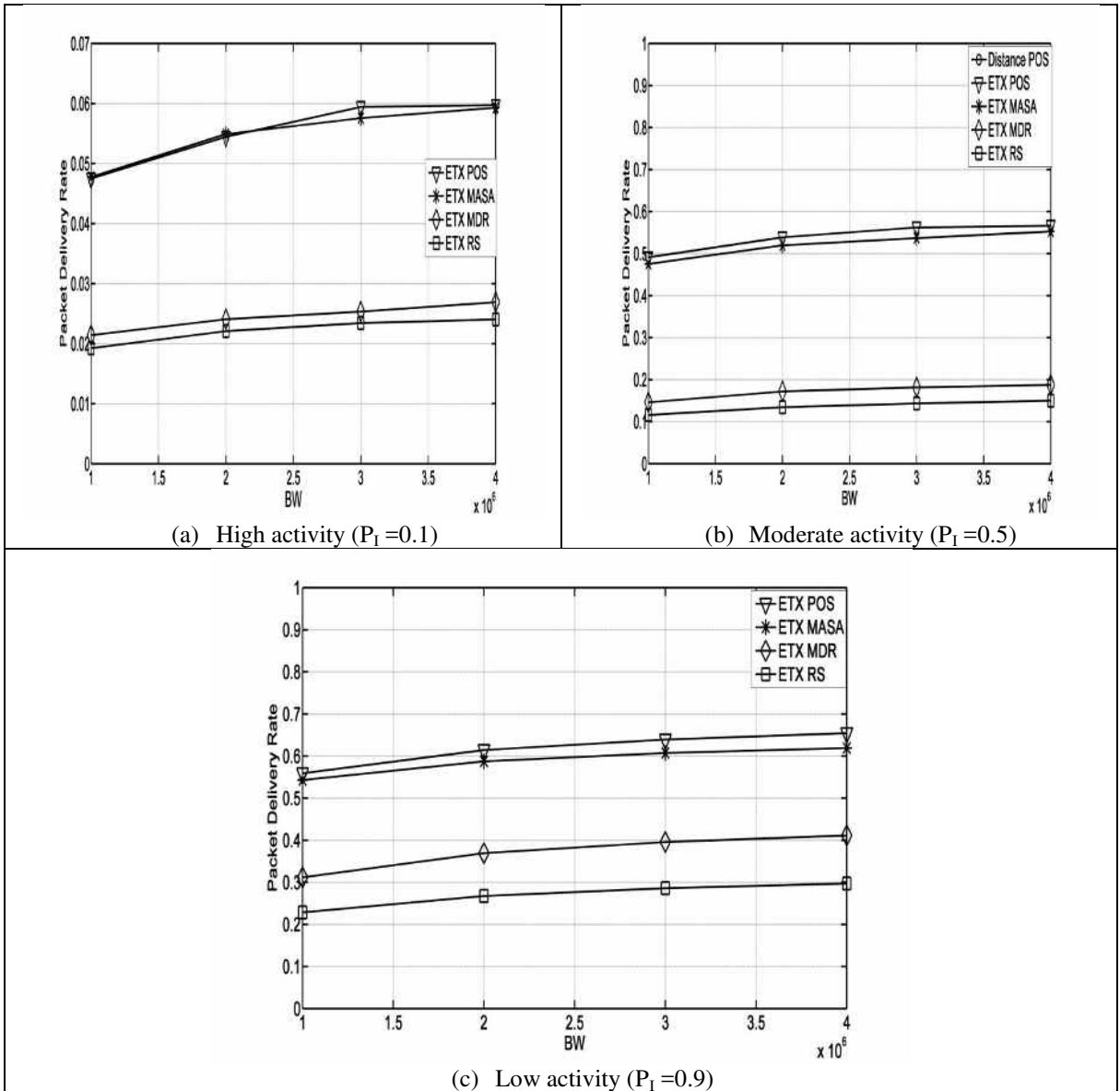

**Figure 3.9:** The PDR vs. channel bandwidth under different PU traffic (SPT).

**3.3.2 Impact of the Packet Size**

We investigate the effects of increasing packet size in terms of throughput, and PDR. We consider the following network conditions: M=40, $M_r$=16, N=20, $P_t$=0.1 W and BW=1 MHZ,

**3.4.2.1 Throughput Performance versus Packet Size**

Figure 3.10 shows network throughput as a function of the packet size D. As the packet size increases, the throughput for all protocols decreases. This expected, which is in line with equations 3. The packet size affects on the actual transmission time needed for a successful transmission. If D increases, the required time increases. Thus, we need to choose channels with more channel availability.

At high activity of PUs $P_I$=0.1, (POS and MASA) outperform MDR, and RS with 102%, and 116.7%, respectively as shown in Figure 3.10(a). At moderate and low activity of PUs, similar performance is reported in Figure 3.10(b) and (c). POS slightly outperforms MASA by 5.2%. At low activity, (MASA and POS) outperform MDR and RS by 49%, and 70.7%, respectively. The improvement increases as PU traffic loads decreases. However, the best improvements achieves at idle probability of 0.9.

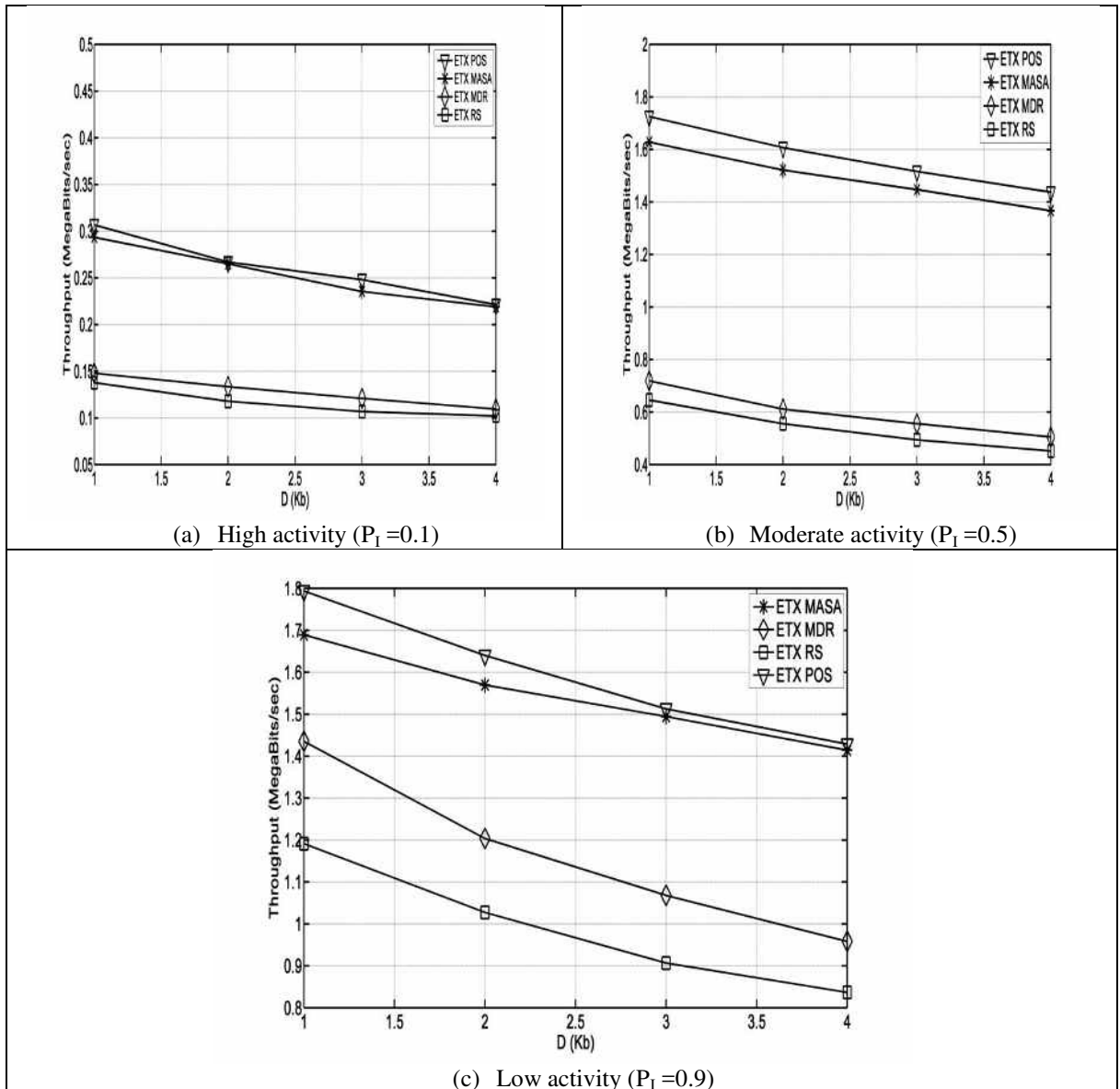

**Figure 3.10:** Throughput vs. channel packet size under different PU traffic (SPT).

**3.4.2.2 The Packet Delivery Rate Performance versus the Packet Size**

Figure 3.11 shows the PDR as a function of D. This Figure shows that, all protocols decrease the performance of PDR as D increases. At $P_I$=0.1, there are slightly difference among all protocols in terms of the PDR. The performance of MASA and POS are comparable and outperform MDR and RS; because the limited number of channel available in the network. The improvements are 110% and 143.5% compared to MDR, and RS, respectively. At $P_I$=0.5, the improvement of POS reaches 3.4% for MASA, 232% for MDR, and 320% for RS. The performance improves as a result of increasing in channel availability. At $P_I=0.9$, 90% of PU channels are occupied. The improvements of (POS and MASA) are 77.4% and 135% compared to the other schemes (Figure 3.11(c)). We note that POS slightly outperforms MASA.

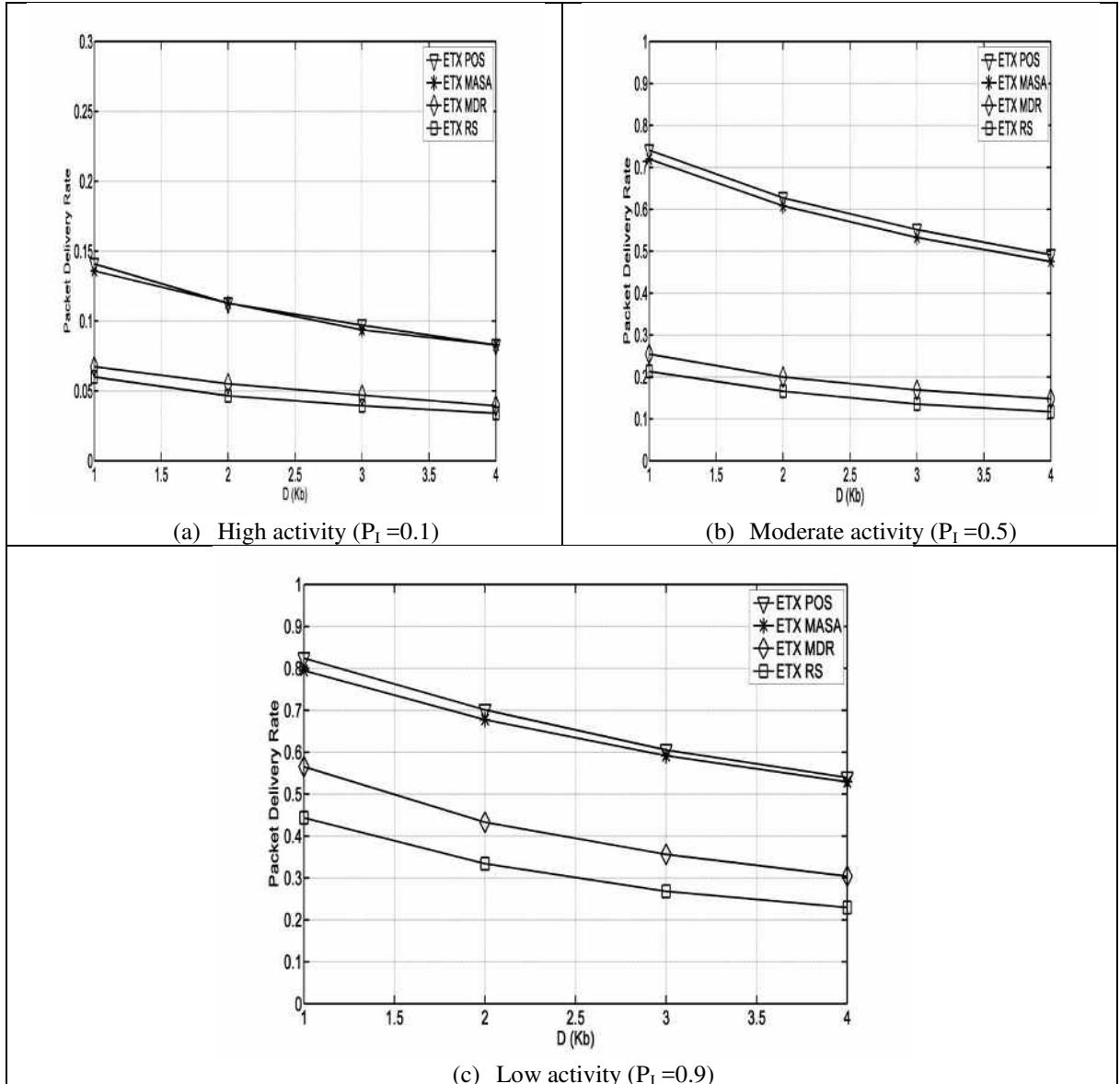

**Figure 3.11:** The PDR vs. Packet size under different PU traffic (SPT).

### 3.3.3 Impact of the Number of PU Channels

In this section, we investigate the performance of the number of PU channels, in terms of throughput, and PDR. We consider the following network conditions: $M=40$, $M_r=16$, $BW=1$ MHZ, $P_t=0.1$ w and $D=4$ KB, under different PUs activity.

**3.4.3.1 Throughput Performance versus the Number of PU Channels**

Figure 3.12, illustrates the effect of increasing the number of primary channels in network performance in terms of throughput, which increases as the channel availability increases. Thus, increasing the number of idle channels provides more chances to select the appropriate channels for transmissions. The proposed protocol POS outperforms the others for all situations. And taken the consideration, the channels range from 10 to 30.

The best improvement reaches by POS shown in Figure 3.12(c), which refers to primary traffic load equals 0.9. According to MASA, its improvement achieves 6.3% when the channels exceed 20 channels. POS outperforms MDR with 64%, and 73% for RS.

Figure 3.12(b) illustrates the throughput evaluation of the proposed protocols for moderate activity $P_I$=0.5. POS and MASA outperform up to 150%, and 186% compared to the other protocols.

In Figure 3.12(a), we check the performance of POS and MASA, they achieves 170% and 180% for MDR, and RS, respectively.

As a result, the improvement ratios decreases as long as the activity of PUs decreases, which indicates that POS has not only enhances its performance, but also all schemes have achieved the best performance over idle-probability=0.9.

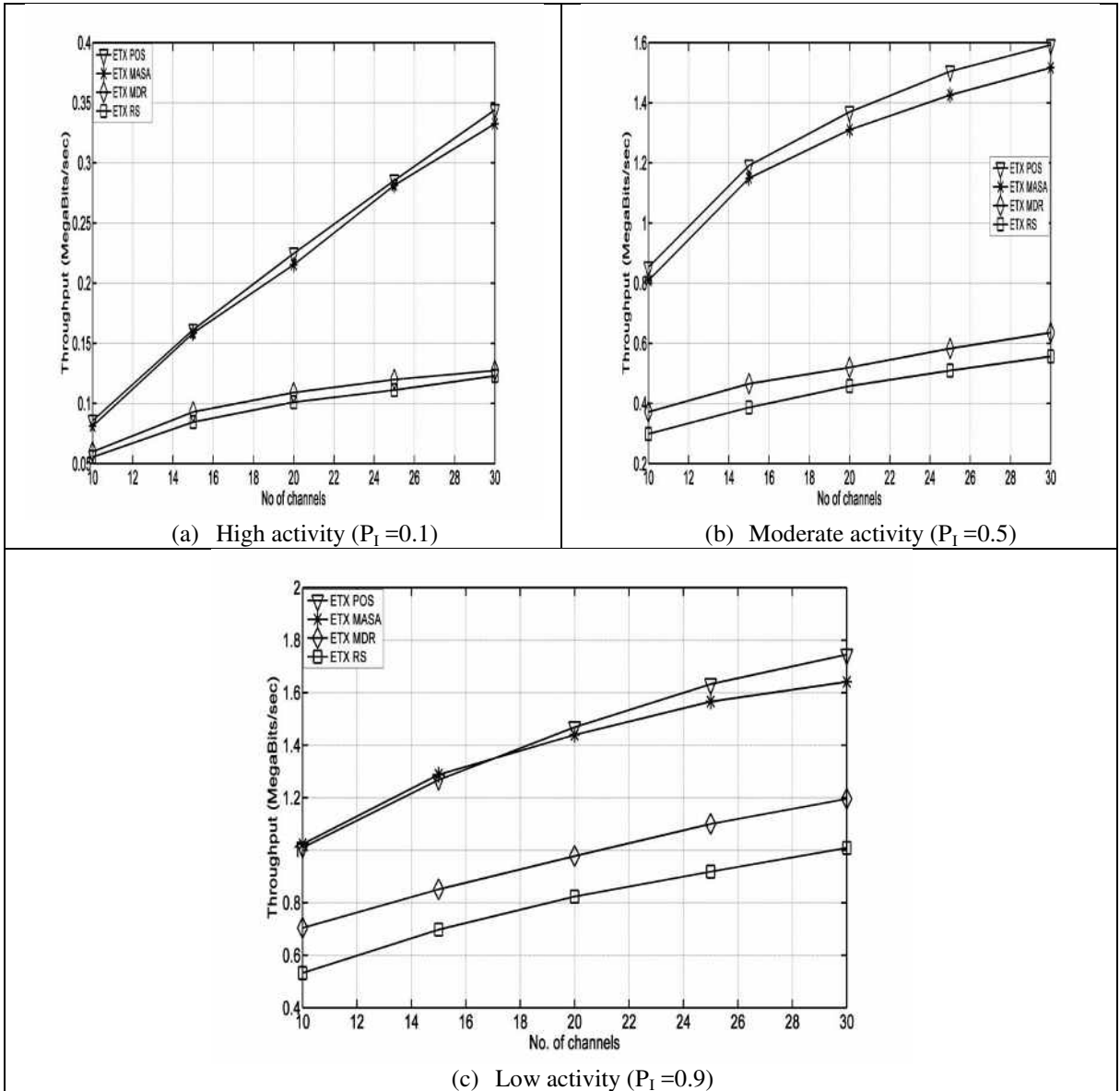
**Figure 3.12:** Throughput vs. number of PU channels under different PU traffic (SPT).

### 3.4.3.2 The PDR Performance versus the Number of PU Channels

Figure 3.13 shows the PDR as a function of increasing the number of primary channels in network performance in terms of PDR, the improvements increase as the channels increase for all protocols. The result achieves at idle-probability equal 0.5 and 0.9 are better than it achieves at idle-probability equals 0.1. The improvement of the proposed algorithm increases as the idle-probability increases from 0.1 to 0.9. Figure 3.13 illustrates this improvement.

(POS and MASA) outperform MDR, and RS in changing the traffic loads over the following ratios: 178.65%, and 215.2% over idle-probability equals 0.1, 191%, and 268.5% at idle-probability equals 0.5, and reaches the best improvement at idle-probability equals 0.9. Slightly improvement has achieved between MASA and POS.

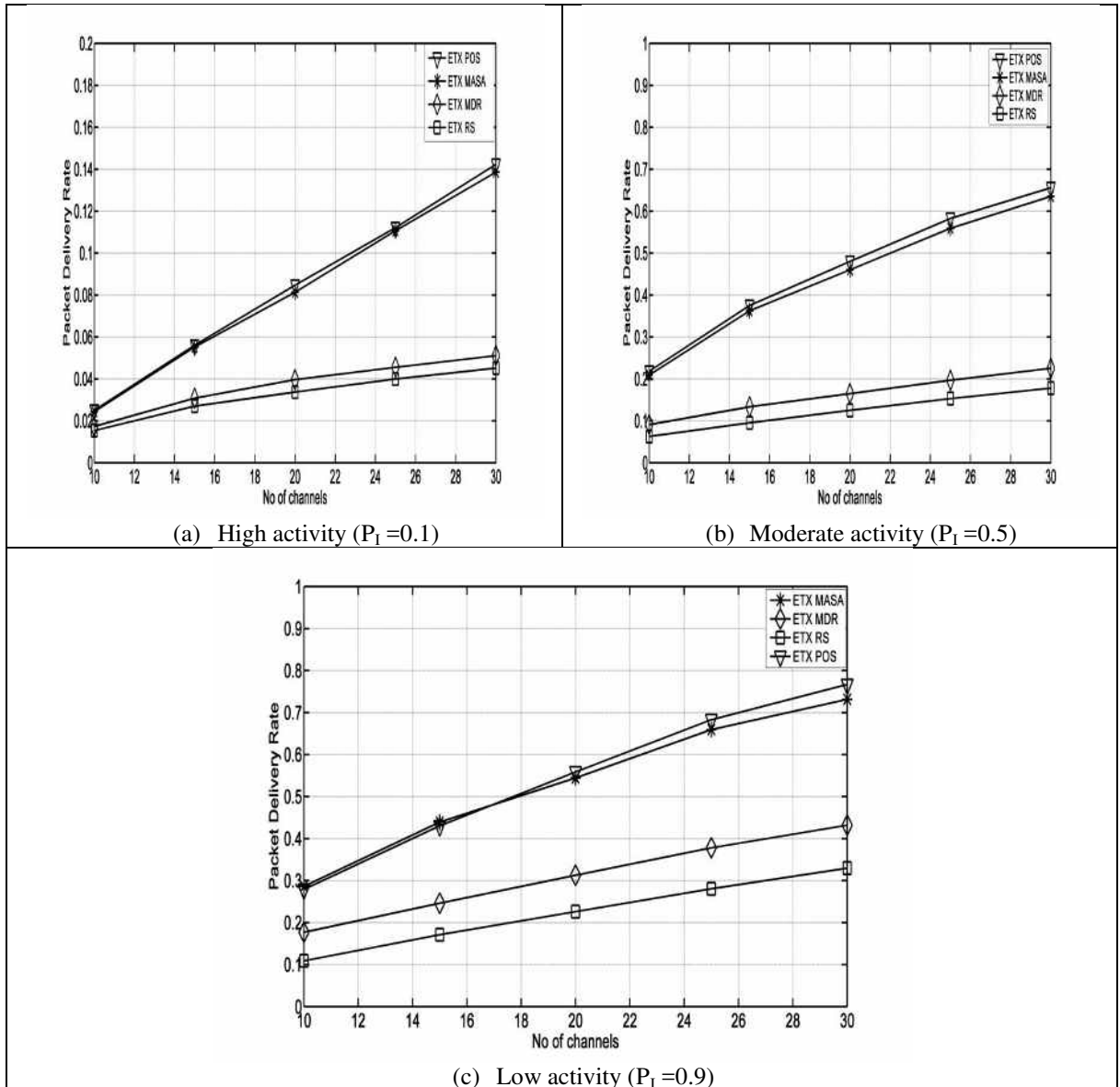

(a) High activity ($P_I$ =0.1)   (b) Moderate activity ($P_I$ =0.5)

(c) Low activity ($P_I$ =0.9)

**Figure 3.13:** The PDR vs. number of PU channels under different PU traffic (SPT).

### 3.3.4 Impact of the Transmissions Power

We investigate the effects of increasing the transmission power in terms of throughput, and PDR. We consider the following network conditions: M=40, $M_r$=16, N=20, BW=1 MHZ and D=4 KB,

**3.4.4.1 Throughput Performance versus the Transmissions Power**

Transmission power is the most network conditions that significantly affects on the performance of the transmission time needed and the data rate of networks. As the transmission power increases, the data rate (throughput) increases and the required transmission time decreases. However, the maximum transmission power that used by CUs are limited, in comparing to PUs use. CU uses a lower power to protect the performance PRNs and to keep the interference between them under certain threshold.

In Figure 3.14 shows the performance of the proposed protocols in terms throughput as a function of increasing the transmission power at various traffic loads. The best performance achieves in $P_I$=0.9. The throughput increases as the transmission power increases. POS outperforms the others with 4.5%, 45.3%, and 75.5%, respectively.

The improvement of (POS and MASA) over $P_I$=0.1 reaches to 99.8%, and 125.3% as compared to MDR, and RS, respectively. The performance of POS and MASA achieve 170.3%, and 208.2% as compared to MDR and RS.

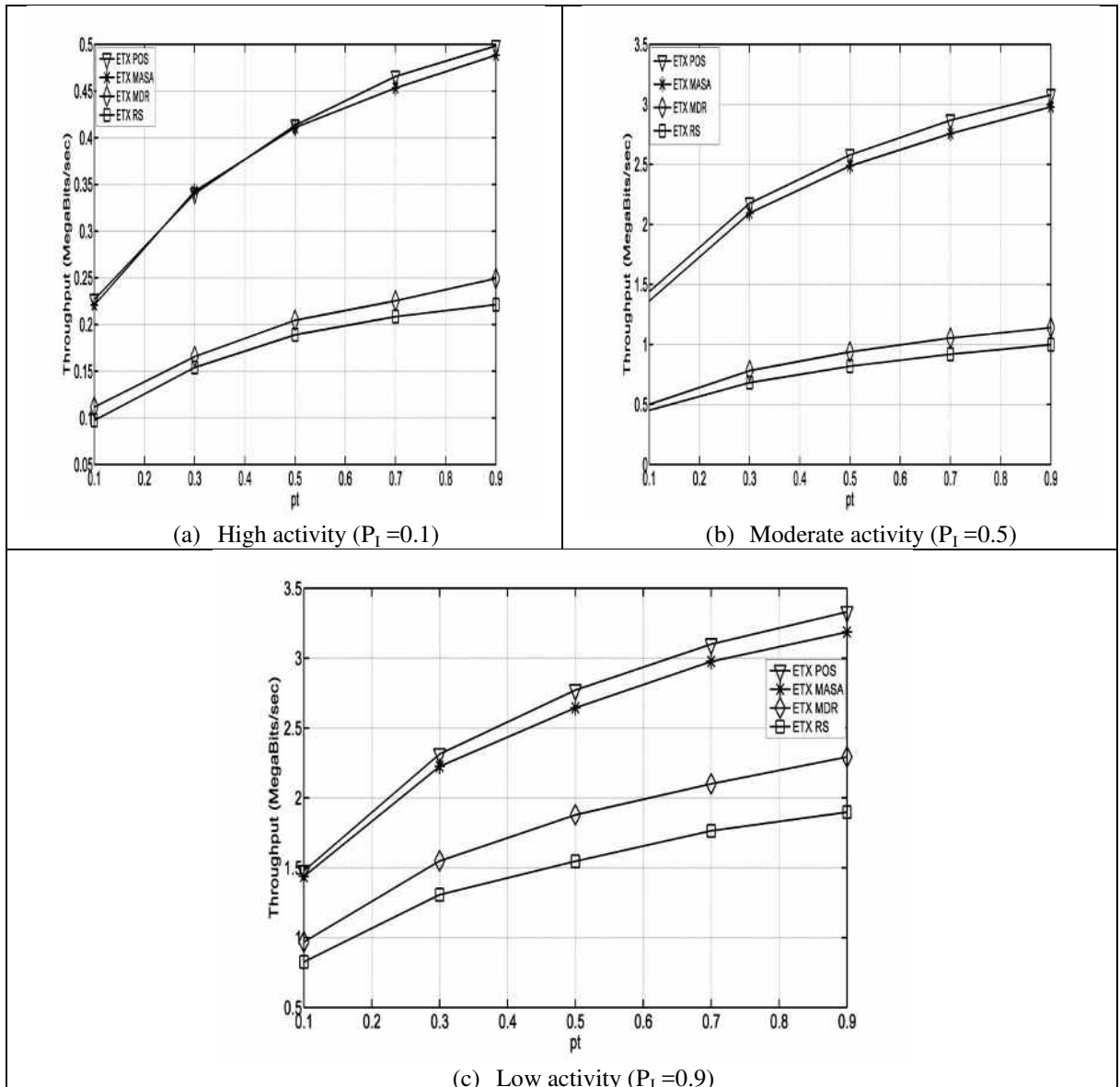

(a) High activity ($P_I$ =0.1)

(b) Moderate activity ($P_I$ =0.5)

(c) Low activity ($P_I$ =0.9)

**Figure 3.14:** Throughput vs. power transmissions under different PU traffic (SPT).

### 3.4.4.2 The PDR Performance versus the Transmissions Power

We illustrate the performance of PDR as a function of transmission power. According to Figure 3.15, the PDR performance increases as the power increases. (POS and MASA) achieve the best performance as compared to the other protocols. With 107%, and 138% over $P_I$=0.1. The improvements of (MASA and POS) are 223% and 274%, which referred to Figure 3.15(b). The improvements in Figure 3.15(c) are 67.4%, and 104% achieved by (POS and MASA) over $P_I$=0.9. We note, the performance of the POS slightly enhances the performance of the MASA.

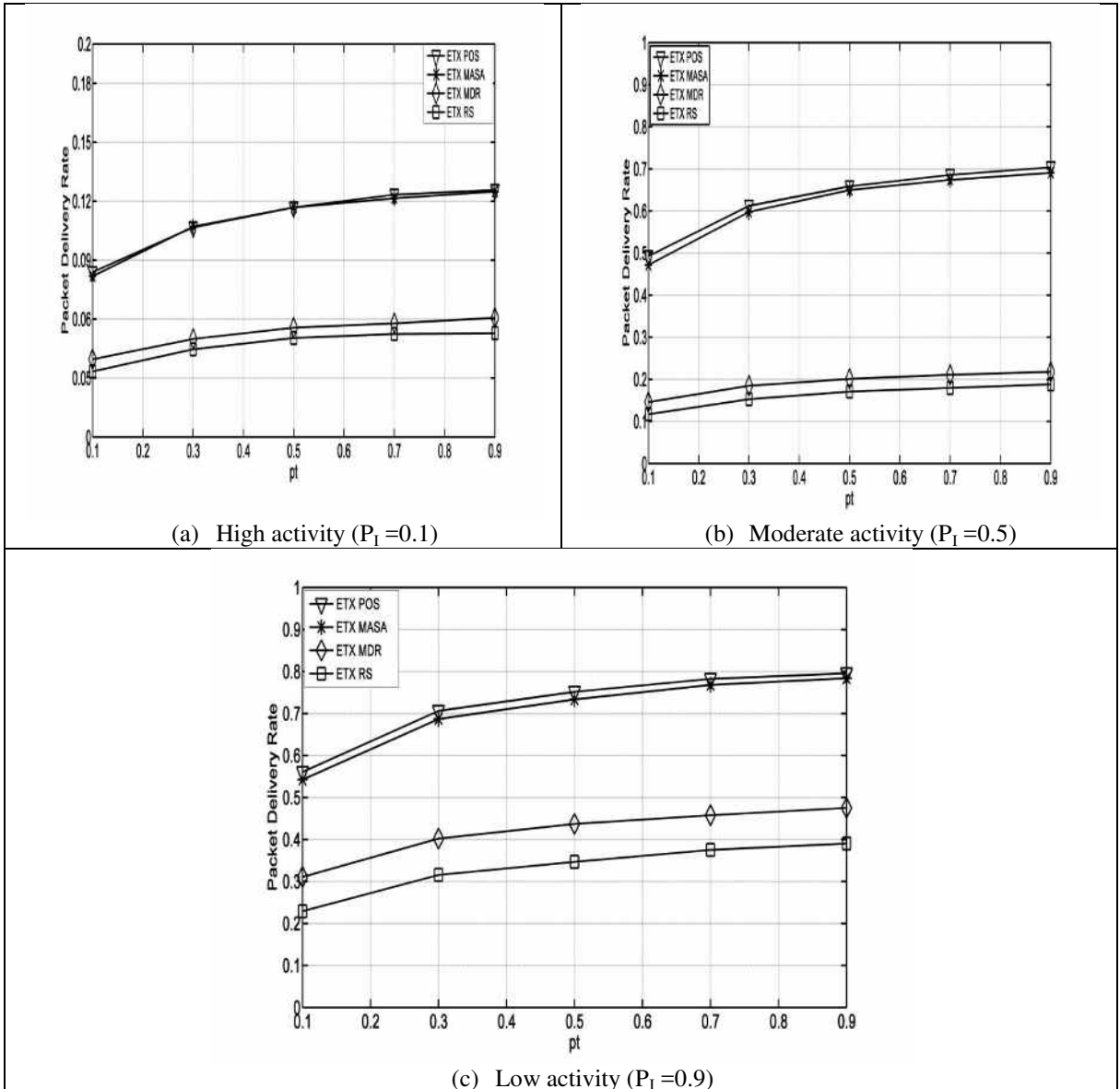

(a) High activity ($P_I$ =0.1)　　(b) Moderate activity ($P_I$ =0.5)

(c) Low activity ($P_I$ =0.9)

**Figure 3.15**: The PDR vs. power transmissions under different PU traffic (SPT).

### 3.3.5 Impact of the Number of Nodes in Network Performance

We investigate the performance of increasing the number of nodes, in terms throughput, and PDR. We consider the following network conditions: $M_r$=16, N=20, BW=1 MHZ, $P_t$=0.1 w and D=4 KB, over PU traffic loads.

### 3.4.5.1 Throughput Performance versus the Number of Nodes

Figure 3.16, illustrates the performance of throughput as a function of increasing the number of nodes that including in the network. The numbers of nodes range from 20 to 100 nodes do not make great differences for all protocols. The network complexity is increasing. POS outperforms all protocols for different traffic loads. At lower $P_I$, (MASA and POS) achieve 115.4%, and 138%. At moderate $P_I$, they achieve 155%, and 210%. At higher $P_I$, they are 5.6%, and 79.6%. These ratios consider the improvements performance of (POS and MASA) over MDR, and RS.

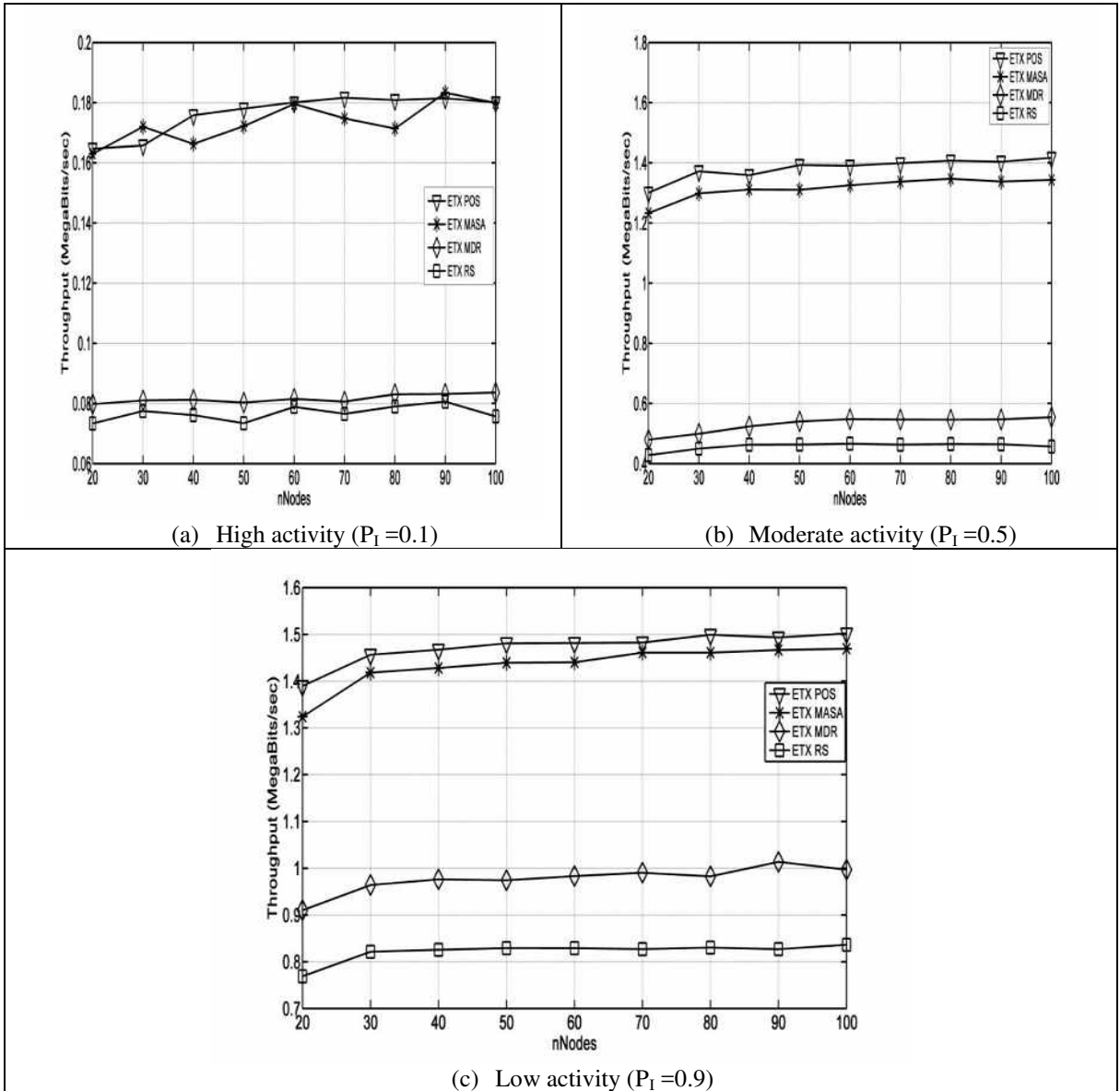

(a) High activity ($P_I$ =0.1)  (b) Moderate activity ($P_I$ =0.5)

(c) Low activity ($P_I$ =0.9)

**Figure 3.16:** Throughput vs. number of nodes in network under different PU traffic (SPT).

## 3.4.5.2 The PDR Performance versus the Number of Nodes

In general, increasing in the number of nodes is not significantly associated increasing the PDR performance. The improvement becomes to be constant as $P_I$ increases. The POS achieves a minimum improvement over the MASA. (MASA and POS) outperform MDR and RS with 129.7%, and 172.7%, at lower $P_I$. They provide improvements up to 173%, and 291.6% at moderate $P_I$. And, it reach 70.6%, and 139.6% at high level. This has shown in Figure 3.17.

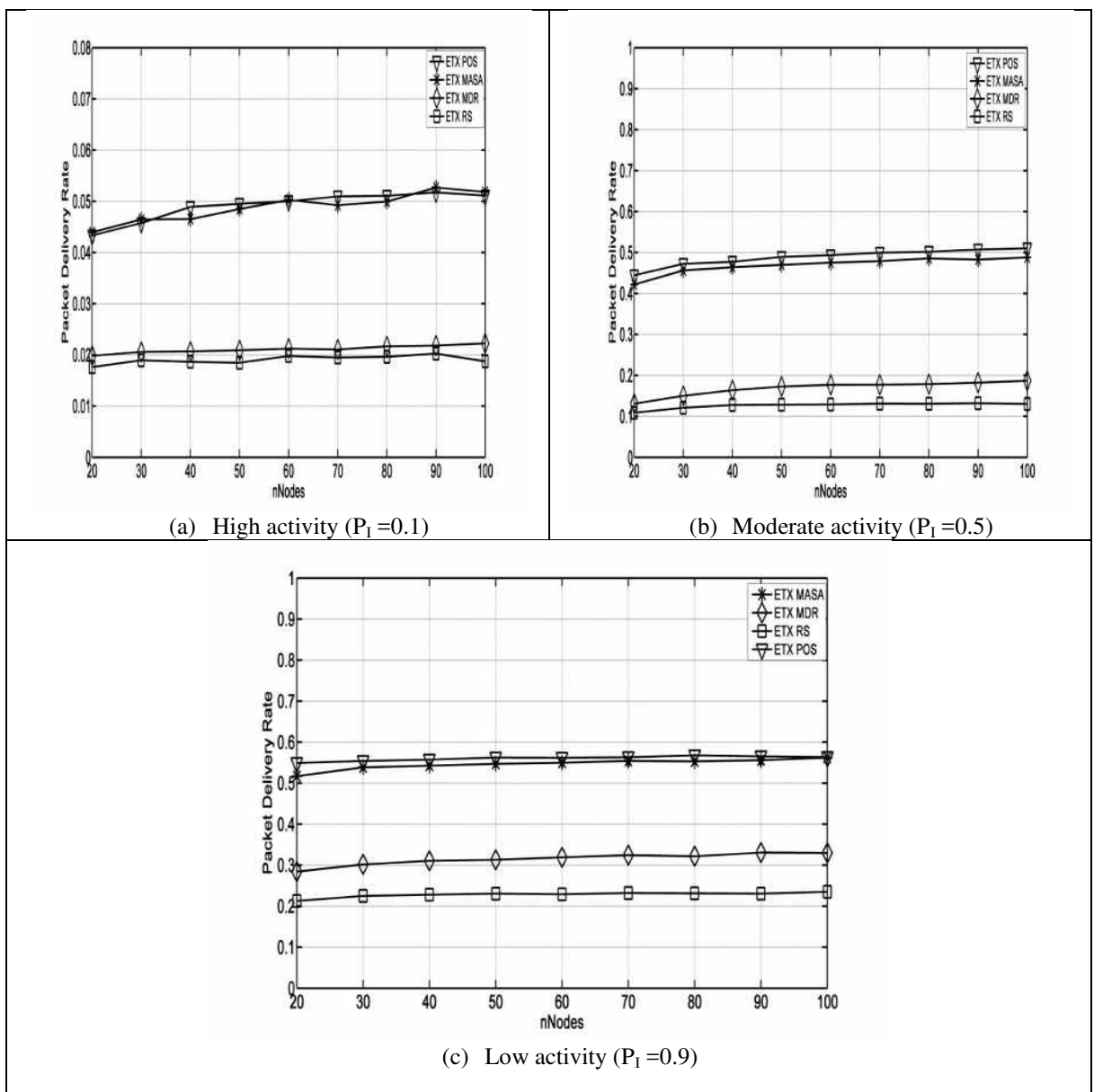

(a) High activity ($P_I$ =0.1)  (b) Moderate activity ($P_I$ =0.5)

(c) Low activity ($P_I$ =0.9)

**Figure 3.17:** The PDR vs. number of nodes in network under different PU traffic (SPT).

### 3.3.6 Impact of the Number of Destinations in the Network

In this section, we illustrate the performance of the number of destinations, in terms throughput, and PDR. We consider the following network conditions: M=40, N=20, BW=1 MHZ, Pt=0.1 w and D=4 KB, under different PUs activity.

### 3.4.6.1 Throughput Performance versus the Number of Destination Nodes

Figure 3.18 summarizes the impact of increasing the number of destination nodes in network throughput. However, when the destination nodes increase, we need more resources. So, we need to increase the number of channel availability, which is difficult in CRN; because of the limitation number of channels that depends only on the traffic load of PUs. Thus, the network goes down as the destinations increases.

At high and low activity of PUs, the network throughput decreases as the destinations increases. Figure 3.18(a), POS compared with other existed protocols at high activity and it achieves 4.84%, 69.2%, and 63.4%, respectively. In addition, it achieves 3.51%, 48.5%, 63.2% over low activity of PU, in Figure 3.18(c). The difference between two Figures, POS downfalls sharply from the maximum value to minimum at $P_I$=0.1. Because of decreasing the channels availability.

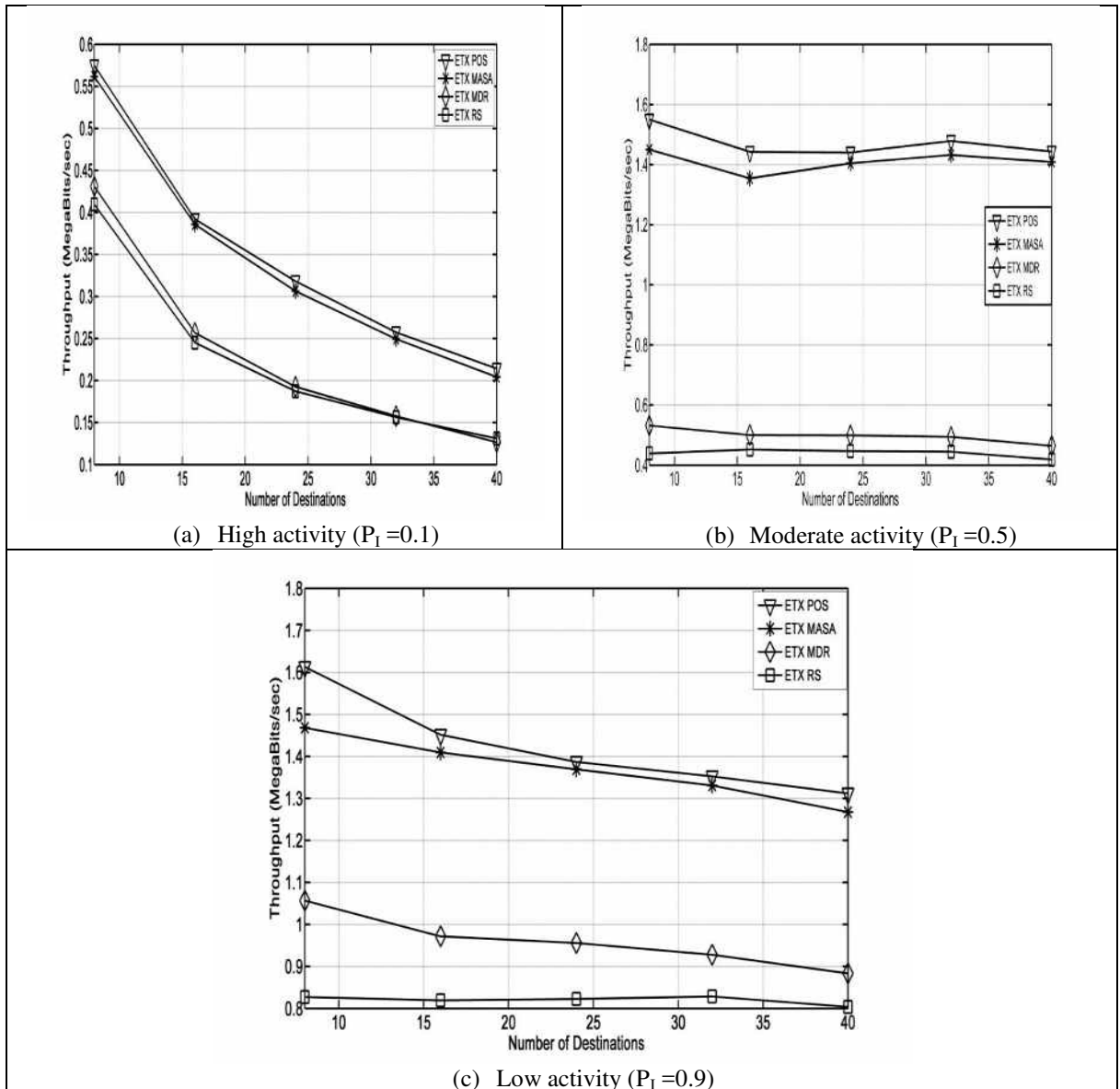

**Figure 3.18:** Throughput vs. number of destinations in network under different PU traffic (SPT).

### 3.4.6.2 The PDR Performance versus the Number of Destinations

In Figure 3.19 shows the PDR performance as a function of increasing the number of destination nodes. The PDR performance decreases as the destination nodes increases. (MASA and POS) outperform MDR and RS for different traffic loads. At lower $P_I$, they achieve 74.2%, and 76.3%. It reaches at moderate $P_I$, 276.8%, and 357.5%. And, at higher $P_I$ 75.8%, and 120.6%.

The performance of POS increases slightly at the destinations less than 16 nodes, and then it becomes constant, at $P_I=0.5$. Whereas $P_I=0.9$, the performance decreases at the nodes less than 16 and then becomes constant.

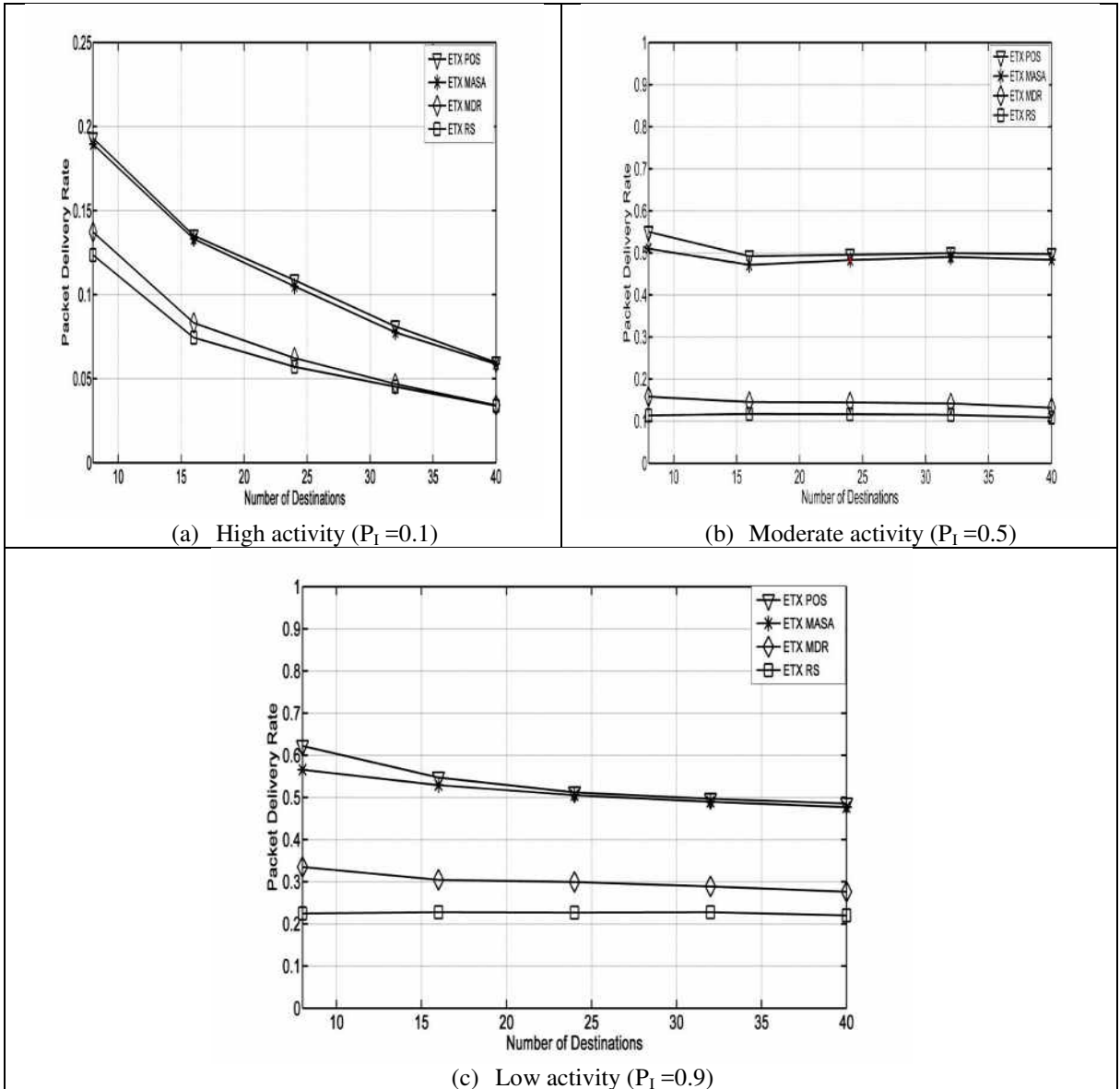

(a) High activity ($P_I =0.1$)  
(b) Moderate activity ($P_I =0.5$)  
(c) Low activity ($P_I =0.9$)  

**Figure 3.19:** The PDR vs. number of destinations in network under different PU traffic (SPT).

## 3.3.7 Impacts of the Maximum Transmission Range on Network Performance

In this section, we investigate the performance of increasing the maximum range between two nodes in the network, in terms network throughput, and the PDR. We consider the following network conditions: M=40, $M_r$=16, N=20, $P_t$=0.1 W and D=4 K, and BW=1 MHZ, under different PU traffic loads.

### 3.4.7.1 Throughput Performance versus the Transmission Range

Figure 3.20 shows the performance of network throughput as a function of increasing the maximum range between two nodes in the network. It has a significant impact on the network. In order to increase the range, we need to increase the transmission power. Thus, the interference with PRN increase. The range is determined based on the environment of networks, also, to avoid wireless problems (for example, hidden and exposed problems).

At lower rate of $P_I$, the network throughput increases as long as the range increases, because the limited number of channels availability and we need to keep the connectivity of network. Then, the throughput becomes constant at moderate of $P_I$. And, the throughput decreases as the range increases, at high rate of $P_I$.

 (MASA and POS) outperform MDR and RS for different traffic loads. At lower $P_I$, achieve 106.7%, and 127.4%. At moderate $P_I$, achieve 167%, and 190.8%. And, at higher $P_I$, are 40.4%, and 62.64%.

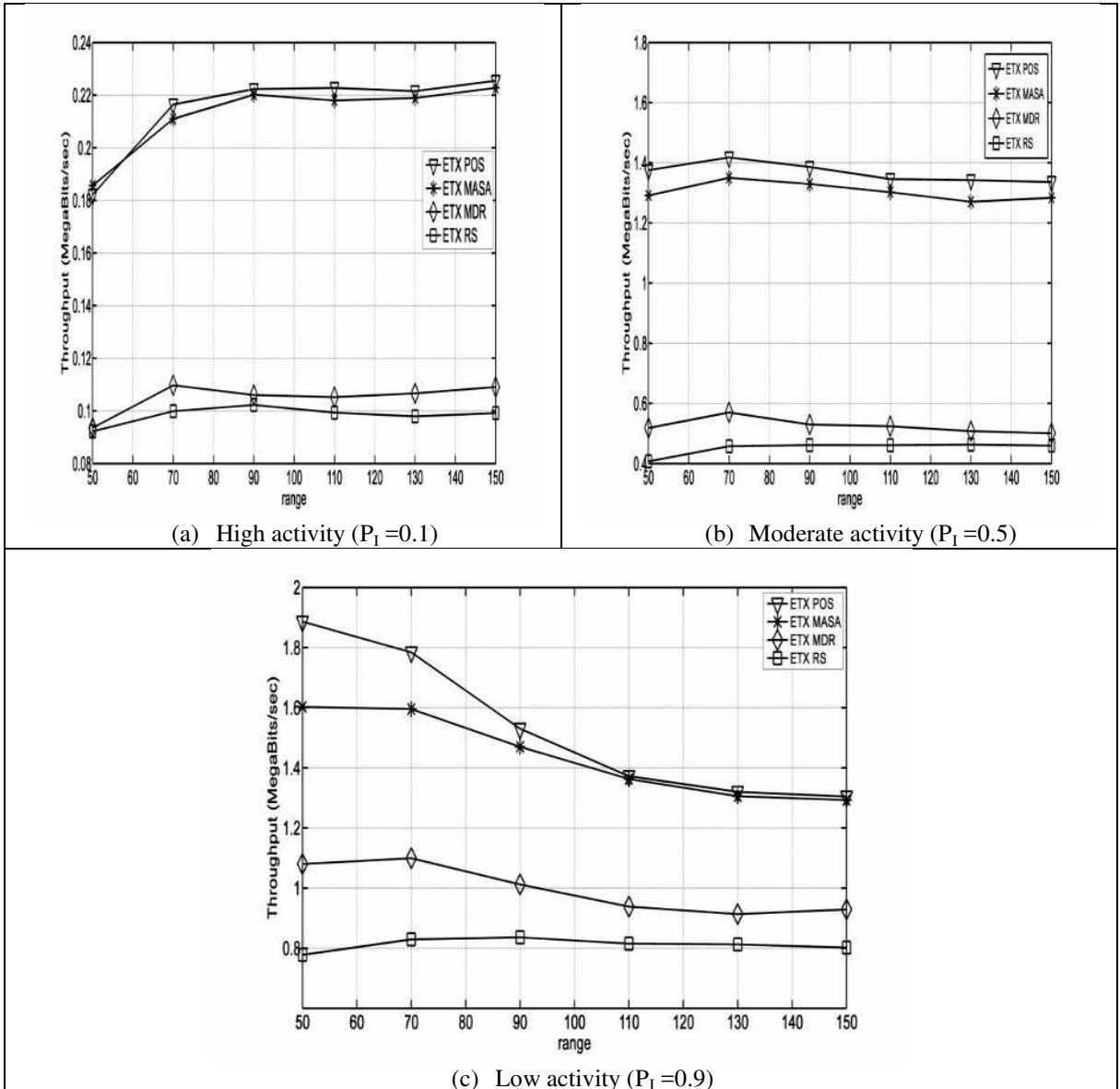

**Figure 3.20:** Throughput vs. range in network under different PU traffic (SPT).

### 3.4.7.2 The PDR Performance versus the Transmission Range

Figure 3.21 summarizes the PDR performance as a function of increasing the range between nodes. The performance of PDR increases as the range increases within a fewer number of channels are available at $P_I$=0.1. as the $P_I$ increases, the available channels also increases.

(POS and MASA) in compare to other protocols, the improvements are (107%, and 138%), (223%, and 274%), and (67.4%, and 104%) by referred to Figures 3.21(b), 3.21(a), and 3.21(c), respectively.

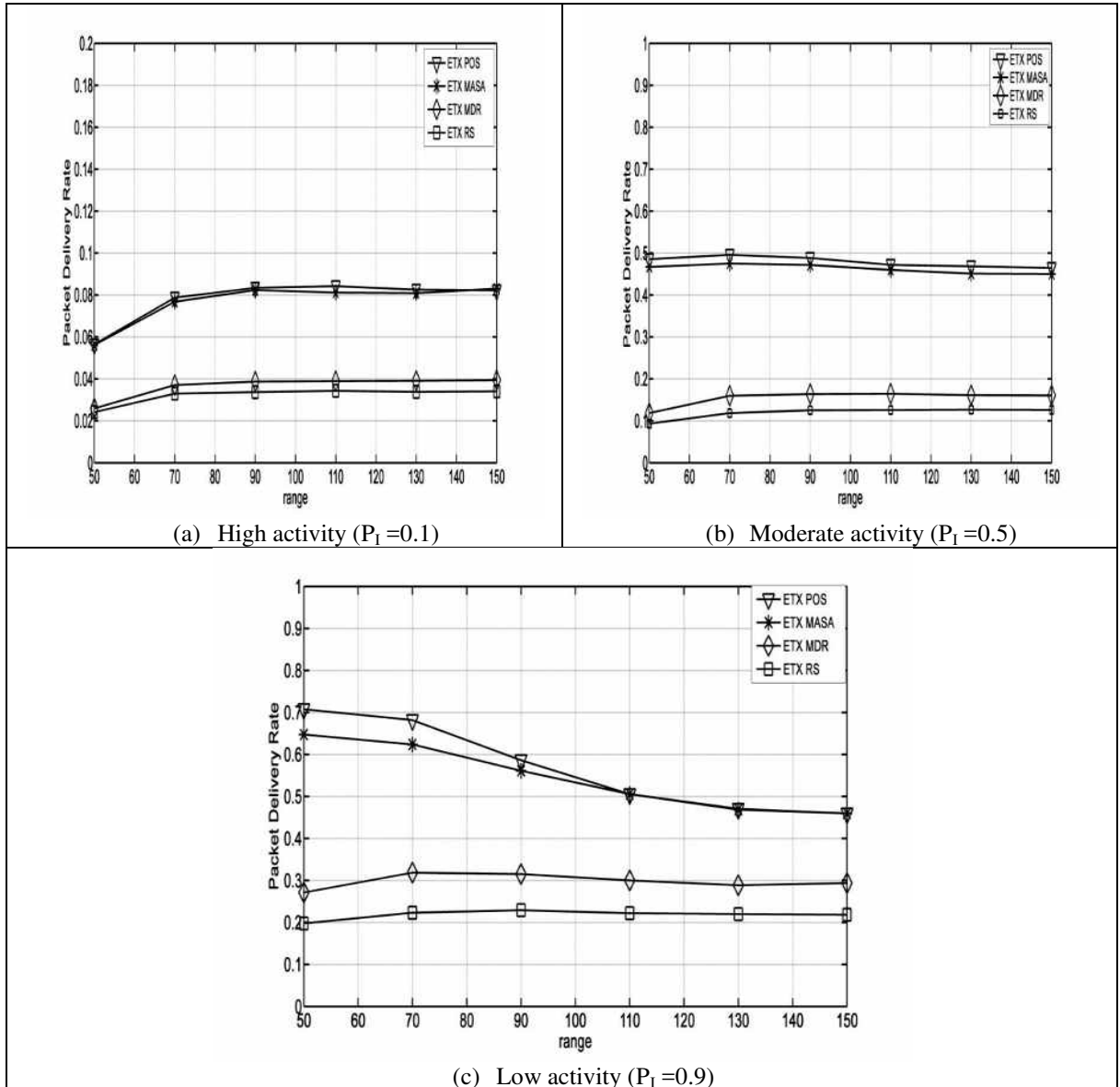

(a) High activity ($P_I$ =0.1)
(b) Moderate activity ($P_I$ =0.5)
(c) Low activity ($P_I$ =0.9)

**Figure 3.21:** The PDR vs. the transmission range under different PU traffic loads (SPT).

### 3.3.8 Impact the Changed in the Area of Network

We investigate the performance of increasing area field in the network, in terms network throughput, and the PDR. We consider the following network conditions: M=40, $M_r$=16, N=20, BW=1 M HZ, $P_t$=0.1 W and D=4 KB, for different PU traffic loads.

### 3.4.8.1 Throughput Performance versus the Field Area

Figure 3.22 shows the throughput as function of area L changed. As L increases, the throughput for all proposed protocols decreases, because of reducing the density of nodes in the network and uses the fixed transmission power. The improvement increases as the idle probability increases. The best improvement achieves by POS. At high activity of PUs; $P_I$=0.1, POS outperforms MASA, MDR, and RS with 13.5%, 104%, and 151.5%, respectively, as shown in Figure 3.22(a). In Figure 3.22(b), the POS scheme outperforms MASA up to 5.1%, MDR with 150.3% and 213.1% for RS. Figure 3.22(c), the POS protocol outperforms MASA, MDR, and RS with 22.2% 73.8%, 142.8%, respectively.

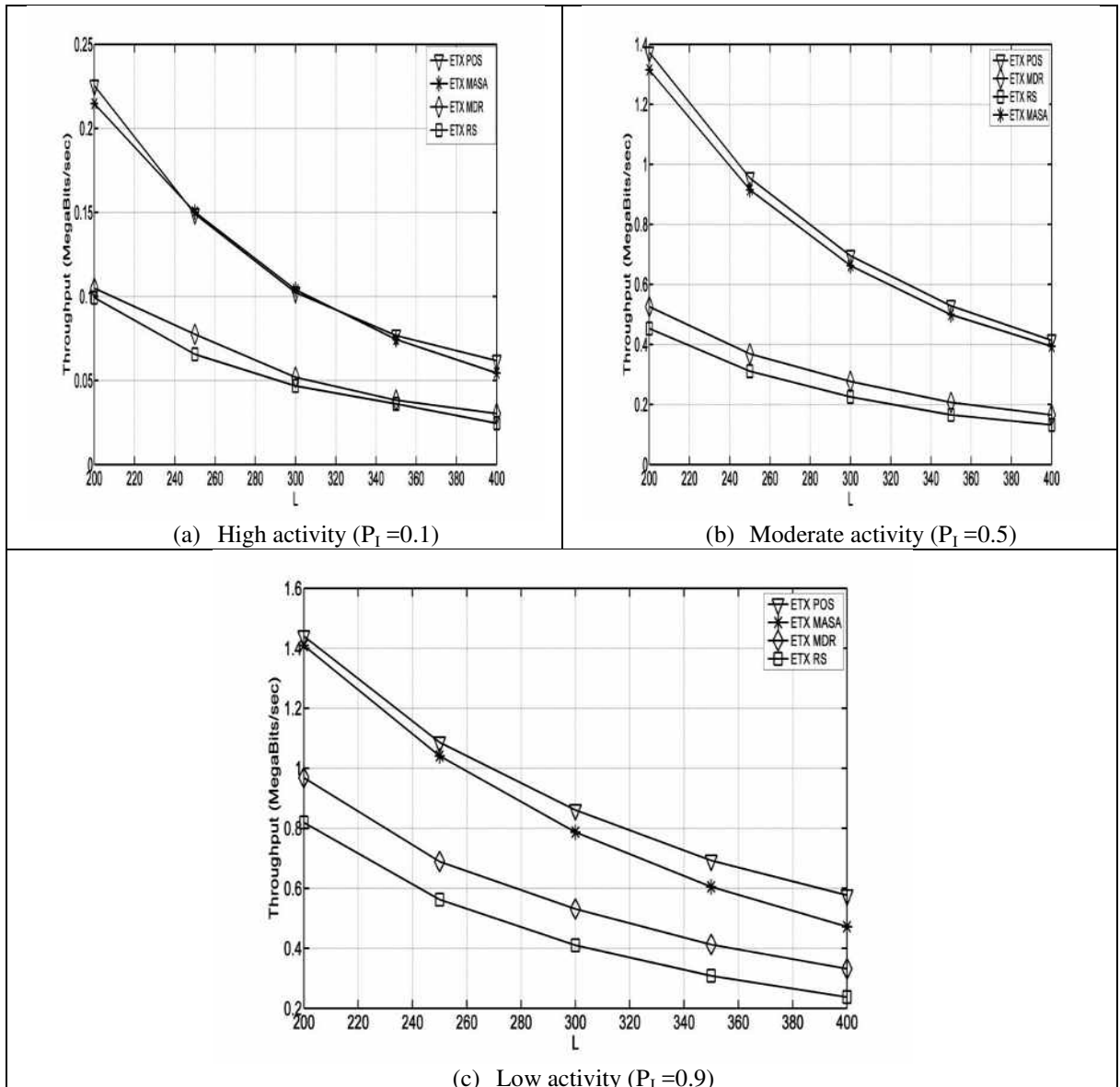
**Figure 3.22:** Throughput vs. field area under different PU traffic (SPT).

**3.4.8.2 The PDR Performance versus the Field Area**

Figure 3.23 shows the PDR as function of L. The performance of proposed protocols decreases over L increases. However, POS scheme outperforms the other protocols. At $P_I$=0.1, the performance of (MASA and POS) are closer and together outperform MDR and RS; because the limited number of channels availability in the network. The POS improvements reaches 10.3%, 120.5%, and 183% compared for MASA, MDR, and RS, respectively, see Figure 3.23(a). At $P_I$=0.5 the improvement of POS arrives 12% for MASA, 235.2% according to MDR, and 417.7% for RS. The performance improves as a result of availability channels increases. This is shown in Figure 3.23 (b), At $P_I$=0.9 the improvement of POS achieves 33.4%, 180.3%, and 393.2% compared to the other protocols.

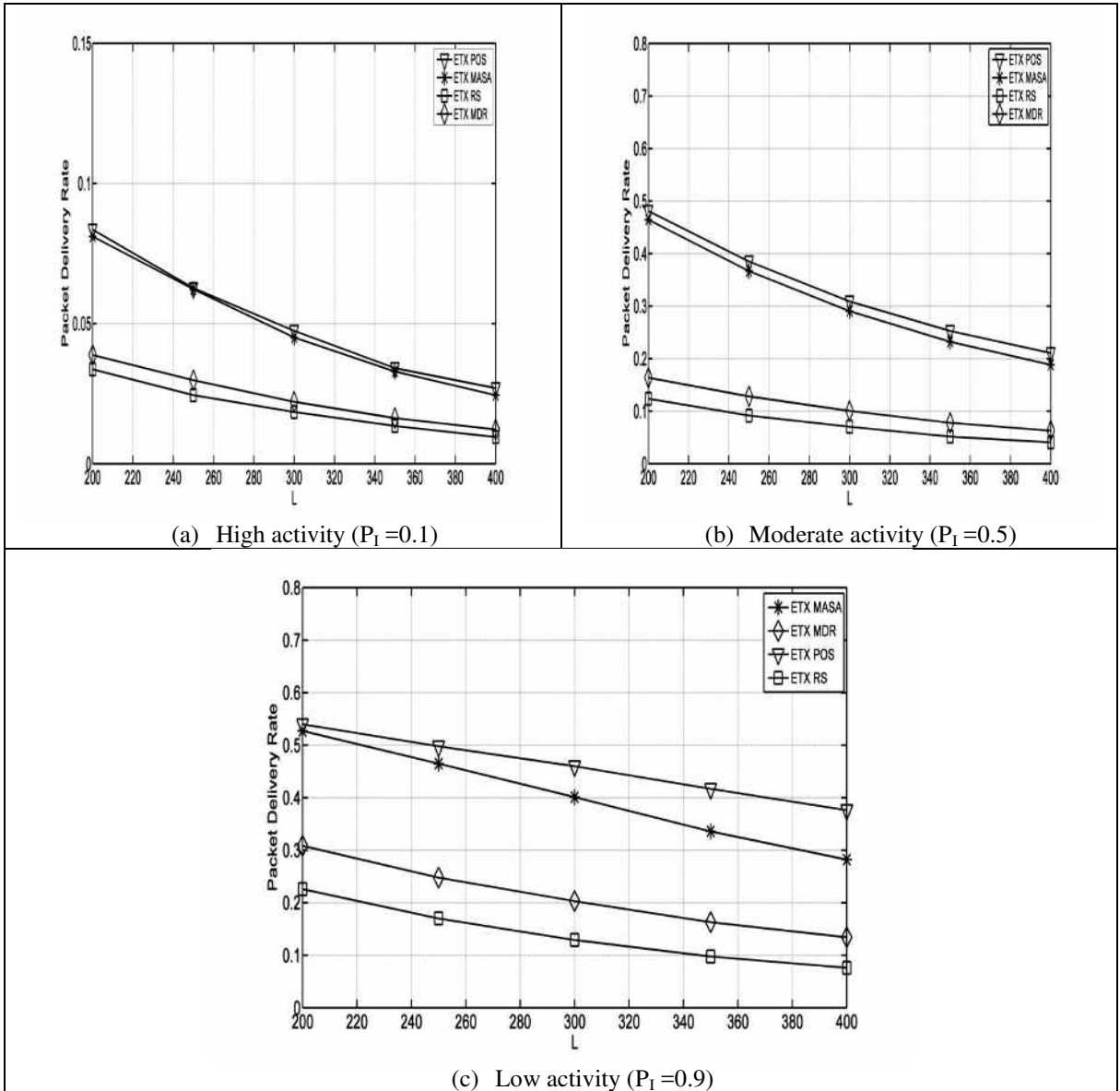

(a) High activity ($P_I$ =0.1)

(b) Moderate activity ($P_I$ =0.5)

(c) Low activity ($P_I$ =0.9)

**Figure 3.23:** The PDR vs. field area under different PU traffic (SPT).

## 3.3.9 Impact of PUs Traffic Load

POS achieves better performance than other protocols in terms network throughput and PDR. As the $P_I$ increases, the performance also increases. The improvement at lower rate ($P_I<=0.4$) is greater than higher rate of $P_I$; because it decreased the opportunities of cutting the transmission as the idle-probability increases, as shown in Figure 3.24, POS outperforms MASA, MDR, and RS protocols in network throughput within 5.7%, 44.1%, and 73%, respectively. The performance achievements of the PDR are 7%, 76.7%, and 144.6%, respectively.

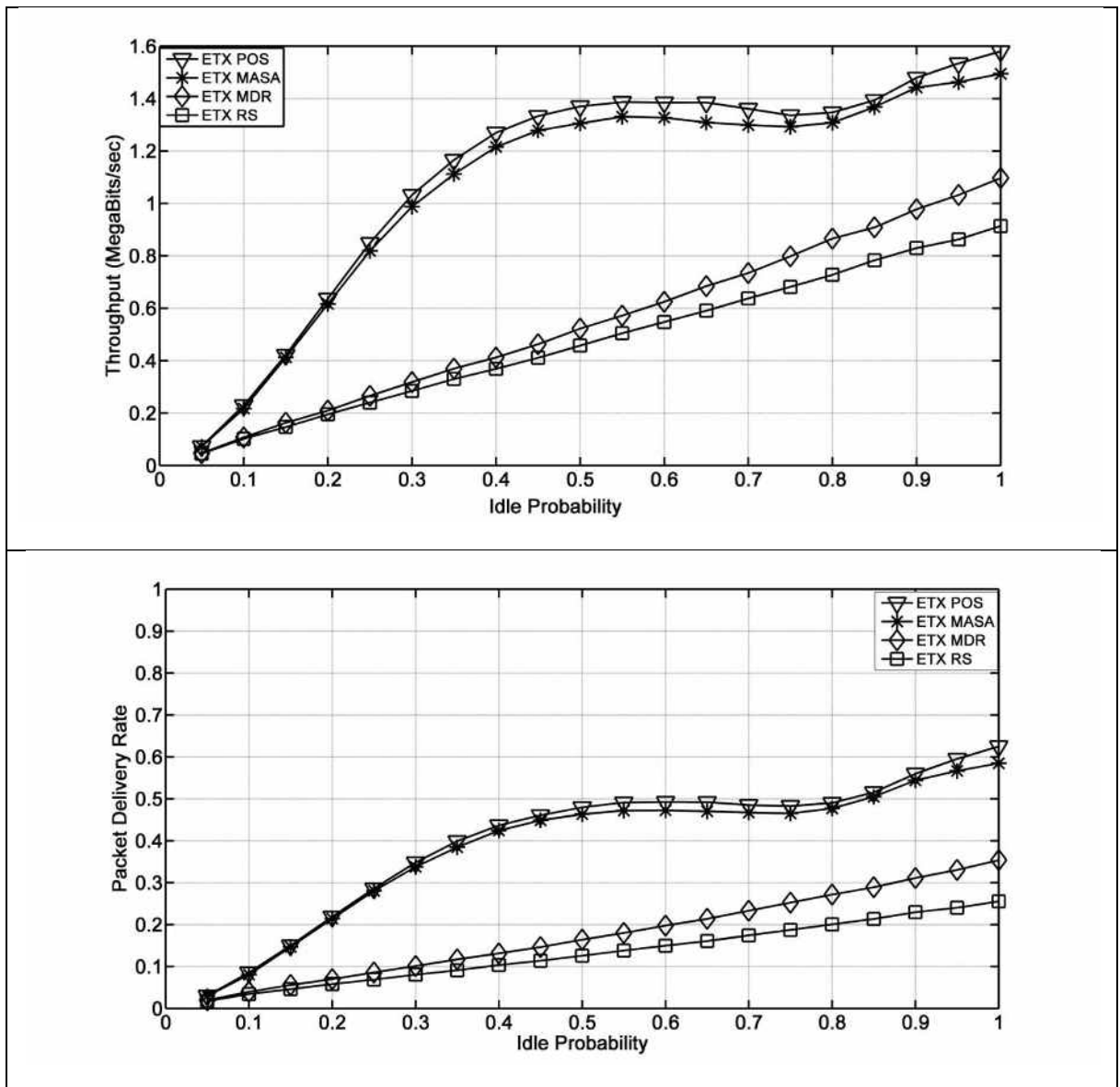

**Figure 3.24:** The PDR and Throughput vs. PU traffic (SPT).

## 3.3.10 Extremely Scenarios Impact of Fixed Data Rate

Figure 3.25 shows the performance of network throughput and the PDR as a function of primary traffic loads; with fixing data rate of all channels 5 MHZ. the performance of POS closer to MASA as compared to MDR. POS and MASA outperform MDR and RS with maximum improvement reaches in terms throughput are 3000% and 73%, respectively. In terms PDR, the improvement achieves to 3800% according to MDR, and 88% according to RS.

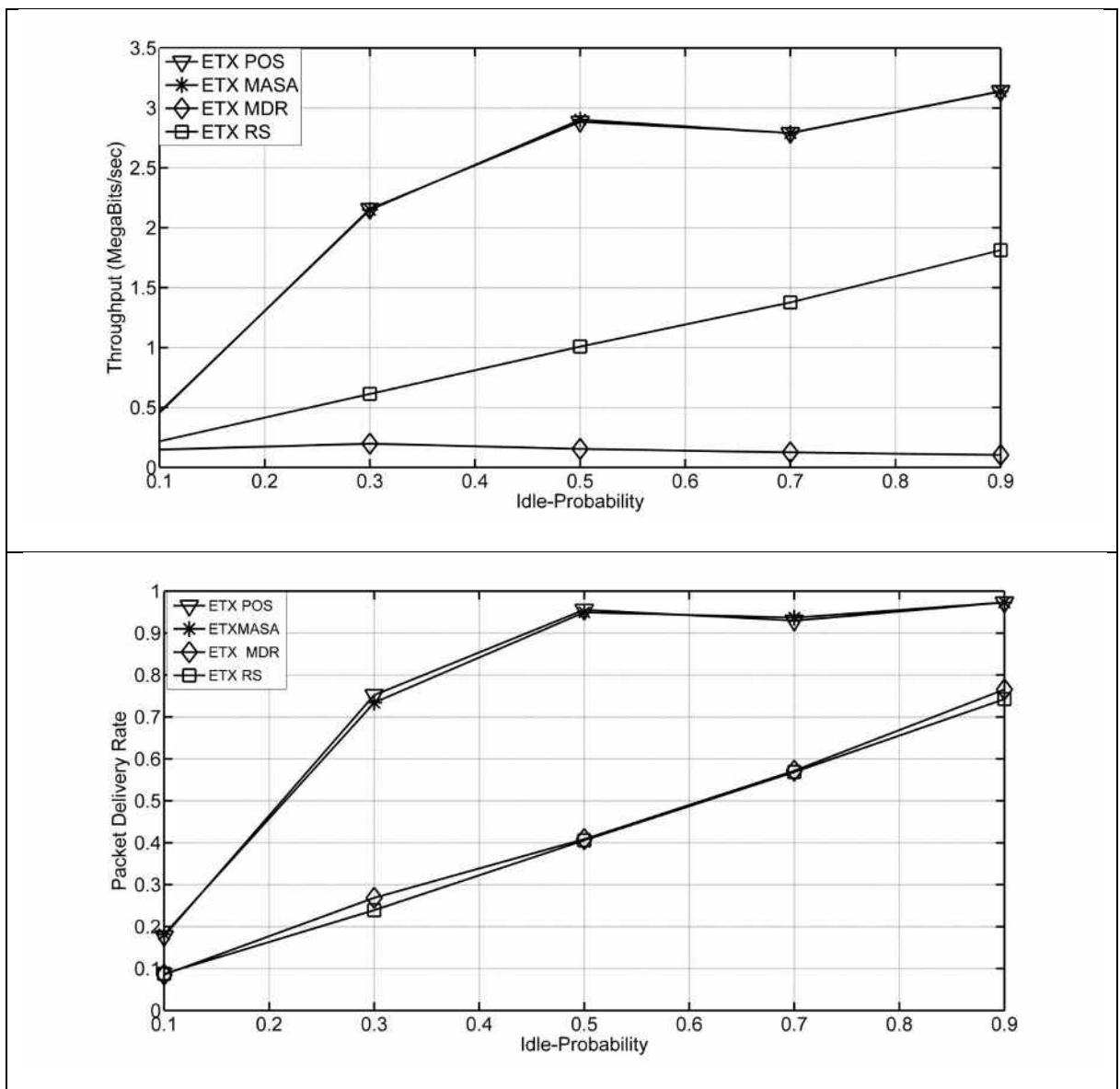

**Figure 3.25:** The PDR and Throughput vs. PU traffic (extremely fixed data rate).

## 3.3.11 Extremely Scenarios Impact of Fixed Available Time

Figure 3.26 shows the performance of network throughput and the PDR as a function of primary traffic loads, with fixing average available time of all channels 70 ms. The performance of MDR is closer to POS than MASA. POS improves MDR with 8%, RS with 32% and MASA with 43%, in terms throughput. According to PDR, POS achieves 11%, 41%, and 56% according to MDR, RS, and MASA, respectively.

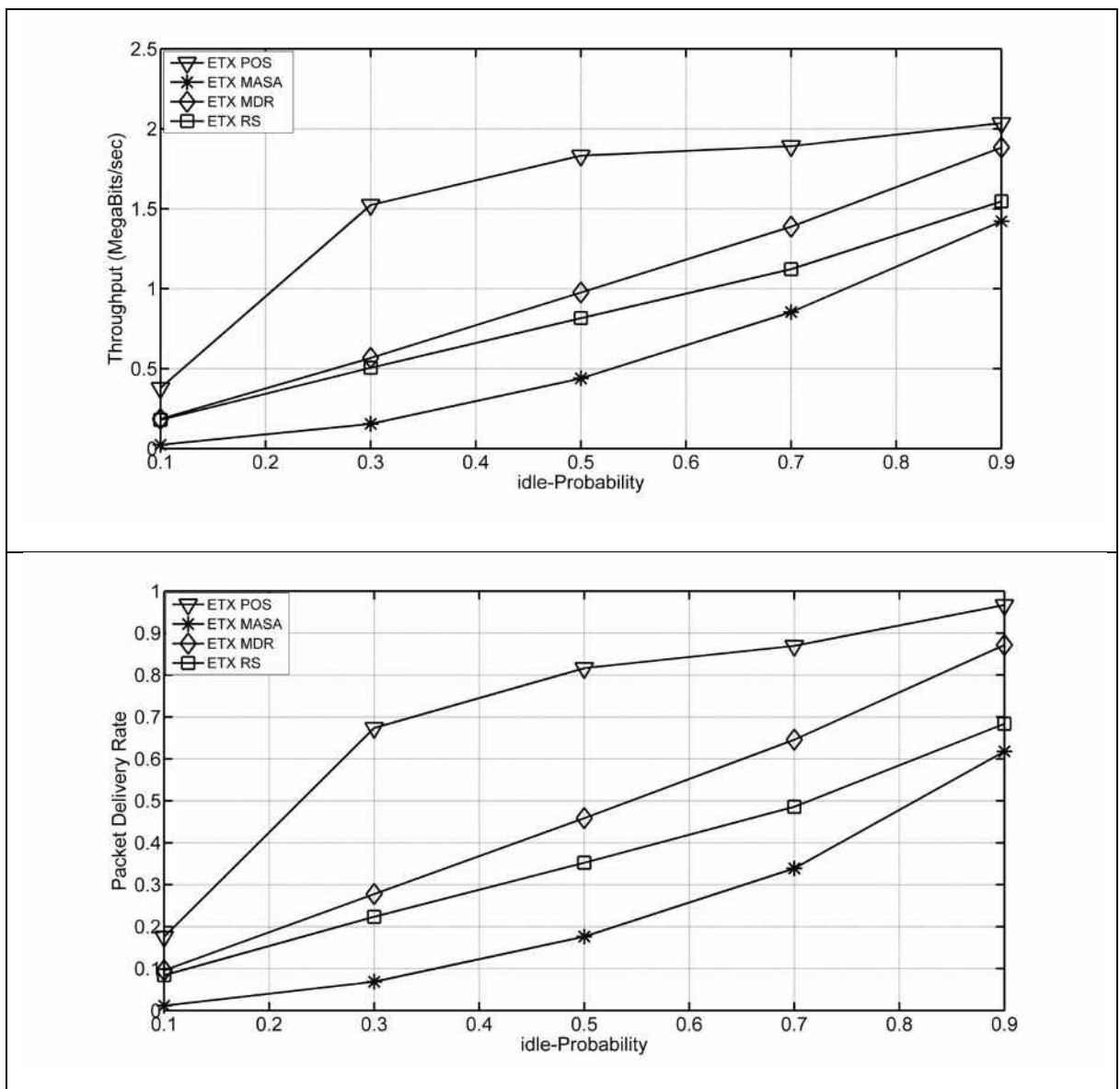

**Figure 3.26:** The PDR and Throughput vs. PU traffic (extremely fixed available time).

## 3.5 Performance Comparison between ETX and Distance as Metric for SPT

In this Section, we present and discuss a brief comparison between ETX and distance schemes in terms of network throughput and packet delivery rate. We consider the total number of nodes (N) is 40 and the total number of destinations (Mr) is 16. The performance of ETX outperforms distance over all network conditions.

### 3.5.1 Impact of Channel Bandwidth

We investigate the throughput and PDR performance as a function of channel bandwidth. We consider the following network conditions; M=40, $M_r$=16, N=20, $P_t$=0.1 W and D=4 KB, at the different PU traffic loads.

### 3.5.1.1 Throughput Performance versus the Channel Bandwidth

Figure 3.27 shows network throughput as a function of the channel Bandwidth (BW). This Figure shows the throughput increases as the BW increases for all traffic loads. However, the ETX-POS protocol outperforms the other protocols for all channel conditions because it uses the better channel assignments than the other protocols. Slightly improvements achieves as a result of comparing ETX with distance. The results are closer to each other, the improvement does not exceed to 10% between ETX and distance for all protocols. However, same improvement observed as the idle-probability increases from 0.1 to 0.9.

The following results have shown in Figure 3.27(a), POS-ETX outperforms POS-distance with 7.2%, MASA-ETX outperforms MASA-distance with 7.3%, MDR-ETX outperforms MDR-distance with 10%, and RS-ETX outperforms RS-distance with 3%.

In Figure 3.27(b), POS-ETX outperforms POS-distance with 5%, MASA-ETX outperforms MASA-distance with 4.6%, MDR-ETX outperforms MDR-distance with 3%, and the small percentage observed regarding RS-ETX and RS-distance.

Regarding Figure 3.27(c), ETX slightly outperforms distance over all proposed protocols.

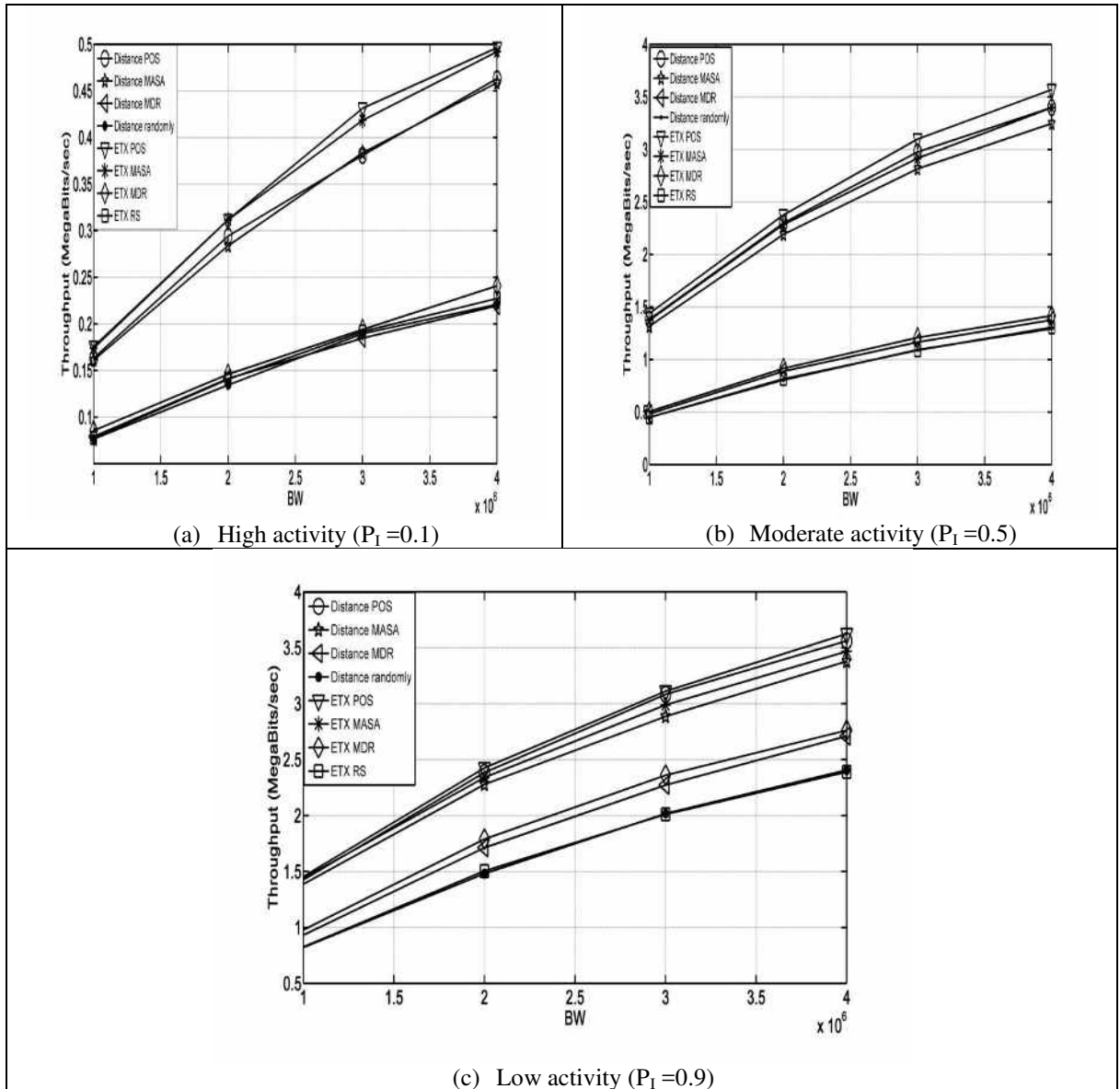

(a) High activity ($P_I$ =0.1)   (b) Moderate activity ($P_I$ =0.5)

(c) Low activity ($P_I$ =0.9)

**Figure 3.27:** Throughput vs. channel bandwidth under different PUs traffic (Distance vs. ETX).

**3.5.1.2 The Packet Delivery Rate Performance versus the Channel Bandwidth**

Figure 3.28 shows the PDR performance as function of the channel Bandwidth. The PDR performance of proposed protocols increases as BW increases. However, the ETX-POS protocols outperform other protocols. The following results have shown in Figure 3.28(a), POS-ETX outperforms POS-distance with 19.3%, MASA-ETX outperforms MASA-distance with 20.3%, MDR-ETX outperforms MDR-distance with 19.3%, and RS-ETX outperforms RS-distance with 7.7%.

In Figure 3.28(b), POS-ETX outperforms POS-distance with 3.5%, MASA-ETX outperforms MASA-distance with 6%, MDR-ETX outperforms MDR-distance with 9%, and RS-ETX and RS-distance with 3%.

Regarding Figure 3.28(c), ETX slightly outperforms distance over all proposed protocols.

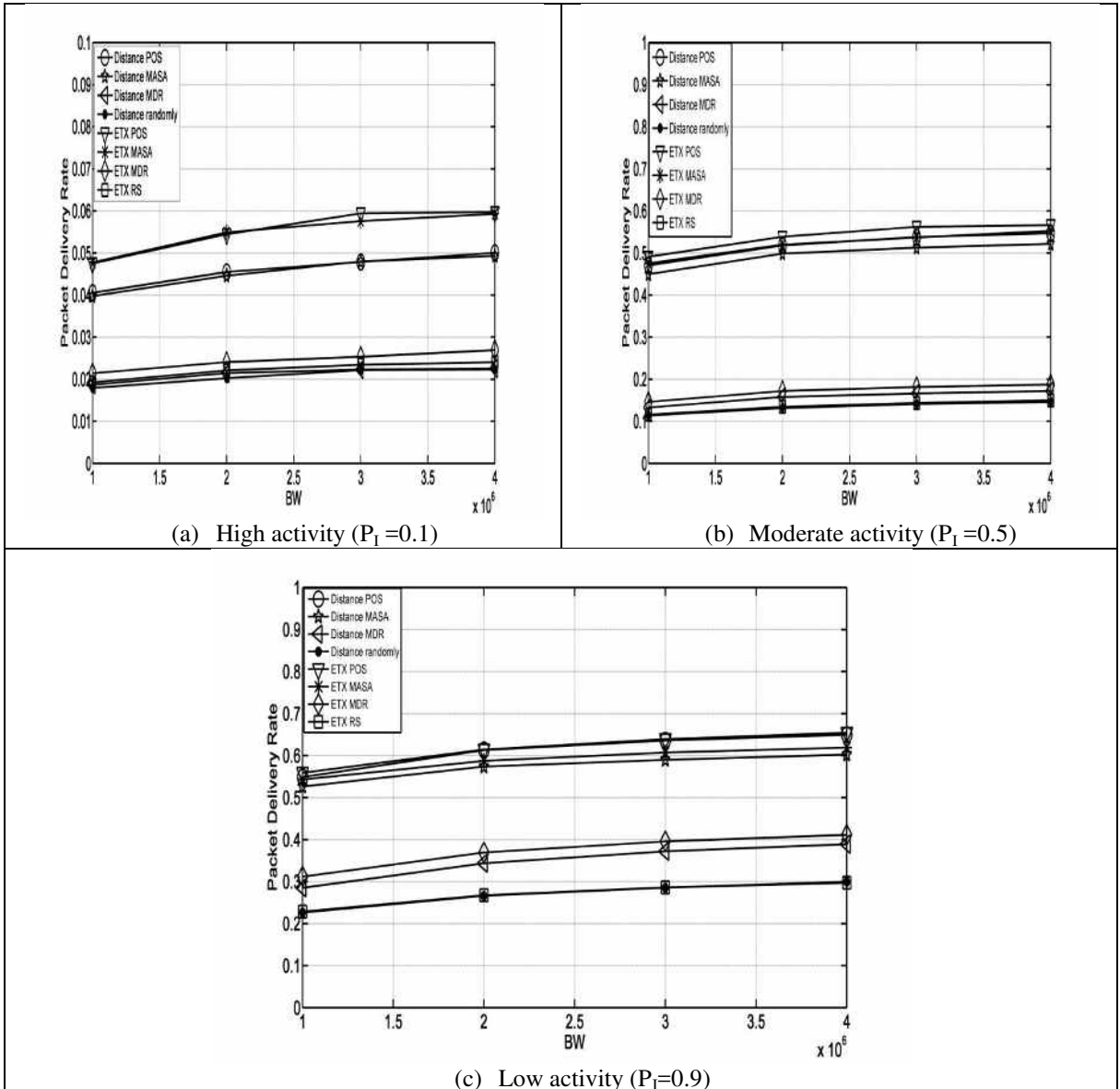

(a) High activity ($P_I$ =0.1)          (b) Moderate activity ($P_I$ =0.5)

(c) Low activity ($P_I$=0.9)

**Figure 3.28:** The PDR vs. channel bandwidth under different PU traffic (Distance vs. ETX).

### 3.5.2 Impact of the Packet Size

We investigate the effects of increasing packet size in terms of throughput, and PDR. We consider the following network conditions; M=40, $M_r$=16, N=20, $P_t$=0.1 W and BW=1 MHZ.

### 3.5.2.1 Throughput Performance versus the Packet Size

Figure 3.29 shows network throughput as a function of the packet size D. As the packet size increases, the throughput for all protocols decreases. At high activity of PUs, ETX outperforms distance with 35%. At moderate and low activity of PUs, it achieves 4% and 2.2%, respectively.

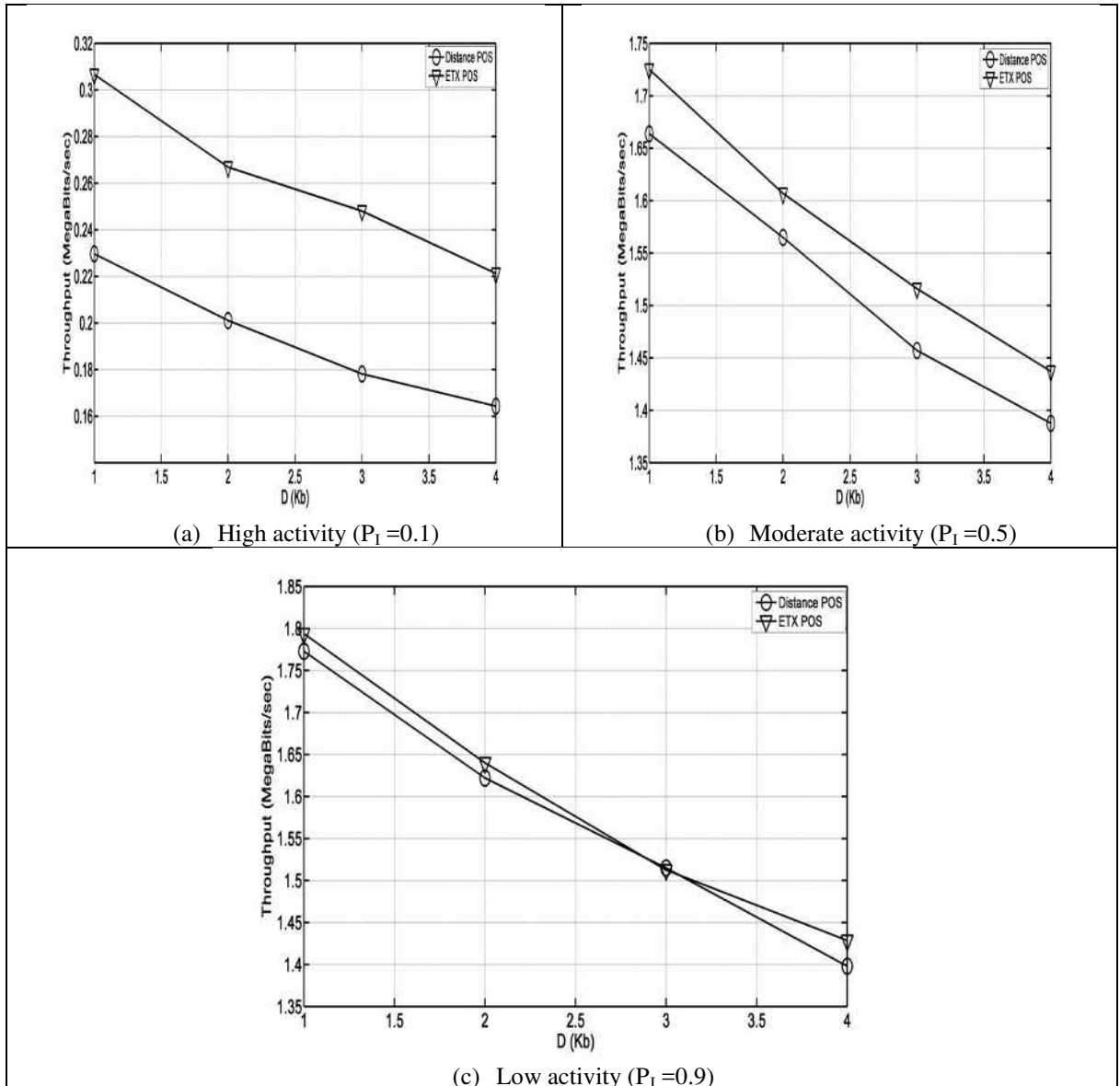

(a) High activity ($P_I$ =0.1)
(b) Moderate activity ($P_I$ =0.5)
(c) Low activity ($P_I$ =0.9)

**Figure 3.29:** Throughput vs. channel packet size under different PU traffic (Distance vs. ETX).

### 3.5.2.2 The PDR Performance versus the Packet Size

Figure 3.30 shows the PDR as a function of D. For all protocols, the PDR decreases as D increases. At $P_I$=0.1, the improvements is 81% when $P_I$=0.5 and $P_I$=0.9, it achieves slightly improvement less to 4%.

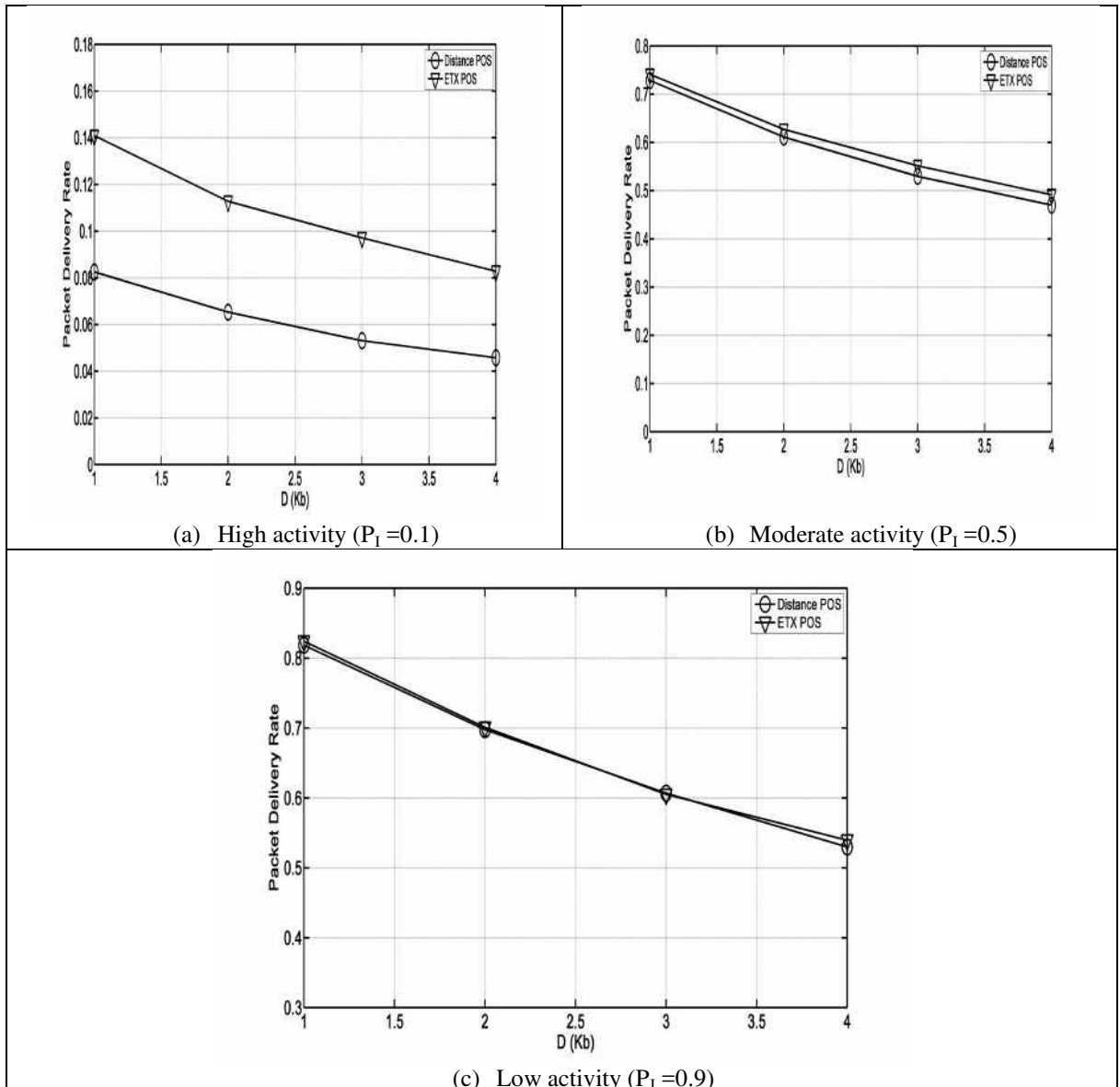

**Figure 3.30:** The PDR vs. Packet size under different PU traffic (Distance vs. ETX).

### 3.5.3 Impact of the Number of PU Channels

We investigate the performance of the number of PU channels, in terms of throughput, and PDR. We consider the following network conditions: M=40, $M_r$=16, BW=1 MHZ, $P_t$=0.1 w and D=4 KB, under different PUs activity.

## 3.5.3.1 Throughput Performance versus the Number of PU Channels

Figure 3.31, illustrates the effect of increasing the number of primary channels in the network performance in terms of throughput, which increases as the channel availability increases. Thus, increasing the number of idle channels provides more chances to select the appropriate channels for transmissions. ETX scheme outperforms distance for all situations. The best improvement reaches at traffic load equals 0.1. The improvement of ETX is 21% in compare to distance. Slightly improvement observed at traffic loads of (0.5 and 0.9).

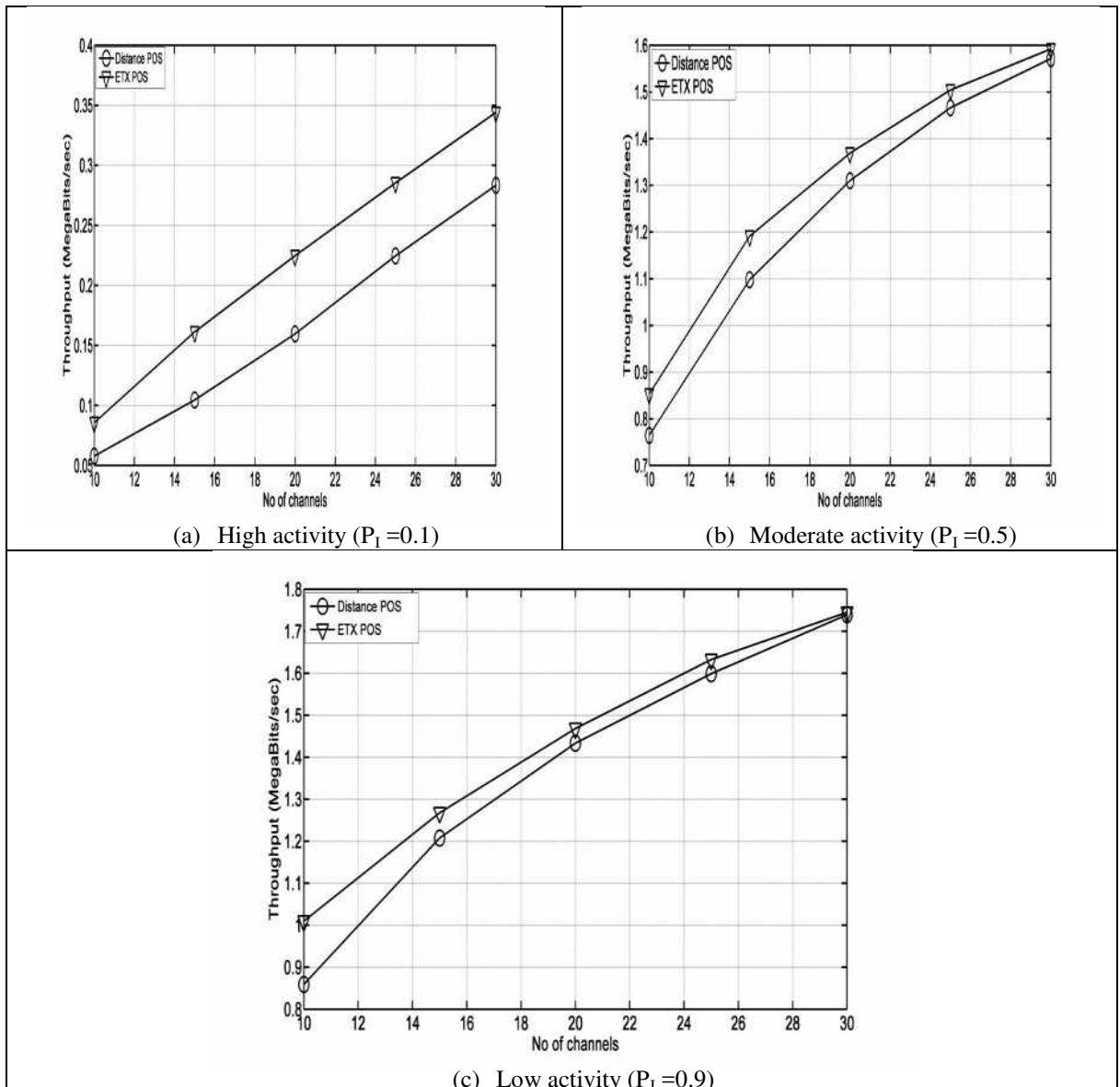

**Figure 3.31:** Throughput vs. number of PU channels under different PU traffic (Distance vs. ETX).

## 3.5.3.2 The PDR Performance versus the Number of PU Channels

This Figure 3.32 shows the PDR as a function of increasing the number of primary channels in network performance of PDR. The improvements increase as the channels increase for all protocols. ETX outperforms distance for all situations. The maximum improvement achieves at idle-probability equals 0.1 is 52%. At traffic loads of (0.5 and 0.9) slightly improvement observed.

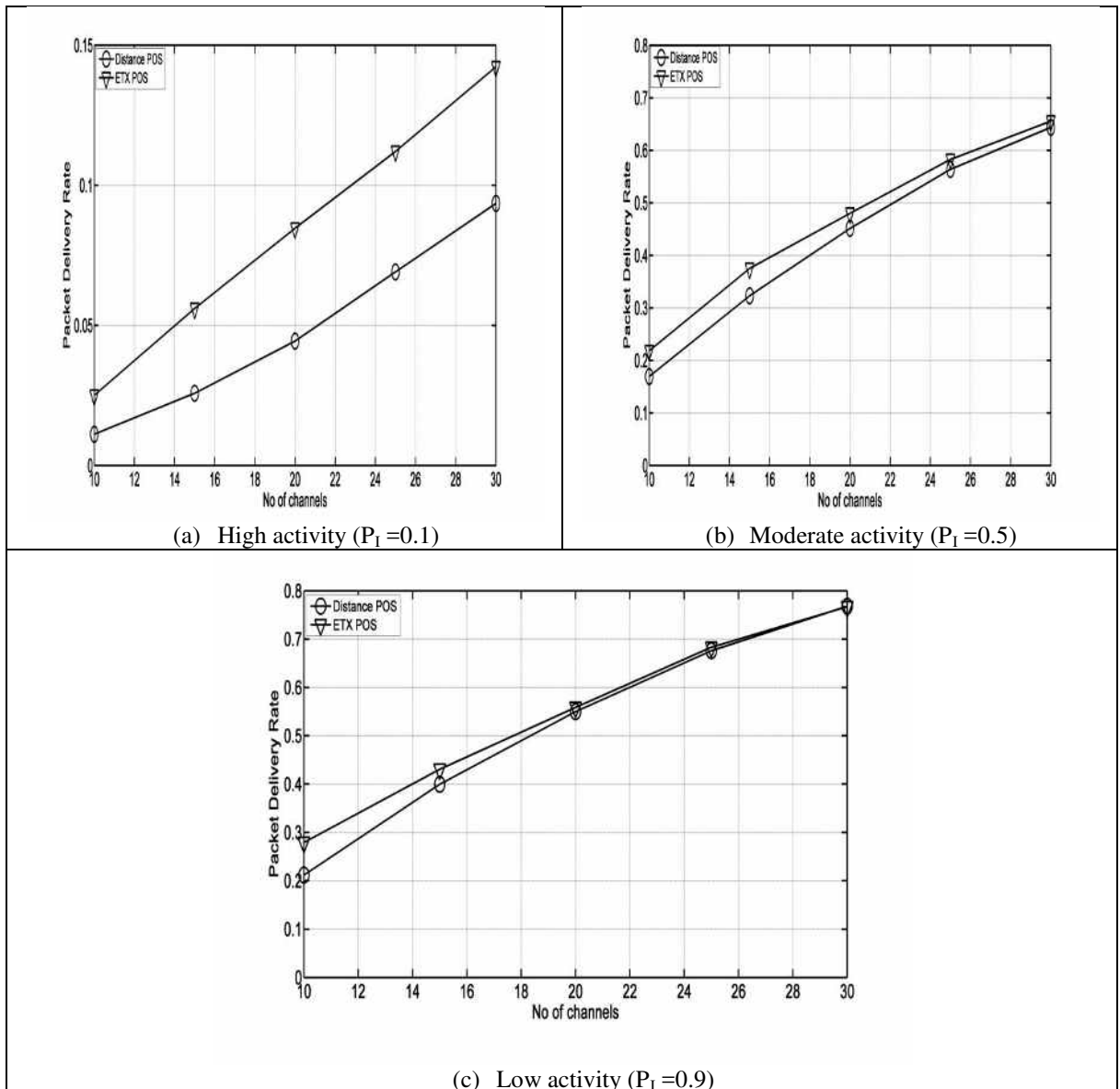

(a) High activity ($P_I$ =0.1)  (b) Moderate activity ($P_I$ =0.5)

(c) Low activity ($P_I$ =0.9)

**Figure 3.32:** The PDR vs. number of PU channels under different PU traffic (Distance vs. ETX).

### 3.5.4 Impact of the Transmissions Power

We investigate the effects of increasing the transmission power in terms of throughput, and PDR. We consider the following network conditions: M=40, $M_r$=16, BW=1 MHZ, N=20 and D=4 KB.

### 3.5.4.1 Throughput Performance versus the Transmissions Power

We illustrate the performance of PDR as a function of transmission power in Figure 3.33. Slightly improvement observed between ETX and distance over the idle-probability increases. The maximum improvement achieves at high activity of PU users is 34%.

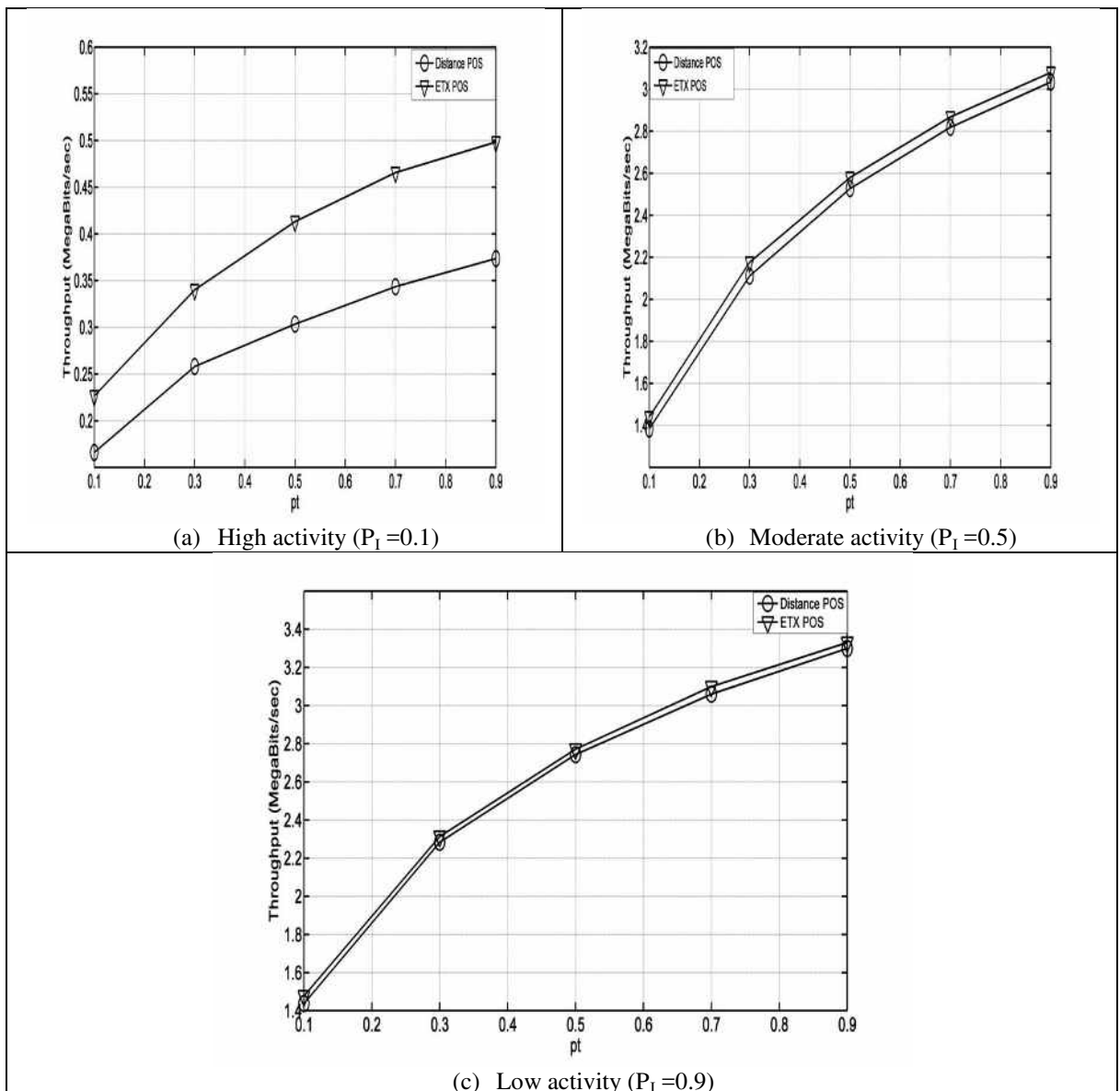

(a) High activity ($P_I$ =0.1)   (b) Moderate activity ($P_I$ =0.5)

(c) Low activity ($P_I$ =0.9)

**Figure 3.33:** Throughput vs. power transmissions under different PU traffic (Distance vs. ETX).

## 3.5.4.2 The PDR Performance versus the Transmissions Power

We illustrate the performance of PDR as a function of transmission power. The PDR increase as the power increases, as it is shown in Figure 3.34. The maximum improvement is 70% at $P_I=0.1$.

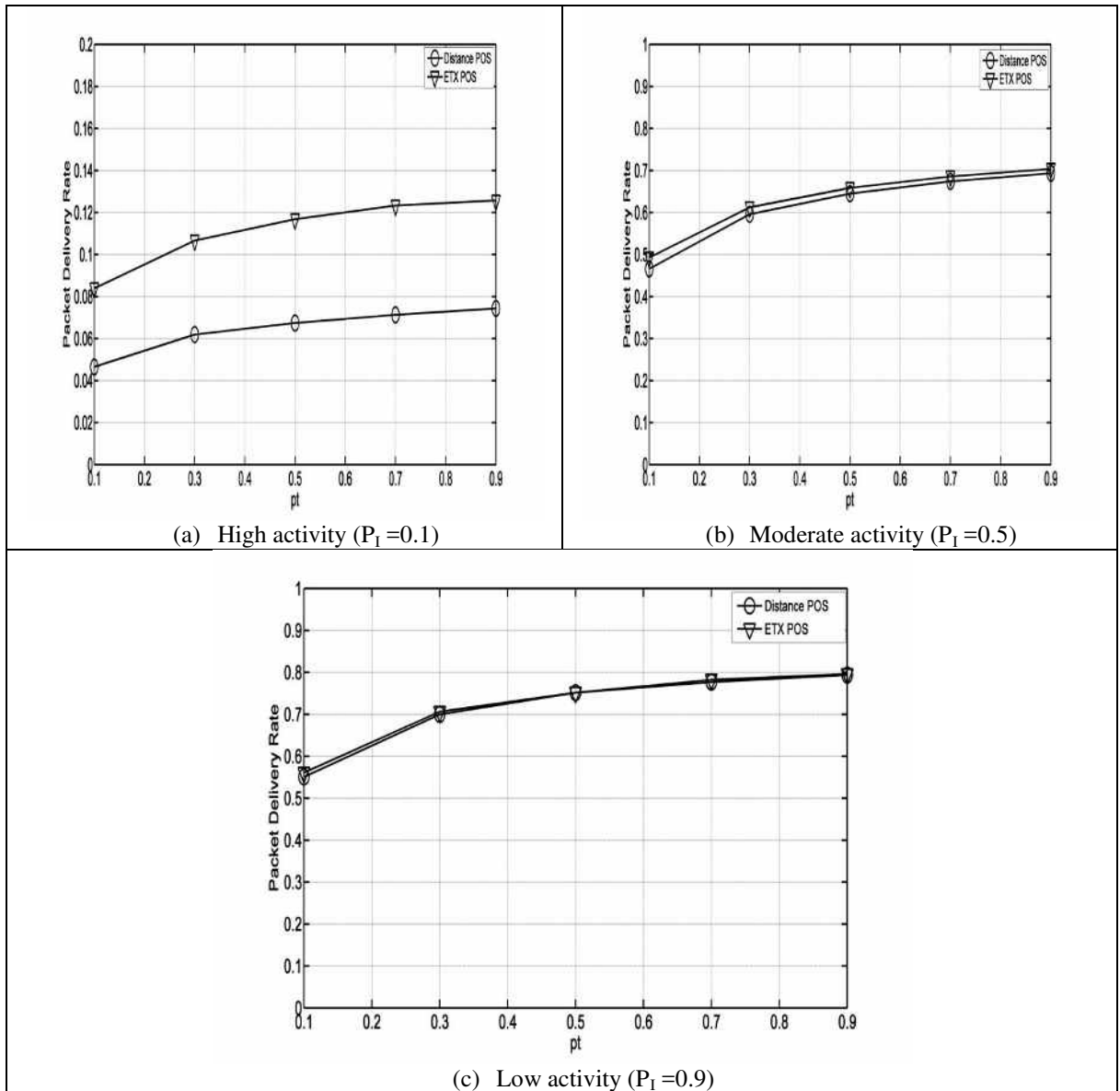

(a) High activity ($P_I$ =0.1)  (b) Moderate activity ($P_I$ =0.5)

(c) Low activity ($P_I$ =0.9)

**Figure 3.34:** The PDR vs. power transmissions under different PU traffic (Distance vs. ETX).

## 3.5.5 Impact of the Number of Nodes in Network Performance

We investigate the performance of increasing the number of nodes, in terms throughput, and PDR. We consider the following network conditions; $M_r=16$, $N=20$, $BW=1$ MHZ, $P_t=0.1$ w and $D=4$ KB, over PU traffic loads.

### 3.5.5.1 Throughput Performance versus the Number of Nodes

Figure 3.35 illustrates the performance of throughput as a function of increasing the node numbers in the network. At lower $P_I$, the improvement is 13.6%. The improvement achieves at $P_I=0.5$ and $P_I=0.9$ are 5.3% and 3.8% respectively.

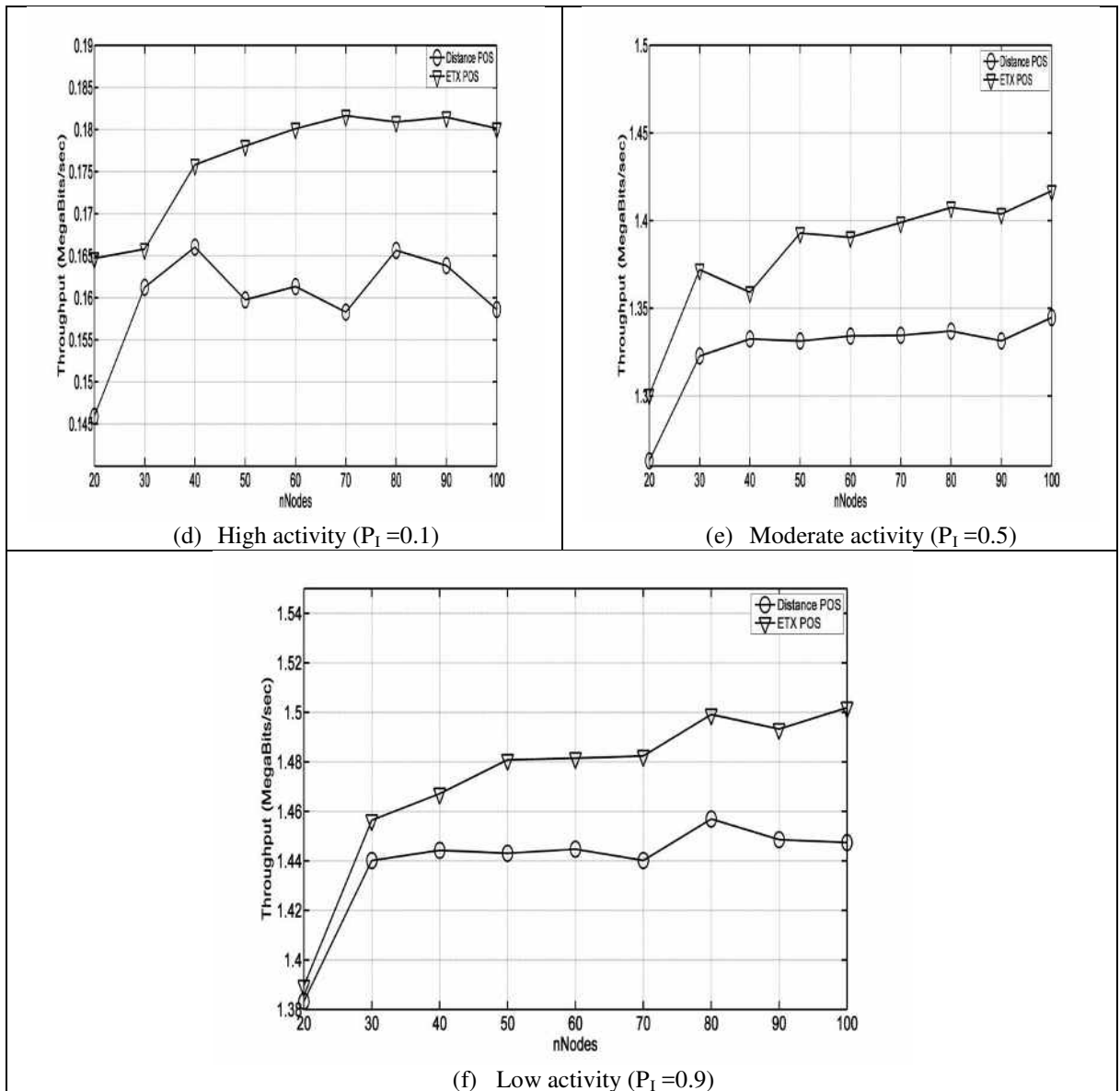

(d) High activity ($P_I =0.1$)      (e) Moderate activity ($P_I =0.5$)

(f) Low activity ($P_I =0.9$)

**Figure 3. 35:** Throughput vs. number of nodes in network under different PU traffic (Distance vs. ETX).

## 3.5.5.2 The PDR Performance versus the Number of Nodes

In general, increasing the number of nodes is not significantly related to increase the PDR. It becomes to be constant as $P_I$ increases. At $P_I=0.1$, the improvements is 28%. At $P_I=0.5$, the improvements is 6% and slightly improvement achieves at $P_I=0.9$ as it is shown in Figure 3.36.

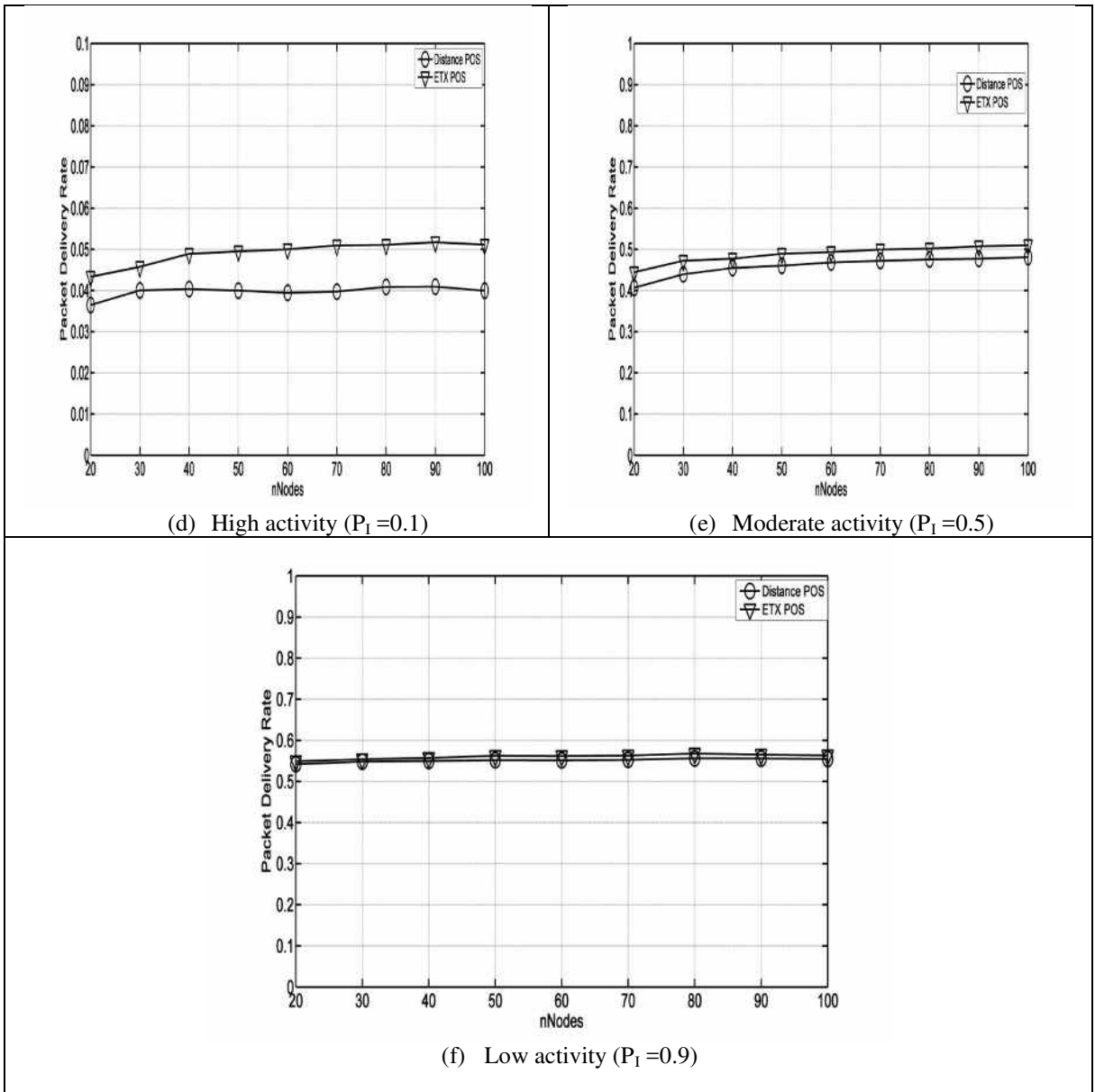

(d) High activity ($P_I=0.1$)   (e) Moderate activity ($P_I=0.5$)

(f) Low activity ($P_I=0.9$)

**Figure 3.36:** The PDR vs. number of nodes in network under different PU traffic (Distance vs. ETX).

## 3.3.6 Impact of the Number of Destinations in the Network

In this section, we illustrate the performance of the number of destinations, in terms throughput and PDR. We consider the following network conditions; N=20, M=40, BW=1 MHZ, $P_t$=0.1 w and D=4 KB, under different PUs activity.

### 3.5.6.1 Throughput Performance versus the Number of Destination Nodes

Figure 3.37 summarizes the impact of increasing the number of destination nodes in network throughput. ETX outperforms distance for all situations. The maximum improvement achieves at idle-probability equals 0.1 and 0.5 is 6%.

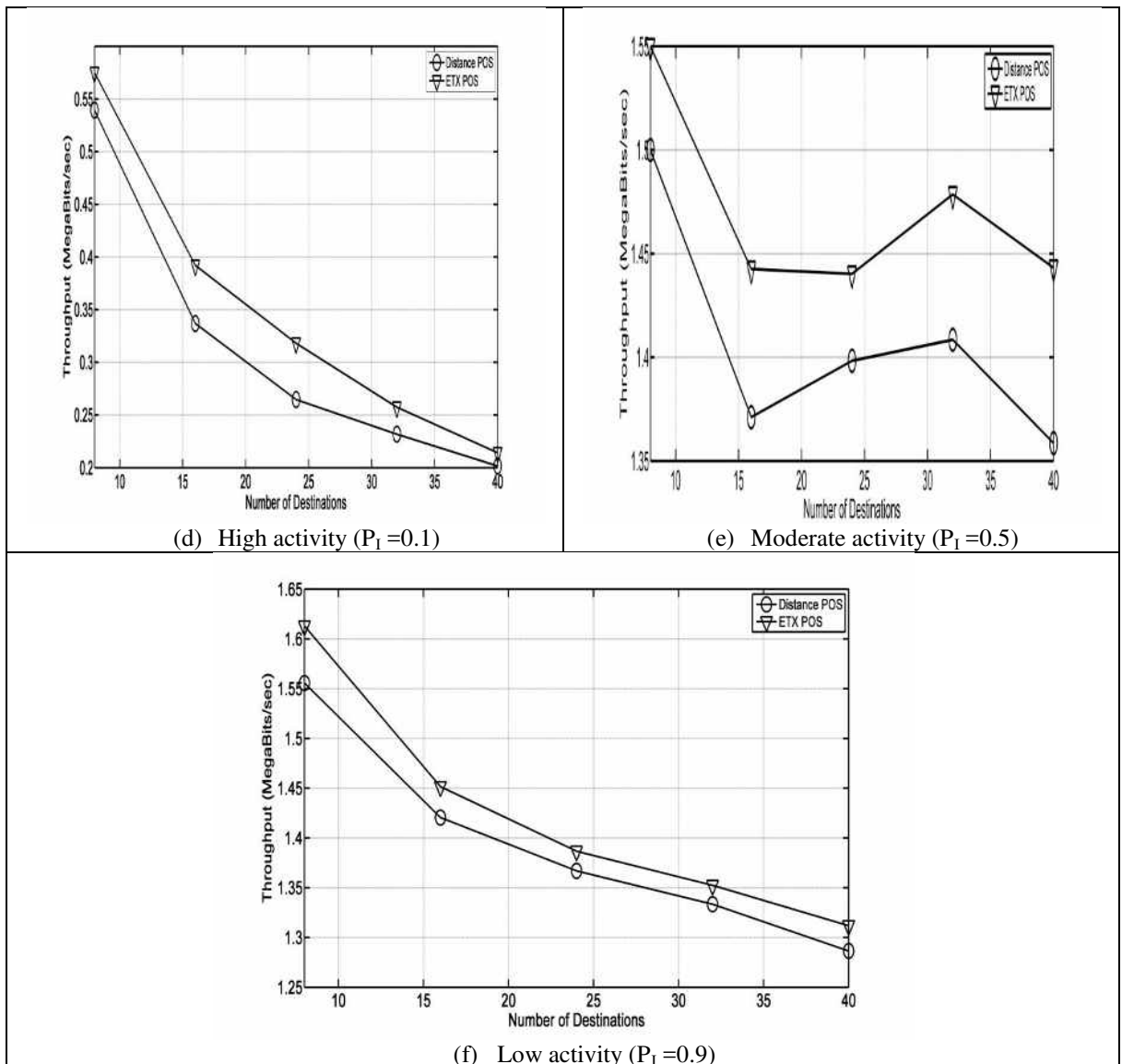

(d) High activity ($P_I$ =0.1)      (e) Moderate activity ($P_I$ =0.5)

(f) Low activity ($P_I$ =0.9)

**Figure 3.37:** Throughput vs. number of destinations in network under different PU traffic (Distance vs. ETX).

## 3.5.6.2 The PDR Performance versus the Number of Destinations

Figure 3.38 summarizes the impact of increasing the number of destination nodes in network throughput. ETX outperforms distance for all situations. The maximum improvement achieves at idle-probability equals 0.1 is 23%, and at idle-probability equals 0.5 is 8%.

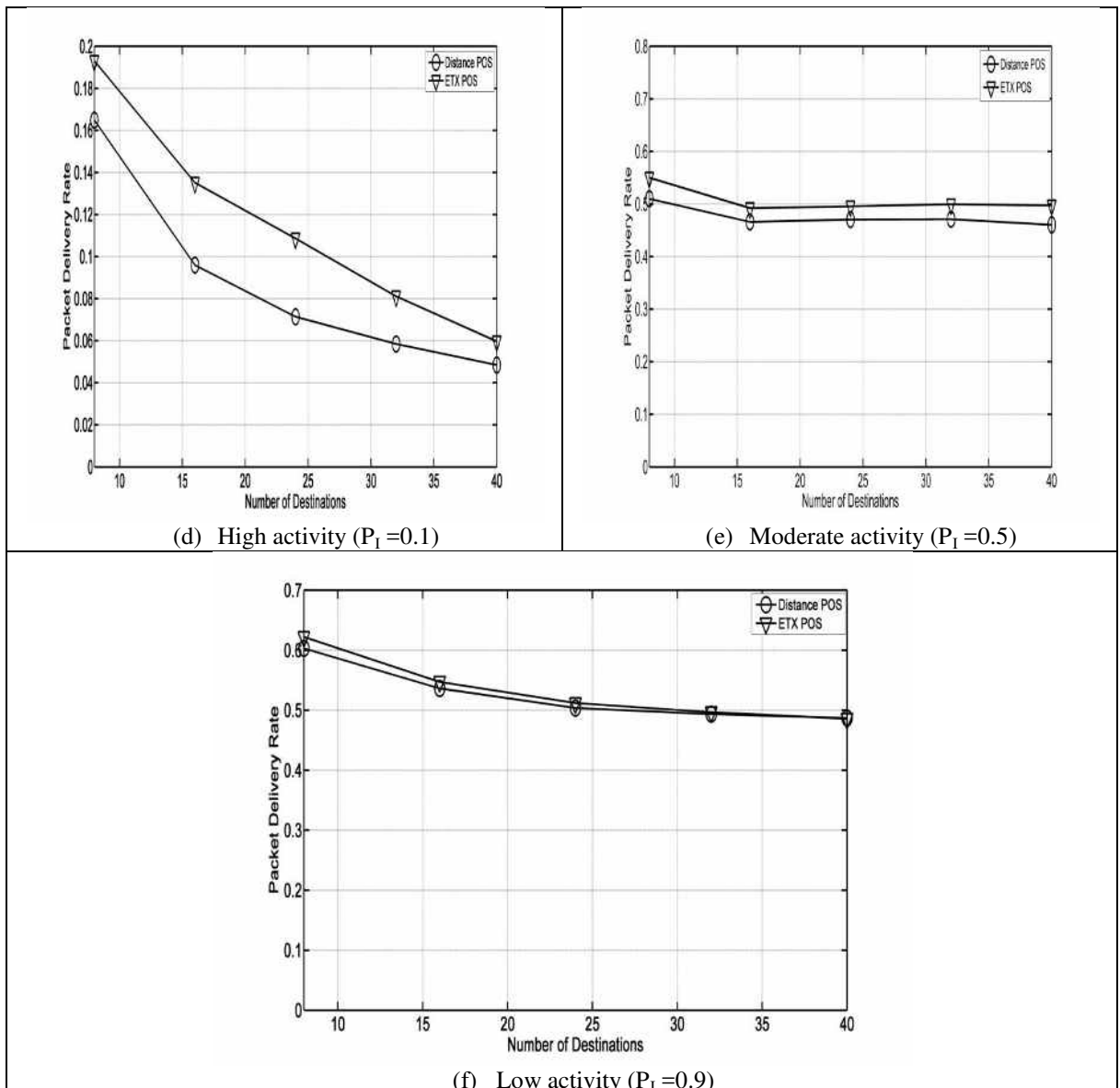

(d) High activity ($P_I$ =0.1)  (e) Moderate activity ($P_I$ =0.5)

(f) Low activity ($P_I$ =0.9)

**Figure 3.38:** The PDR vs. number of destinations in network under different PU traffic (Distance vs. ETX).

### 3.3.7 Impacts of the Maximum Transmission Range on Network Performance

In this section, we investigate the performance of increasing the maximum range between two nodes in the network, in terms network throughput, and the PDR. We consider the following network conditions; M=40, $M_r$=16, N=20, BW=1 MHZ, $P_t$=0.1 W and D=4 KB, under different PU traffic loads.

### 3.5.7.1 Throughput Performance versus the Transmission Range

Figure 3.39 shows the performance of network throughput as a function of increasing the maximum range between two nodes in the network. At lower rate of $P_I$ reaches 37%, at moderate and high rate of $P_I$, achieves 7%.

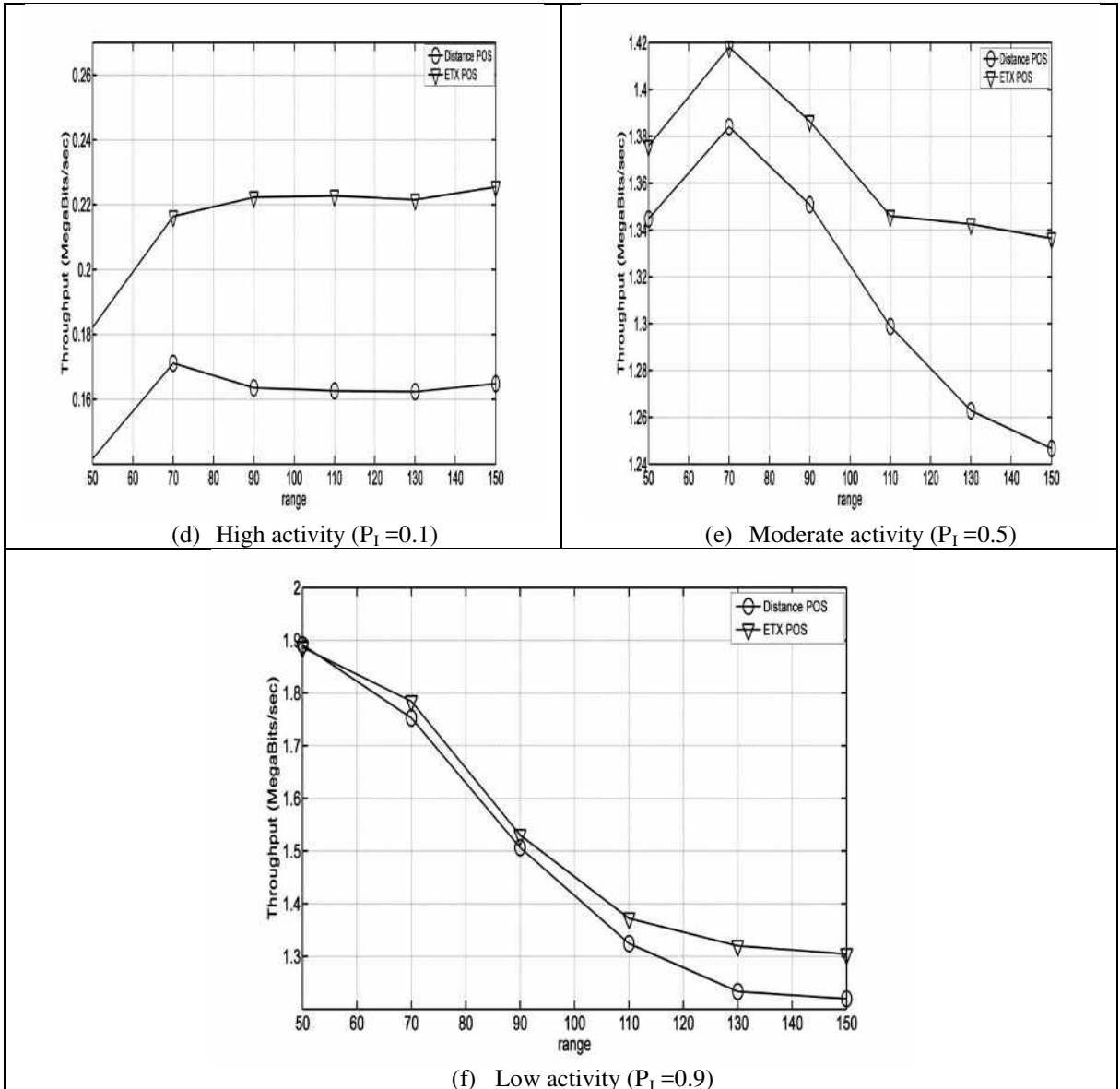

(d) High activity ($P_I$ =0.1)        (e) Moderate activity ($P_I$ =0.5)

(f) Low activity ($P_I$ =0.9)

**Figure 3.39:** Throughput vs. range in network under different PU traffic (Distance vs. ETX).

### 3.5.7.2 The PDR Performance versus the Transmission Range

Figure 3.40 summarizes the PDR performance as a function of increased the range between nodes. ETX outperforms distance for all situations. The maximum improvement achieves at idle-probability equals 0.1 is 89%, at idle-probability equal 0.5 is 19%, and at idle-probability equals 0.9 is 17%.

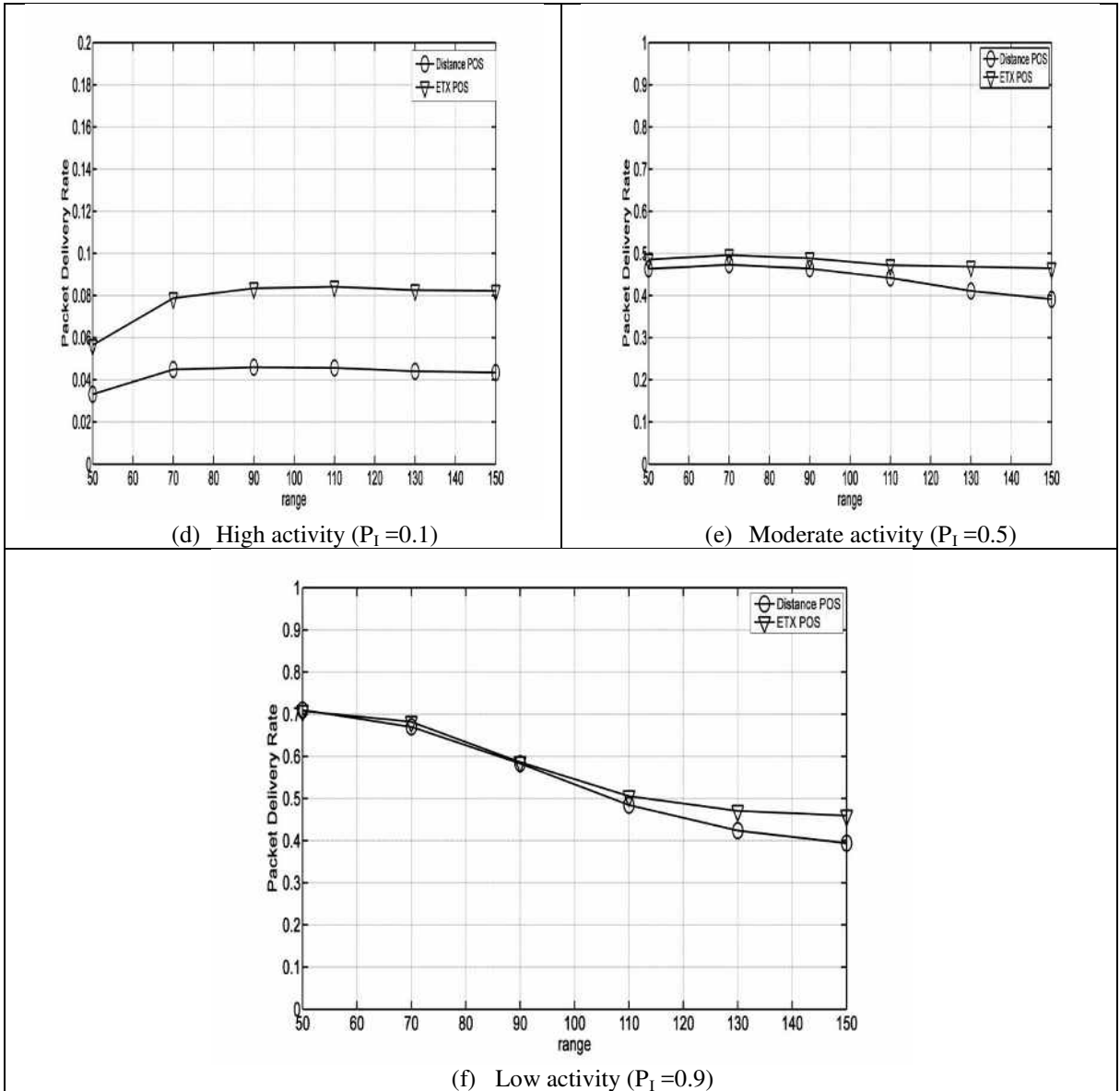

**Figure 3.40:** The PDR vs. the transmission range under different PU traffic loads (Distance vs. ETX).

### 3.3.8 Impact the Changed in the Area of Network

We investigate the performance of increasing the maximum range between two nodes in the network, in terms network throughput, and the PDR. We consider the following network conditions; M=40, $M_r$=16, N=20, BW=1 MHZ, $P_t$=0.1 W and D=4 KB, for different PU traffic loads.

## 3.5.8.1 Throughput Performance versus the Field Area

Figure 3.41 shows the throughput as function of area L changed. As L increases, the throughput for all schemes decreases. ETX outperforms distance for all situations. The maximum improvement achieves at idle-probability equals 0.1 is 34%, at idle-probability equals 0.5 is 5%, and at idle-probability equals 0.9 is 4%.

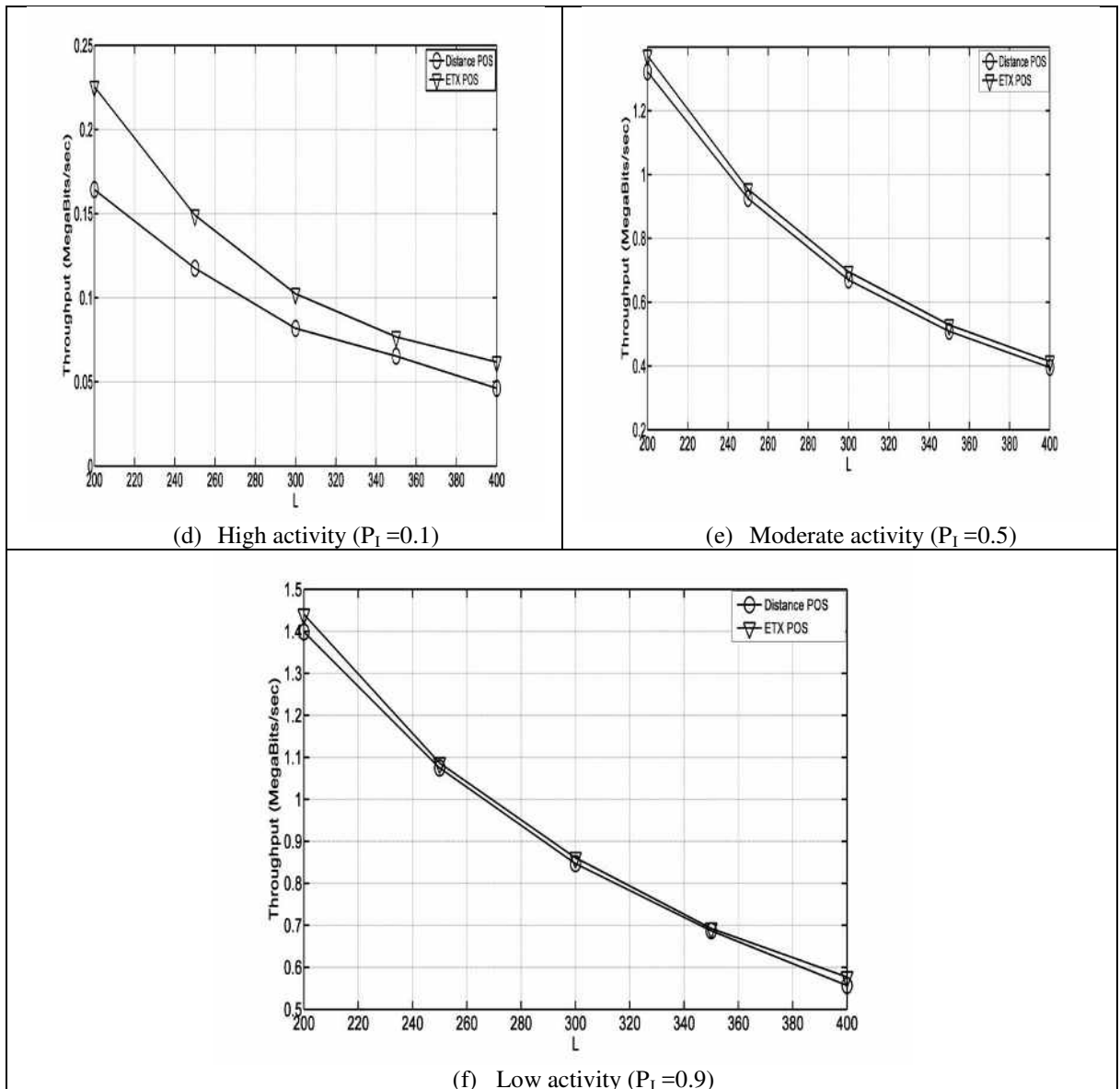

(d) High activity ($P_I = 0.1$)   (e) Moderate activity ($P_I = 0.5$)

(f) Low activity ($P_I = 0.9$)

**Figure 3.41:** Throughput vs. field area under different PU traffic (SPT).

## 3.5.8.2 The PDR Performance versus the Field Area

Figure 3.42 shows the PDR as function of L. The results decrease as L increases. ETX outperforms distance for all situations. The maximum improvement achieves at idle-probability equals 0.1 is 69.4%, at idle-probability equals 0.5 is 8.2%, and at idle-probability equals 0.9 is 3.2%.

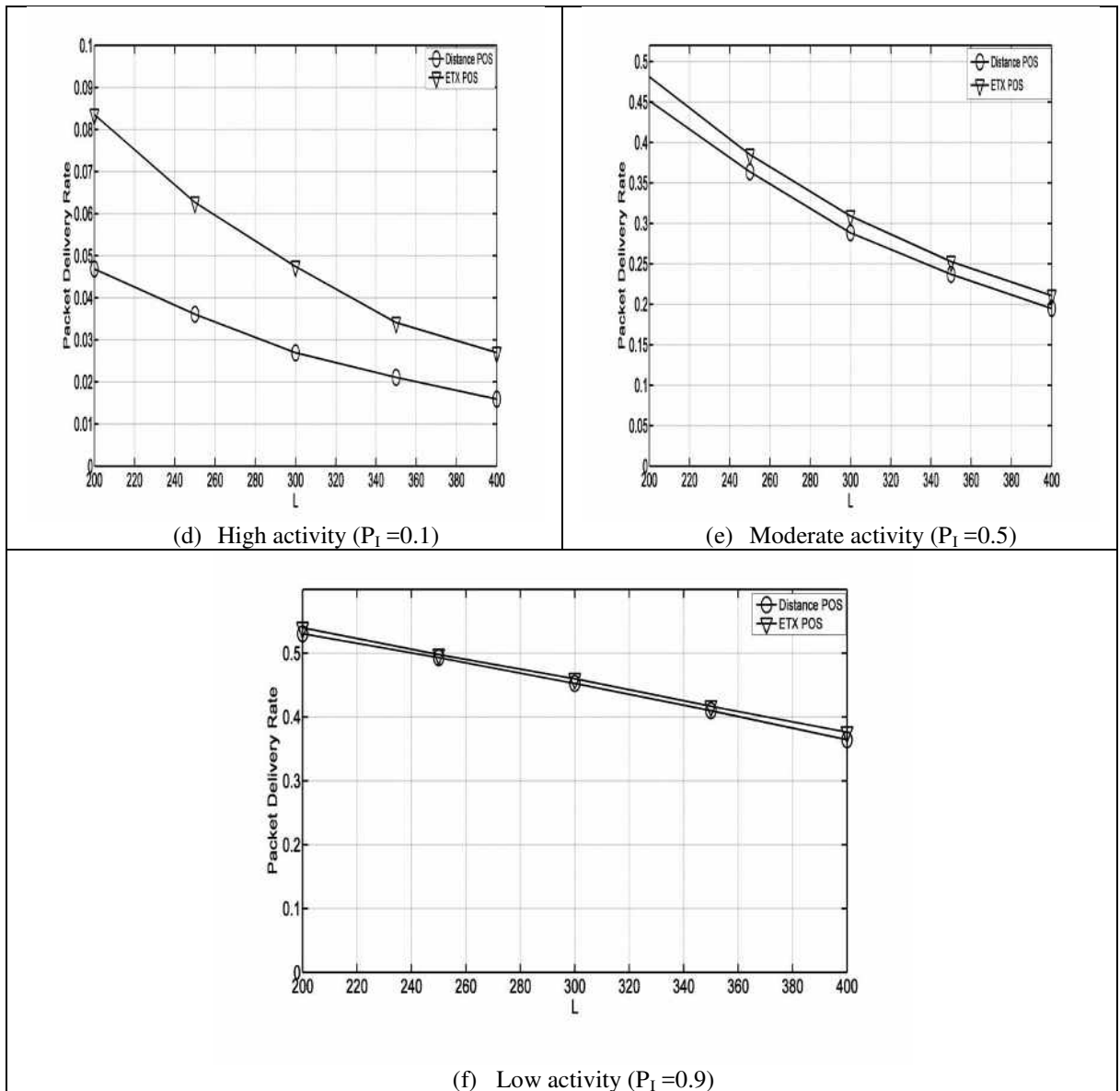

(d) High activity ($P_I$ =0.1)  (e) Moderate activity ($P_I$ =0.5)

(f) Low activity ($P_I$ =0.9)

**Figure 3.42:** The PDR vs. field area under different PU traffic (Distance vs. ETX).

## 3.5.9  Impact of PUs Traffic Loads

Figure 3.43 illustrates the performance of increased traffic loads on network throughput and PDR. Slightly improvement has observed regarding comparing ETX with distance as $P_I$ increases.

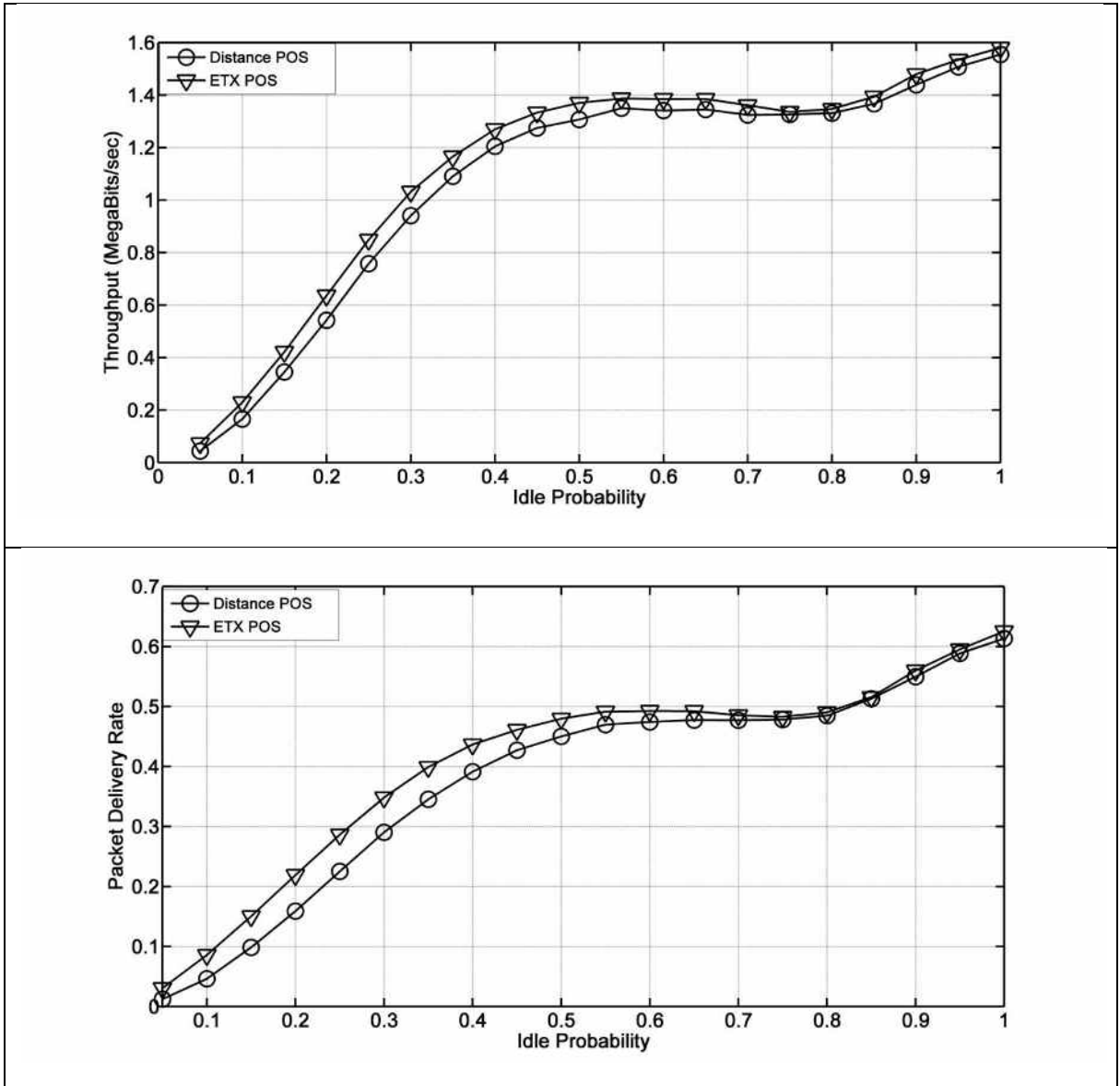

**Figure 3.43:** The PDR and Throughput vs. PU traffic (Distance vs. ETX).

# Chapter 4: Conclusion and Future Work

Routing and channel assignment design are challenging problems in multi-hop mobile Ad Hoc CRNs. Many attempts have been made to design efficient routing protocols, but none of them considers cross layer multi-hop routing protocol for mobile Ad Hoc CRNs. In this thesis, the multi-hop multicast routing protocol for mobile Ad Hoc CRNs using the shortest path tree (SPT) and minimum spanning tree (MST) in the path selection process and POS employs for the channel assignment has investigated. The main technical parameters of multicast and their advantages in increasing the throughput have reviewed.

The proposed algorithm has investigated with considering different network parameters that presents in section 3.1. Extensive simulation has carried out over all different scenarios that are considered in the proposed protocol.

Simulation results have summarized the following achievements:

- The proposed protocol that is implemented based on SPT tree protocol to protocol that implemented based on MST tree, the performance improvement has achieved in throughput by 40% and in the PDR performance by 118%, at higher activity of primary users.

- The proposed protocols POS, MASA, MDR, and RS, compared in terms of throughput and PDR. The maximum improvement has achieved at lower activity of primary users regarding to POS protocol. The POS achieves 6.3%, 64%, 73% and 5%, 78%, and 133% over throughput and PDR, respectively.

- The proposed protocol compared to previous protocol that are also based on joint path routing process and channel assignment, the improvement of throughput arrives to 34%, 21.4%, 33.4%, for increasing packet size, number of primary channels and transmission power, respectively. The PDR performance is 81%, 52%, 69%, at increasing packet size, number of primary channels and transmission power, respectively.

SPT outperforms MST and POS outperforms other protocols in terms of throughput and PDR, at all network parameters and for high activity of primary users. This becomes clear from the investigations presented in Chapter 3.

# الخوارزميه الديناميكيه للارسال المتعدد للشبكات الانتهازيه: باستخدام اداة قياس عدد مرات الارسال المتوقعه.

اعداد:

رشا زياد محمد ابوسمره (2013976009).

المشرف: د. هيثم بني سلامه.

## الملخص

الإدراك الراديوي (CR) تكنولوجيا ذكيه تمكن نظام الاتصال اللاسلكي من توفير حل فعال لاستخدام الطيف الراديوي الغير مستغل بشكل فعال عن طريق الوصول للطيف بشكل ديناميكي وانتهازي. لتصميم شبكات CR، يجب زيادة سرعة وأداء الشبكة الغير مرخصه مع حماية أداء الشبكات الأولية المرخصه (PRNs) والحفاظ على التداخل بين المستخدمين المرخصين (PUs) والمستخدمين الغير مرخصين (CUs) ضمن حد معين. في هذا العمل، طورنا خوارزمية التوجيه المتعدد التي تقوم على قياس العدد المتوقع لمرات حدوث الارسال واستخدمنا (ETX) كمقياس لايجاد شجرة (MST) وشجرة (SPT) لتحديد طريقة اختيار المسار المناسب لاحمال مختلفه في (CRN) واستخدمنا احتمال النجاح (POS) لاختيار القناة والذي يعتمد على عاملين مهمين: اولا الزمن اللازم لارسال المعلومه و ثانيا الزمن المتوقع للقناة ان تكون موجوده. والهدف الرئيسي من هذه الخوارزمية هو تقليل عدد مرات الارسال المتوقعه (مع إعادة الإرسال) اللازمة لارسال حزم البيانات إلى مجموعة محددة، وضمان وصول هذه البيانات عبر القناة التي تم اختيارها بنجاح، هذا المقياس لديه القدره على التلائم مع بيئة (CRNs) الحيويه ودائمة التغيير وذلك بسبب النشاط الغير متوقع للشبكه المرخصه فتتغير القنوات المتوفره بشكل دائم ومستمر. ويقترح البروتوكول طريقه لزيادة الانتاجيه لشبكة (CRN) عن طريق تقليل الزمن اللازم لارسال البيانات من خلال تحديد أفضل مسار بين جميع المسارات المتوفره بين المصدر و الأماكن اللتي نريد ايصال البيانات لها. وللتحقق من فعالية الخوارزميه المقترحه، أجريت تجارب محاكاة واسعه باستخدام الماتلاب. بينت النتائج ان الخوارزميه المقترحه تحقق تحسين كبير في الإنتاجية من حيث سرعة نقل وإيصال الحزمة مقارنة بالبروتوكولات التوجيه المتعدد الأخرى الموجوده.

كلمات مفتاحيه: الراديوات الادراكيه، الشبكات اللاسلكيه، توزيع القنوات، الارسال المتعدد، توقع عدد مرات الارسال.